\pdfoutput=1
\documentclass[useAMS,usenatbib]{mn2e}
\usepackage{subfig}
\usepackage{graphicx}
\usepackage{caption}
\usepackage{float}
\usepackage{morefloats}
\usepackage{amsmath}
\usepackage{amssymb}
\usepackage{footnote}

\title[ISW recovery]{On the recovery of ISW fluctuations using large-scale structure tracers and CMB temperature and polarization anisotropies}
\author[Bonavera et al.]{L. Bonavera\thanks{E-mail: laurabonavera@gmail.com;}, R. B. Barreiro, A. Marcos-Caballero, P. Vielva\\
Instituto de F{\'\i}sica de Cantabria (CSIC-Univ. Cantabria), Avda. de Los Castros s/n, E-39005 - Santander, Spain}
\begin{document}

\date{}


\maketitle

\label{firstpage}

\begin{abstract}
In this work we present a method to extract the signal induced by the integrated Sachs-Wolfe (ISW) effect in the cosmic microwave background (CMB). It makes use of the Linear Covariance-Based filter introduced by Barreiro et al.,  and combines CMB data with any number of large-scale structure (LSS) surveys and lensing information. It also exploits CMB polarization to reduce cosmic variance. The performance of the method has been thoroughly tested with simulations taking into account the impact of non-ideal conditions such as incomplete sky coverage or the presence of noise. 
In particular, three galaxy surveys are simulated, whose redshift distributions peak at low ($z \simeq 0.3$), intermediate ($z \simeq 0.6$) and high redshift ($z \simeq 0.9$). The contribution of each of the considered data sets as well as the effect of a mask and noise in the reconstructed ISW map is studied in detail. When combining all the considered data sets (CMB temperature and polarization, the three galaxy surveys and the lensing map), the proposed filter successfully reconstructs a map of the weak ISW signal, finding a perfect correlation with the input signal for the ideal case and around 80 per cent, on average, in the presence of noise and incomplete sky coverage. 
We find that including CMB polarization improves the correlation between input and reconstruction although only at a small level. Nonetheless, given the weakness of the ISW signal, even modest improvements can be of importance. In particular, in realistic situations, in which less information is available from the LSS tracers, the effect of including polarisation is larger. 
For instance, for the case in which the ISW signal is recovered from CMB plus only one survey, and taking into account the presence of noise and incomplete sky coverage, the improvement in the correlation coefficient can be as large as 10 per cent.

\end{abstract}


\section{Introduction}
\label{sec:intro}
Measurements of the cosmic microwave background (CMB), galaxy clusters, Type Ia supernovae or baryonic acoustic oscillations (BAOs) indicate an accelerated expansion of the Universe \citep[see][for a review]{WEI13} caused by the dark energy \citep{PEE03}. The essence of dark energy is one of the most intriguing problems in modern cosmology. Up to date the observations are in good agreement with the prediction of the existence of a cosmological constant, whose equation of state is given by $p=-\rho$.

The accelerated expansion causes the decaying of potentials, resulting in the distortion of the CMB temperature, called the integrated Sachs-Wolfe effect \citep[ISW;][]{SAC67}: the CMB photons entering overdense regions are blue-shifted and those entering underdense regions are red-shifted. For this reason the ISW effect provides a complementary tool to probe dark energy or models of modified gravity \citep[e.g.][]{SON07,POG08}. The ISW signal is too small to be directly identified in the CMB spectrum, but it can be measured by correlating the CMB anisotropies and the large-scale structure \citep[LSS;][]{CRI96}: maps of the distributions of galaxies are correlated with the ISW signal because they are tracers of the gravitational potential causing the ISW effect.

The measurement of the ISW effect became possible with the Wilkinson Microwave Anisotropy Probe \citep[WMAP][]{BEN13} data, with many works reporting its detection \citep[e.g.,][]{BOU04,FOS03,VIE06,HO08,GIA08,SCH12}. More recent works use {\it Planck} data \citep[e.g.,][]{PLA_ISW,PLA_ISW2}. The ISW can also be studied through the bispectrum of the ISW-lensing \citep{HU02}, which was applied for the first time in \cite{PLA_ISW}. Finally, the ISW can also be studied by stacking the CMB fluctuations on the positions of known large-scale superclusters \citep[e.g.,][]{GRA08, PLA_ISW}. In fact these staking analysis indicate some tensions with $\Lambda$CDM \citep[e.g.,][]{NAD12,ILI13}

Complementarily to these techniques which detect the ISW statistically, more recent works have presented methods to reconstruct an actual map of the ISW contribution to the CMB sky. 
This is especially important to understand the universe at large scales, where the ISW effect has its more relevant contribution. In particular, if we can disentangle the primordial CMB anisotropies from those generated at late times, this could shed light on the origin of the CMB anomalies at large scales.
With this aim, \cite{BAR08} proposed the so-called Linear Covariance-Based (LCB) filter to reconstruct the ISW signal exploiting the correlation between the CMB temperature map and one survey tracing the LSS. 
This filter was applied to LSS and CMB data by \cite{BAR13} on WMAP and by \cite{PLA_ISW} on Planck. More recently, \cite{MAN14} have developed an extension of the LCB filter to reconstruct the ISW effect from CMB and several surveys simultaneously. The technique was applied to obtain an ISW map from CMB, lensing and the NVSS galaxy survey. Other approaches, where only the map of a galaxy number density field traced by a particular Large Scale Structure (LSS) survey is used, have also been proposed: \cite{GRA08} on LRGs from SDSS-DR6, or \cite{FRA10} and \cite{DUP11} on 2MASS.

In this work we present a further generalization of the LCB filter to extract a map of the ISW signal, which allows not only the inclusion of an arbitrary number of surveys but also of CMB polarization information.
We perform a thorough study of the performance of the method to reconstruct the ISW effect using a total of three simulated LSS surveys, lensing information and CMB intensity and polarization. In particular, we investigate the degradation in the reconstruction due to noise and/or masking, as well as the results obtained for different combinations of the available data. This method has already been applied to the most recent Planck CMB temperature data in combination with several LSS surveys to produce maps of the ISW effect \citep{PLA_ISW2}, which are publicly available at the Planck Legacy Archive\footnote{http://pla.esac.esa.int/pla/}.

The paper is organized as follows. In Section \ref{sec:met} we present the generalisation of the LCB filter while in Section \ref{sec:simu} we describe the considered CMB and LSS simulations. The results regarding the reconstruction of the ISW are given in Section \ref{sec:rec}. Finally, discussion and conclusions are presented in Section \ref{sec:con}.

\section{Methodology}
\label{sec:met}

In order to recover the weak ISW effect with the best possible signal-to-noise ratio, it becomes necessary to develop a method that combines all the available information about this signal, using the CMB and LSS tracers. With this aim, we present an extension of the LCB filter given in \citet{BAR08}, which originally combined CMB temperature data and one LSS tracer. We extend the method in two ways: first by considering any number of available LSS tracers (including a lensing potential map derived from CMB data) and second by including CMB polarization data.

Before presenting the extended method, we will briefly summarise the original LCB filter (see \citealt{BAR08,BAR13} for details). Note that in order to construct this filter, the covariance matrix $\mathbfss{C}(\ell)$ for the ISW and LSS tracer is assumed to be known. The estimated ISW map $\hat{s}(\ell,m)$ at each harmonic mode is given by:
\begin{equation}
\label{eq:rec}
\begin{split}
\hat{s}(\ell,m)=\frac{L_{12}}{L_{11}}g(\ell,m)+\\
\frac{L_{22}^2}{L_{22}^2+C_{\ell}^n}\left( d(\ell,m)-\frac{L_{12}}{L_{11}}g(\ell,m)\right),
\end{split}
\end{equation}
where $d(\ell,m)$ and $g(\ell,m)$ are the harmonic coefficients of the CMB map and LSS tracer respectively, and $C_{\ell}^n$ is the power spectrum of the CMB signal without including the ISW. The matrix $\mathbfss{L}$ is the Cholesky decomposition of the covariance matrix $\mathbfss{C}$, that is: $\mathbfss{C}(\ell)=\mathbfss{L}(\ell)\mathbfss{L}^T(\ell)$, where $\mathbfss{L}(\ell)$ is a lower triangular matrix.  Note that the first term of the equation corresponds to the part of the LSS tracer which is correlated with the ISW, while the second term is given by applying a Wiener filter to a modified CMB map. Note that, by 
construction, this modified map does not have correlations with the considered LSS tracer.

The expected value of the power spectrum of the estimated ISW is given by:
\begin{equation}
\label{eq:cl_lcb}
\left< C_\ell^{\hat{s}} \right>= \frac{(C_\ell^{sg})^2\left(\left|\mathbf{C}(\ell)\right|+C_\ell^g
C_\ell^n\right)+\left|
\mathbf{C}(\ell)\right|^2}{C_\ell^g\left(\left|\mathbf{C}(\ell)\right|+C_\ell^g C_\ell^n\right)},
\end{equation}
where $C_\ell^g$ is the auto-spectrum of the considered LSS tracer (including a possible contribution from Poissonian noise), $C_\ell^{sg}$ is the cross-power between the ISW and the catalogue of galaxies and $\left| \mathbfss{C}(\ell)\right|$ is the determinant of the covariance matrix.

It is well known that the Wiener filter introduces a bias in the power spectrum of the reconstruction towards values lower than those of the true signal. Therefore, the proposed filter will also be biased \citep{BAR08}. The bias will be smaller for larger cross-correlations between the ISW signal and the considered LSS tracers, since in this case the relative weight of the Wiener filter is smaller. Obviously, since we assume knowledge of the correlation model, the bias can be simply estimated as $C{_\ell}^s - \left< C_\ell^{\hat{s}} \right>$, where $C{_\ell}^s$ is the theoretical power spectrum of  the ISW signal, and therefore corrected in the reconstructed power spectrum.

\subsection{LCB filter generalization to n tracers}
\label{sec:LBC}

\begin{figure*}
 \centering
  \subfloat{\includegraphics[width=8.0cm]{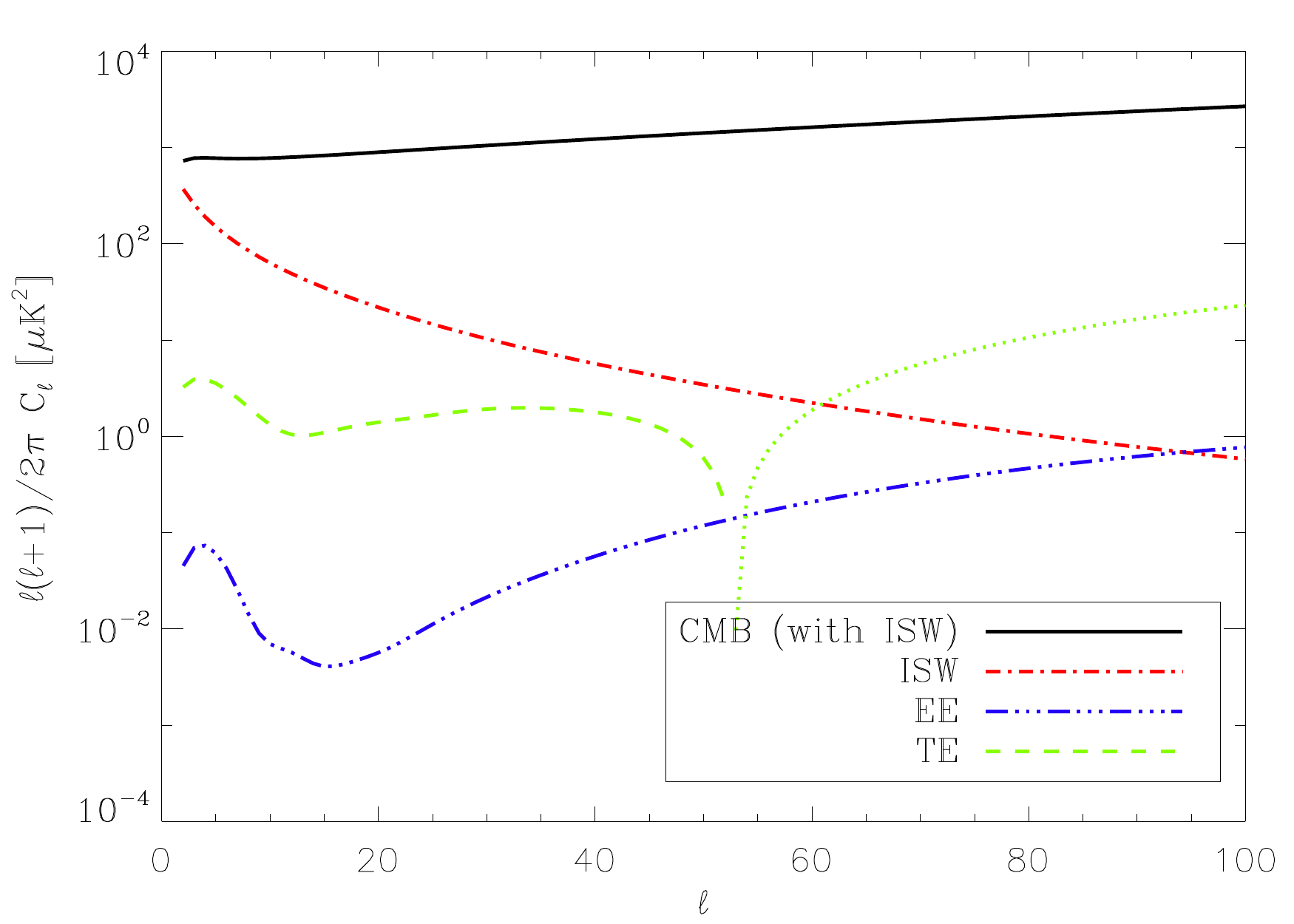}}\quad
  \subfloat{\includegraphics[width=8.0cm]{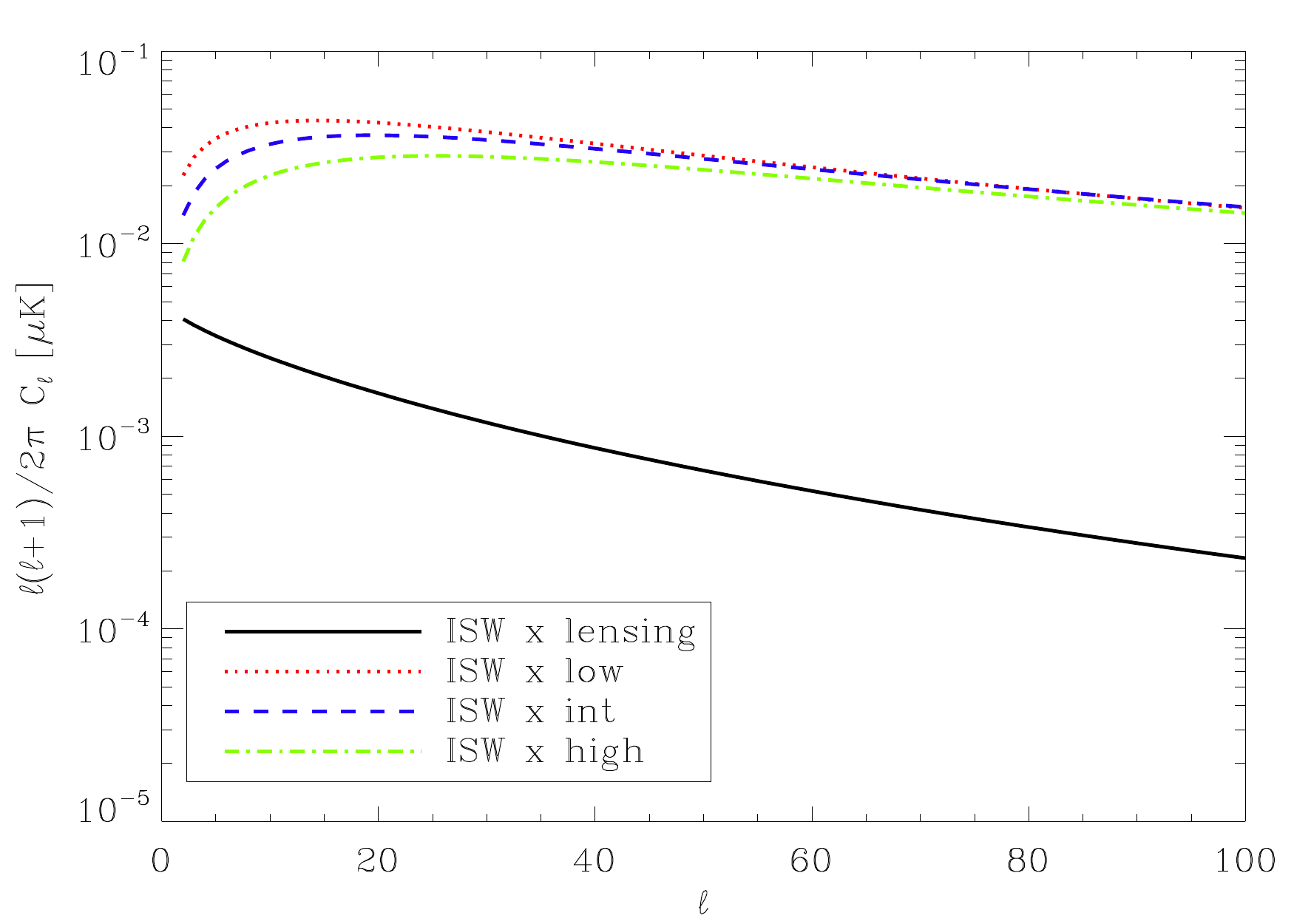}}\\
  \subfloat{\includegraphics[width=8.0cm]{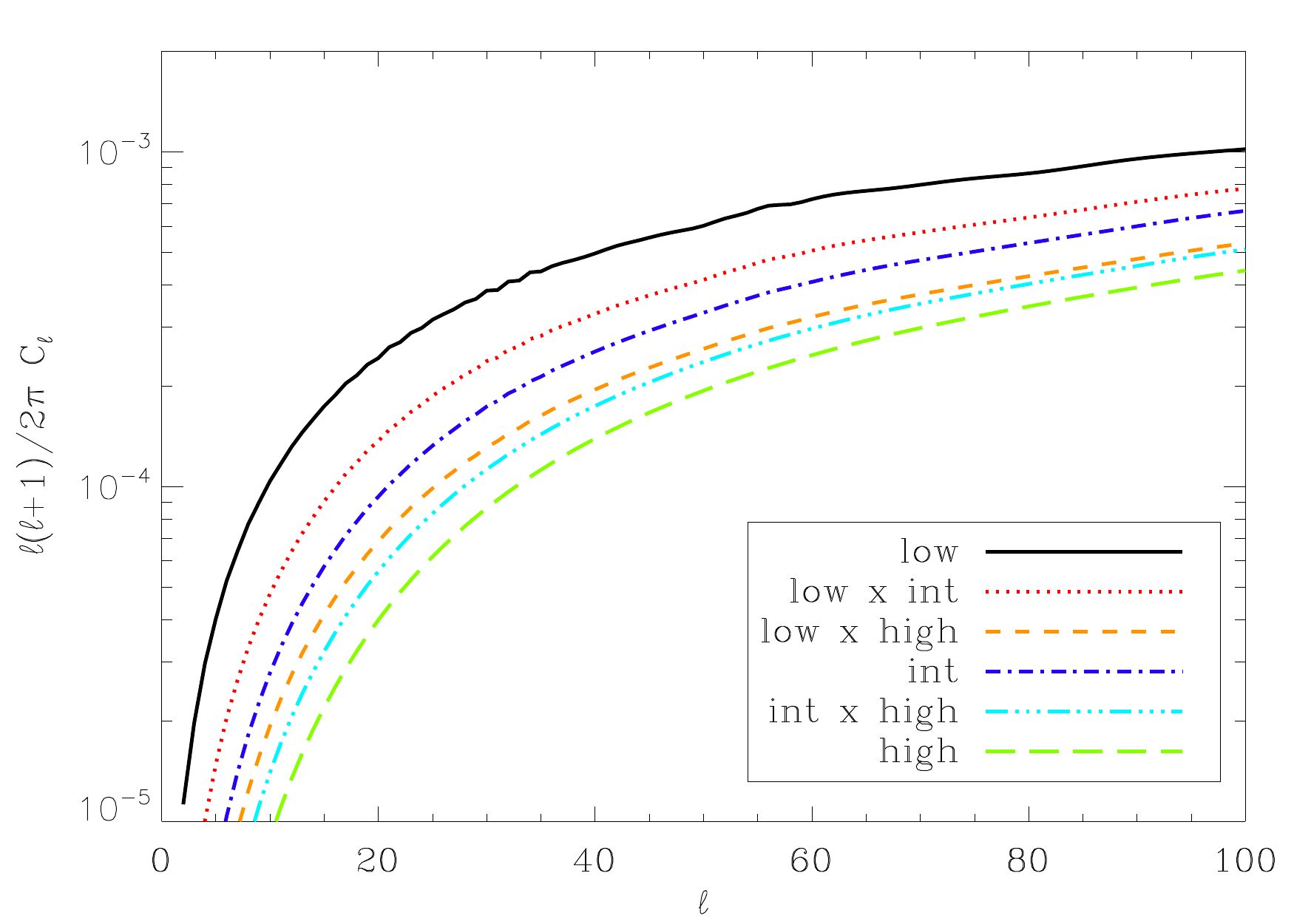}}\quad
  \subfloat{\includegraphics[width=8.0cm]{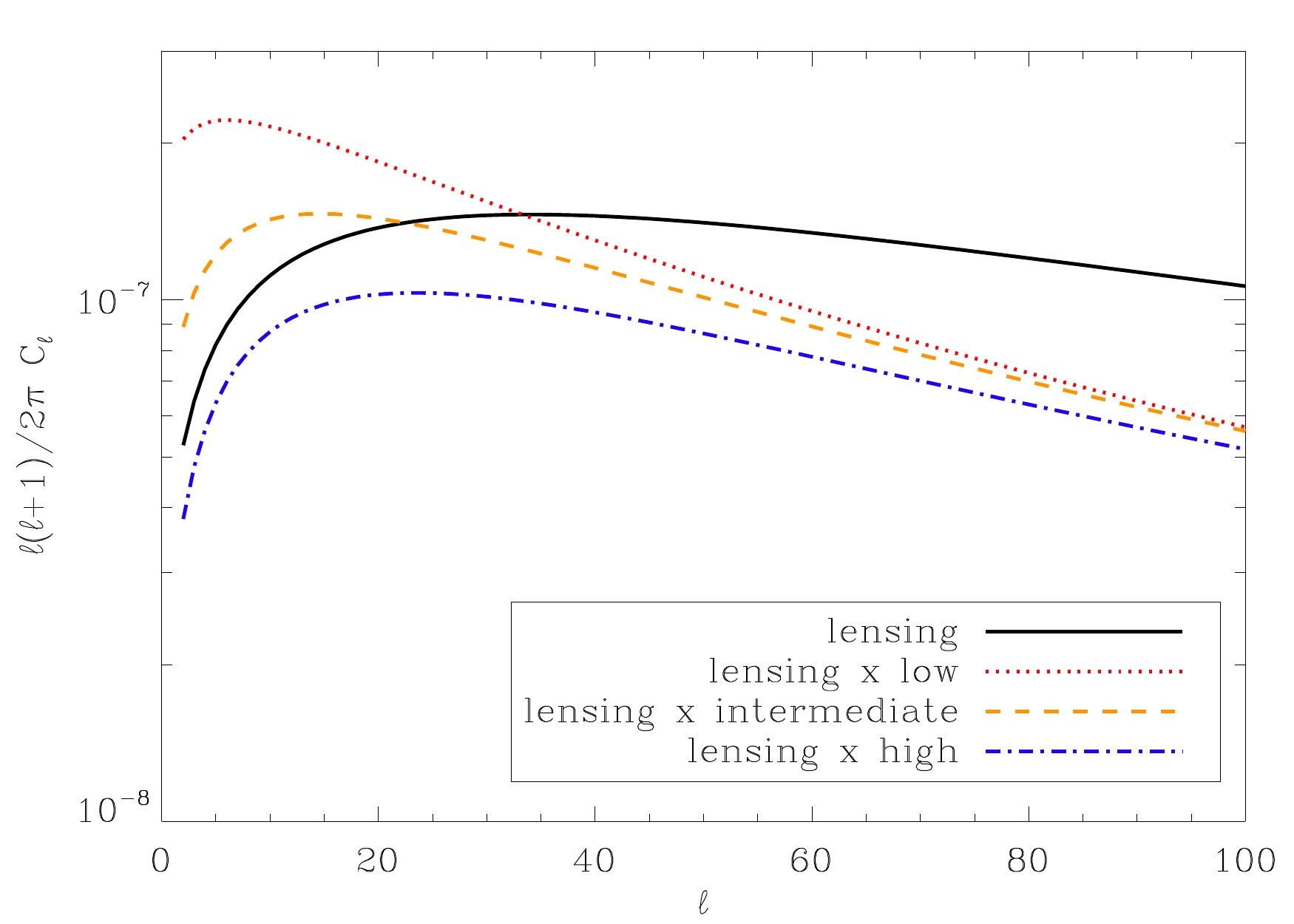}}\\
 \caption{Power-spectra used for the simulations. Top left: CMB related spectra (in particular the green line is the absolute value of the TE power spectrum where the dotted line stands for the negative part), top right: cross-spectra between ISW and surveys; bottom left: auto and cross-spectra for LSS components; bottom right: auto and cross-spectra for lensing.}
 \label{fig:cls_teo}
\end{figure*}

In this section we present the generalization of the LCB filter to use any number of tracers to reconstruct the ISW (see also \citealt{MAN14}, and \citealt{PLA_ISW2}).
As for the previous case, the covariance matrix $\mathbfss{C}(\ell)$ of the ISW and the LSS surveys is assumed to be known. It is convenient to arrange the elements of the covariance matrix in the following order: the first $n$ signals are the LSS surveys and the last one is the ISW. We then construct the $\mathbfss{L}(\ell)$ lower triangular matrix obtained with the Cholesky decomposition of $\mathbfss{C}(\ell)$.

In the case of $n$ surveys, the analogues of Eq. \ref{eq:rec} becomes:
\begin{equation}
\label{eq:rec_n}
\begin{split}
\hat{s}(\ell,m) = \sum\limits_{i=1}^{n} \left[ L_{it} \left(\sum\limits_{j=1}^{n}(L^{-1})_{ij}g_j(\ell,m)\right)\right]  + \frac{L_{tt}^2}{L_{tt}^2+C_{\ell}^n}\\
\times \left\lbrace  d(\ell,m)-\sum\limits_{i=1}^{n}\left[ L_{it} \left( \sum\limits_{j=1}^{n}(L^{-1})_{ij}g_j(\ell,m)\right)\right] \right\rbrace ,
\end{split}
\end{equation}
where: $t=n+1$ is the order of the $\mathbfss{C}(\ell)$ and $\mathbfss{L}(\ell)$ matrices, $g_{j=1,n}(\ell,m)$ are the harmonic coefficient of the $n$ surveys and $d(\ell,m)$ are the harmonic coefficients of the CMB temperature map. Note that, as expected, when $n=1$, the previous equation defaults to the case of the original LCB filter (given by Eq.~\ref{eq:rec}).

Note that non-idealities as Poissonian noise or incomplete sky can also be taken into account in the previous filter. If Poissonian noise needs to be considered in any of the surveys, its contribution is simply added into the corresponding auto-spectrum. If any or several of the data have incomplete sky coverage, the corresponding auto- and cross-spectra are modified {\it a la} MASTER \citep{HIV02}, in order to include the presence of a mask, i.e., {\it masked} power spectra are introduced in Eq.~(\ref{eq:rec_n}).

The expected value of the power spectrum of the reconstruction is now given by
\begin{align}
\label{eq:cl_lbc_n}
\left< C_\ell^{\hat{s}} \right> &= \sum\limits_{j=1}^{n}\sum\limits_{k=1}^{n}\left[ \left( \sum\limits_{i=1}^{n}L_{it}(L^{-1})_{ji}\right)\times \right. \nonumber \\
&\qquad {} \left. \left( \sum\limits_{i=1}^{n}L_{it}(L^{-1})_{ki}\right) C_{\ell}^{g_jg_k} \right]+\frac{\left( L_{tt}^2\right)^2 }{L_{tt}^2+C_{\ell}^n}
\end{align}
%
\begin{figure}
 \centering
  \subfloat{\includegraphics[width=8.0cm]{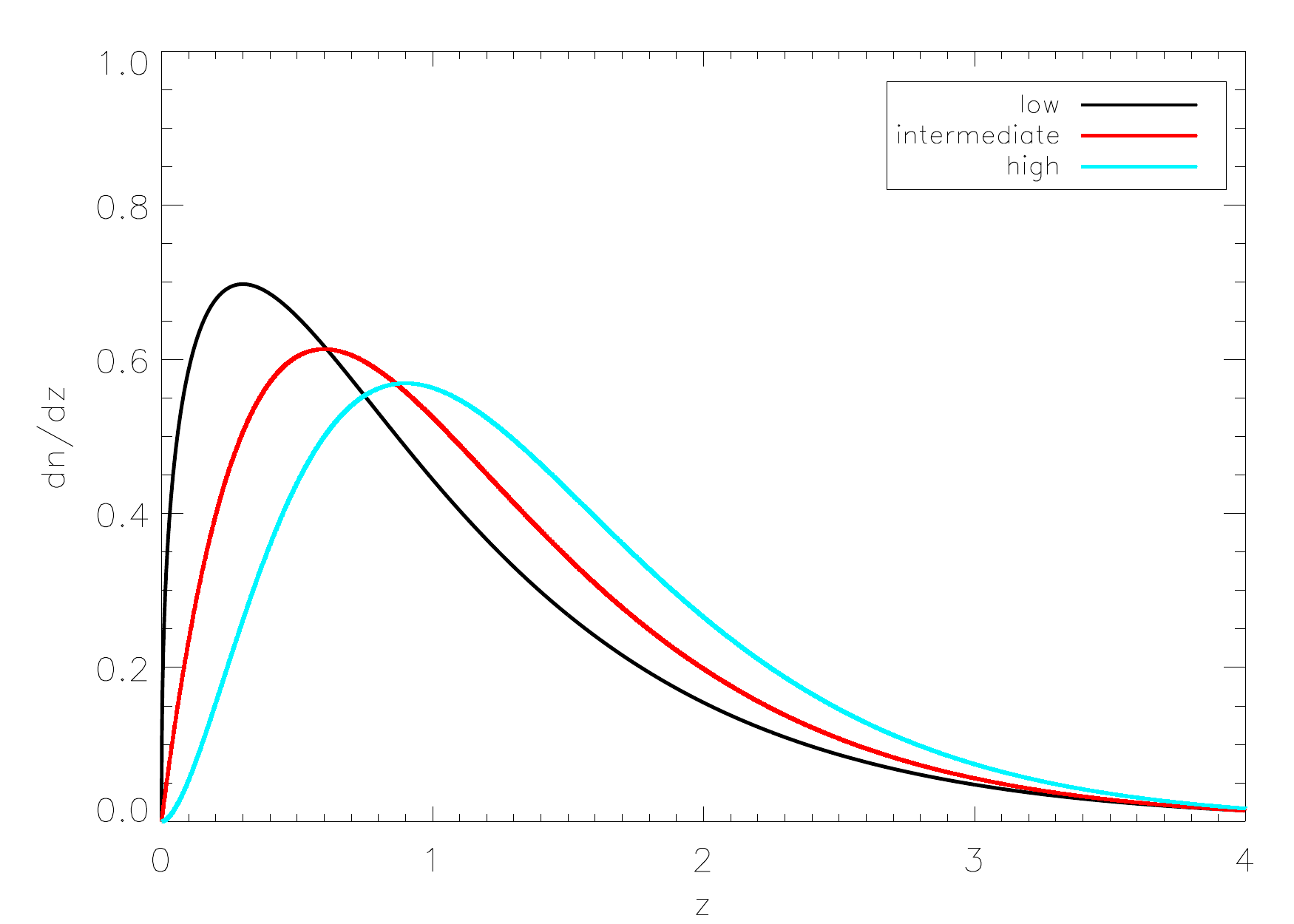}}\\
 \caption{Galaxy redshift distributions (dn/dz) adopted for each galaxy survey: peaking at low-z (black), intermediate-z (red) and high-z (blue).}
 \label{fig:redshift}
\end{figure}
which, as before, is biased with respect to the expected value and can be corrected if needed.
In this formalism the lensing information is handled as an additional LSS tracer.

\subsection{Exploiting polarization information}
\label{sec:pol}

 Although some level of correlation could, in principle, be present between the E-mode of CMB polarization and the ISW effect, this has been proved to be small \citep{COO06}. Therefore, in a first approximation, we can consider both signals to be uncorrelated. This fact, together with the correlation between the E-mode of polarization and the CMB temperature map,  allows one to make use of CMB polarization information in order to improve the recovery of the ISW signal. Following \citet{FRO09} we need to subtract from the data the information about the temperature fluctuations correlated with the polarization data. This is given by:
\begin{equation}
\label{eq:TE}
t_E(\ell,m)=\frac{C_{\ell}^{TE}}{C_{\ell}^{EE}}e(\ell,m)
\end{equation}
where $C_{\ell}^{EE}$ and $C_{\ell}^{TE}$ are the E-mode power spectrum and the temperature and E-mode cross-correlation of the CMB, respectively, and $e(\ell,m)$ are the harmonic coefficients of the E map. Note that a possible instrumental noise contribution would be included in $C_{\ell}^{EE}$, since the large scales of the T map can be assumed to be noise-free, and correlation between the T and E noise are expected to be zero.
When including polarization information, Eq. \ref{eq:rec_n} becomes:
\begin{equation}
\label{eq:rec_n_E}
\begin{split}
\hat{s}(\ell,m) = \sum\limits_{i=1}^{n}\left[ L_{it} \left( \sum\limits_{j=1}^{n}(L^{-1})_{ij}g_j(\ell,m)\right)\right]\\
+ \frac{L_{tt}^2}{L_{tt}^2+C_{\ell}^n-\frac{\left( C_{\ell}^{TE}\right) ^2}{C_{\ell}^{EE}}}\\
\times \left\lbrace  d(\ell,m)-\sum\limits_{i=1}^{n}\left[ L_{it} \left( \sum\limits_{j=1}^{n}(L^{-1})_{ij}g_j(\ell,m)\right)\right]-t_E(\ell,m)\right\rbrace, 
\end{split}
\end{equation}
and Eq. \ref{eq:cl_lbc_n} reads:
\begin{align}
\label{eq:cl_lbc_n_E}
\left< C_\ell^{\hat{s}} \right> &= \sum\limits_{j=1}^{n}\sum\limits_{k=1}^{n}\left[ \left( \sum\limits_{i=1}^{n}L_{it}(L^{-1})_{ji}\right) \times  \right. \\
&\qquad {} \left. \left( \sum\limits_{i=1}^{n}L_{it}(L^{-1})_{ki}\right) C_{\ell}^{g_kg_j}\right]+\frac{\left( L_{tt}^2\right)^2 }{L_{tt}^2+C_{\ell}^n-\frac{\left( C_{\ell}^{TE}\right) ^2}{C_{\ell}^{EE}}}. \nonumber 
\end{align}
Note that by subtracting $t_E$ from the data, we obtain a modified CMB temperature map with lower intrinsic variance than the original map. Therefore, including polarization information should reduce the error in the estimation of the ISW effect, although at a moderate level \citep[see][]{FRO09}.

\section[]{Simulated maps}
\label{sec:simu}
\begin{figure*}
 \centering
   \subfloat{\includegraphics[width=5.0cm, angle=90]{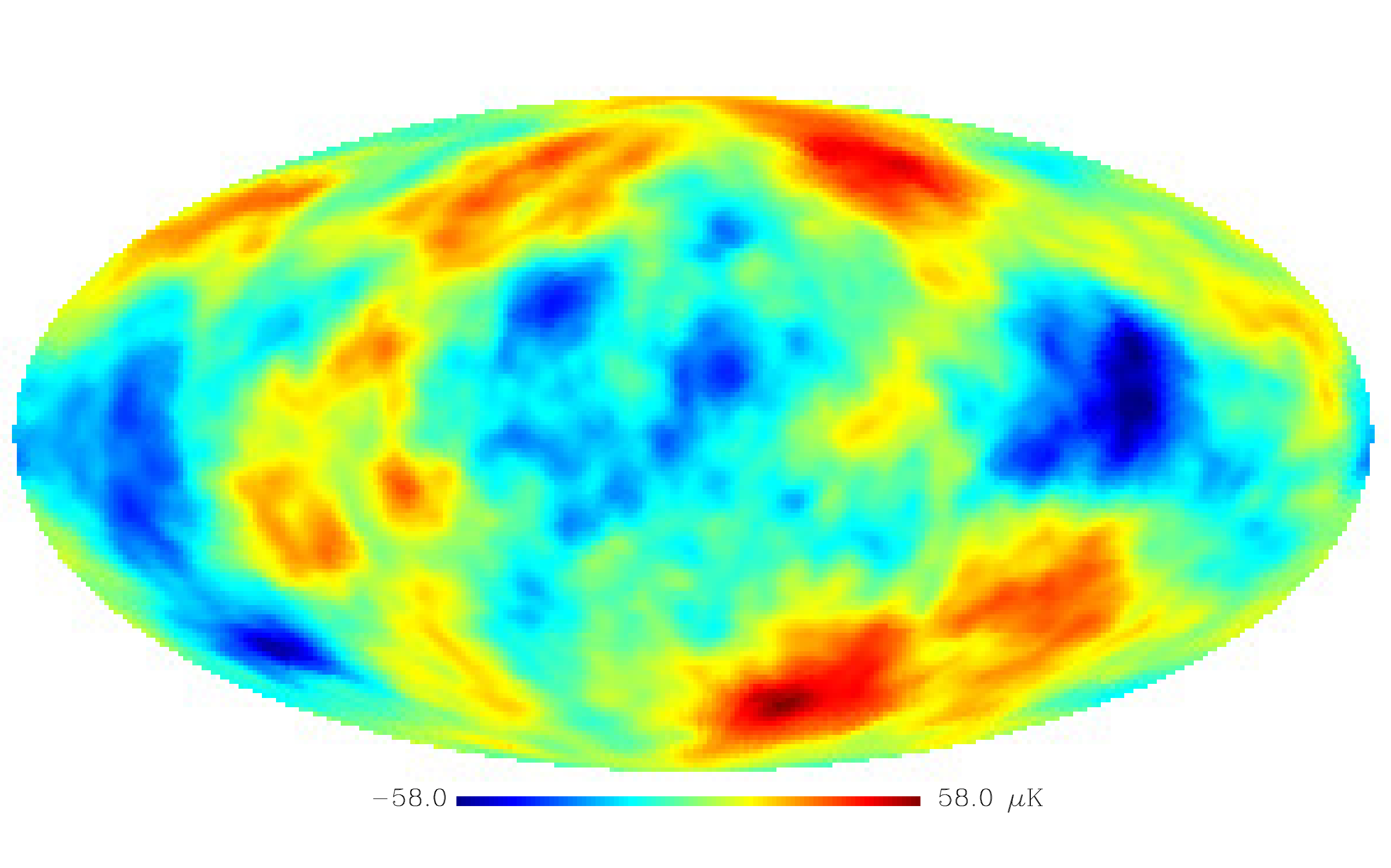}}\\
   \subfloat{\includegraphics[width=5.0cm, angle=90]{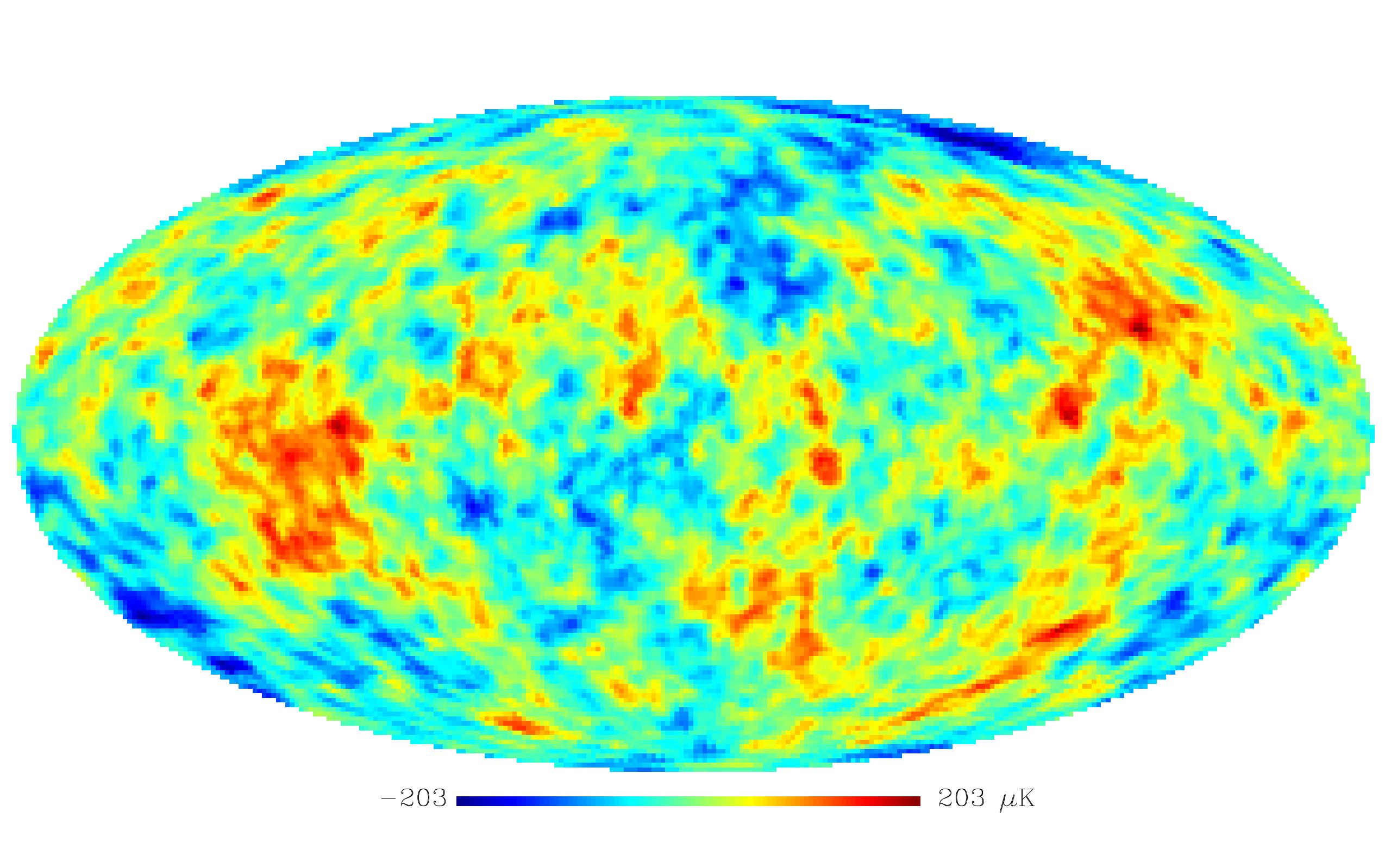}}\quad
   \subfloat{\includegraphics[width=5.0cm, angle=90]{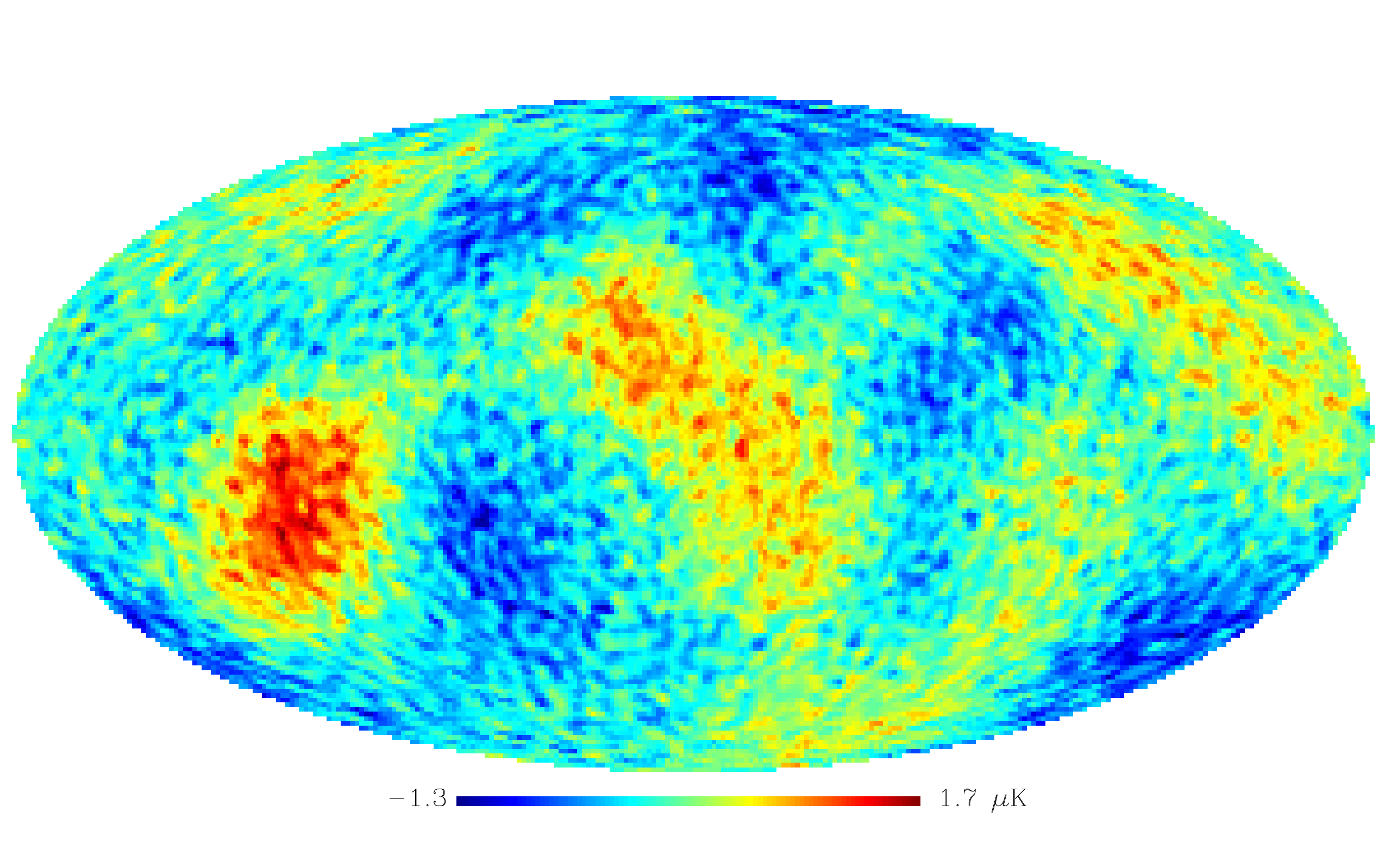}}\\
   \subfloat{\includegraphics[width=5.0cm, angle=90]{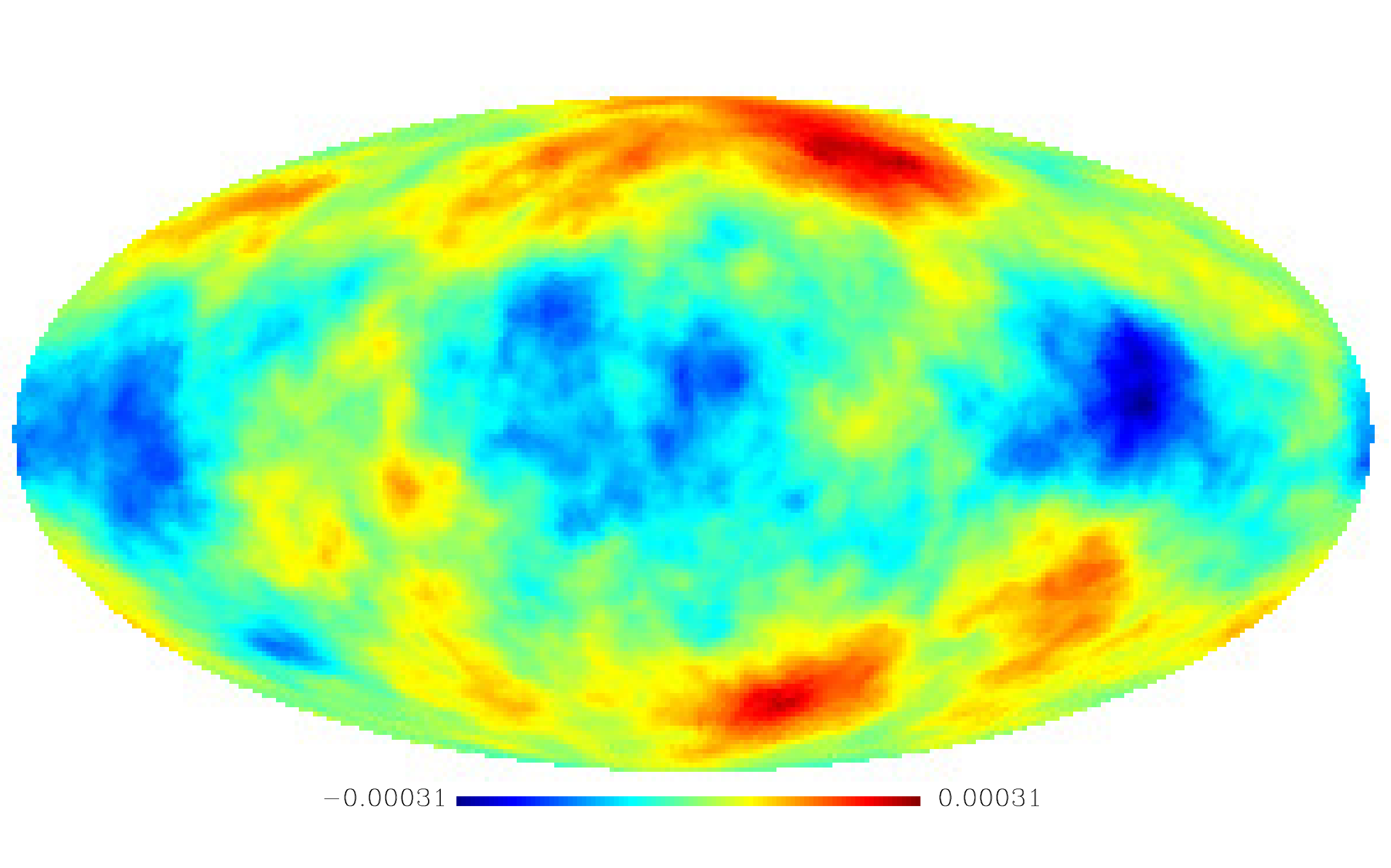}}\quad
   \subfloat{\includegraphics[width=5.0cm, angle=90]{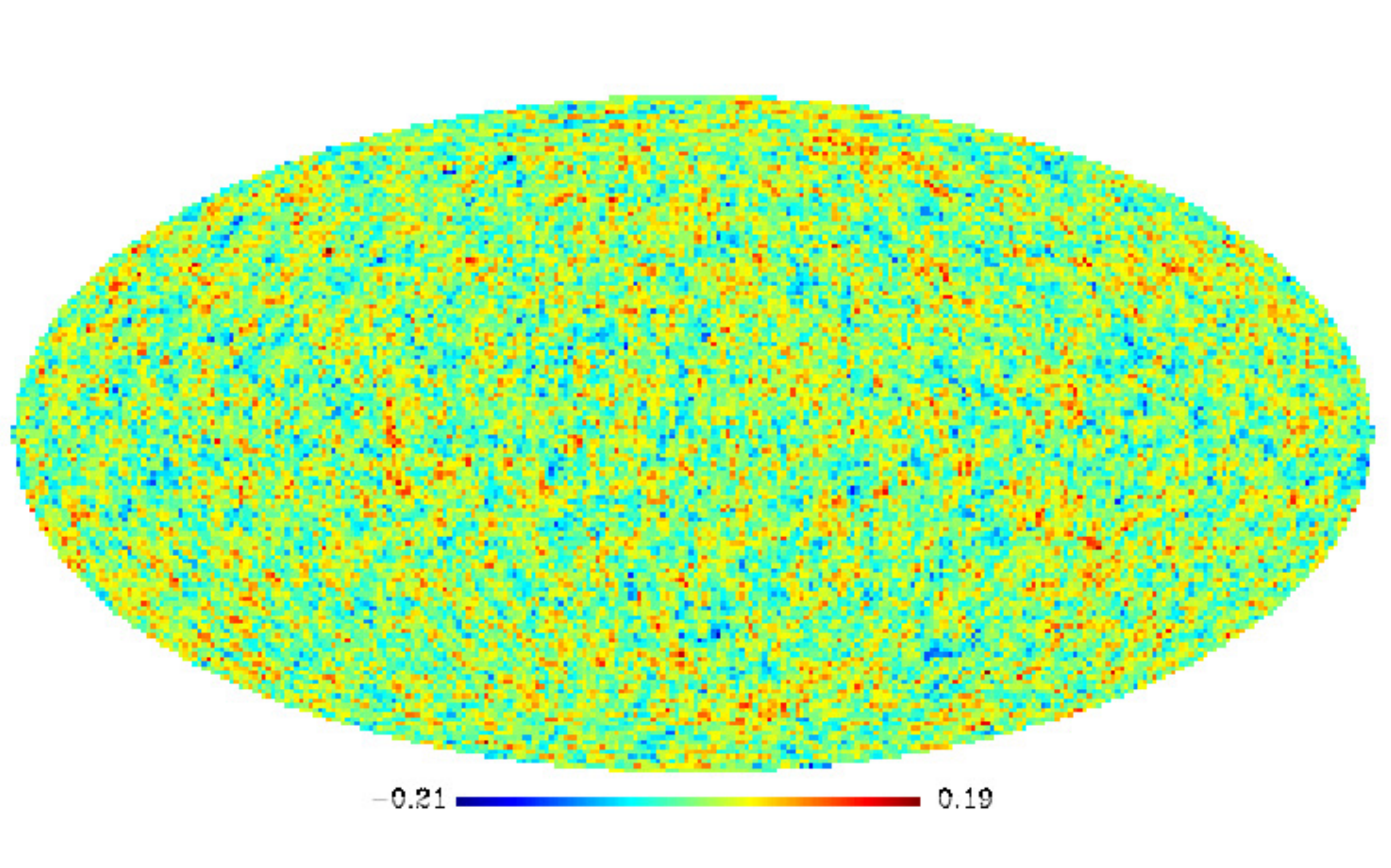}}\\
   \subfloat{\includegraphics[width=5.0cm, angle=90]{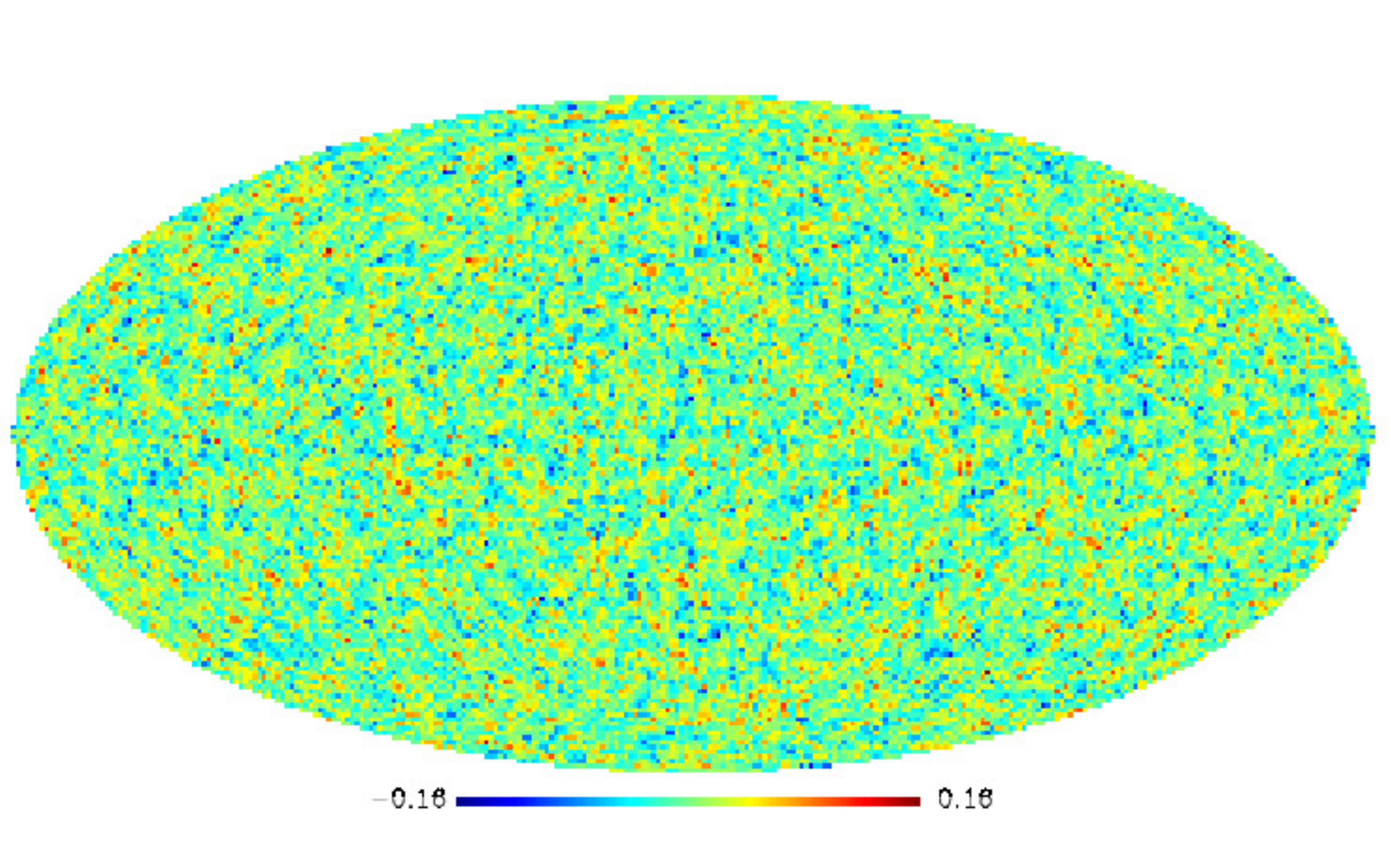}}\quad
   \subfloat{\includegraphics[width=5.0cm, angle=90]{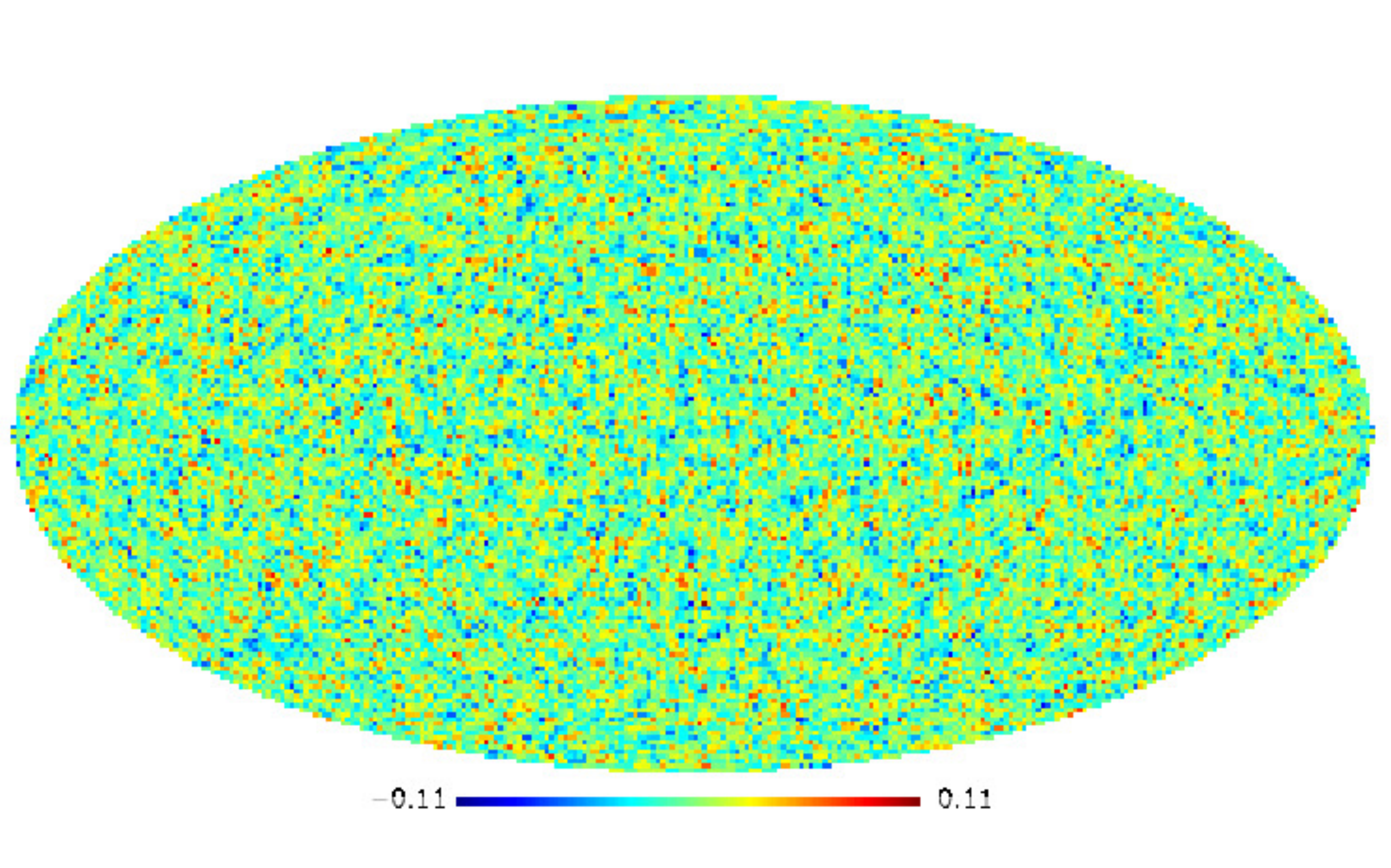}}\\
 \caption{Reference simulated maps for the ideal case (i.e., full-sky and without noise). Top: ISW signal. Left column (from top to bottom): CMB temperature map (including the ISW contribution), lensing potential map, survey with density distribution peaking at intermediate redshift. Right column: E-mode CMB map, low- and high-z peaking distribution surveys. CMB (intensity and polarization) and ISW maps are in units of $\mu$K while the other maps are dimensionless. The survey maps represent the galaxy density number fluctuations.}
 \label{fig:simu_maps}
\end{figure*}

\begin{savenotes}
\begin{table}
 \centering
 \begin{tabular}{@{}lcccc@{}}
 \hline
      &  $f_{sky}$ (\%) & Noise & $ z_0 $ & $\alpha$ \\
\hline
CMB (T) & 78.2 & - & - & - \\
CMB (E) & 78.2 & $3.3\times10^{-4}$ & - & - \\
\hline 
lensing & 78.2 & $1.2\times10^{-4}$ & - & - \\
low & 50.0   & $2.0\times10^{-7}$ & 0.3 & 0.4 \\
intermediate & 50.0 & $1.1\times10^{-7}$ & 0.6 & 1.0 \\
high & 70.4 & $6.7\times10^{-6}$ & 0.9 & 1.8 \\
\hline
\end{tabular}
  \caption{Summary of the characteristics of the simulated maps. From left to right: fraction of the sky available after masking, amplitude of noise and the $z_0$ and $\alpha$ parameters of the redshift distribution function of each survey.
The noise column gives the (constant) amplitude of the noise power spectrum for polarization --in units of $(\mu$K$)^2$ -- and for the surveys (dimensionless), while for the lensing case the dispersion of the noise map is shown (dimensionless). }
   \label{tab:poiss}
 \end{table}
 \end{savenotes}

The performance of the method described in the previous section has been studied on coherent simulations of the CMB (temperature and polarization) anisotropies, three different galaxy surveys (whose redshift distribution peaks at low, intermediate and high redshift) and the lensing potential. For the simulations, we have assumed the cosmological parameters given by the Planck fiducial $\Lambda$CDM model \citep{PLA_PAR} and the different power spectra (see Fig.~\ref{fig:cls_teo}) have been calculated through a modified version of the CAMB\footnote{http://camb.info} code \citep{LEW00}. In particular, 10000 simulations have been generated with a resolution given by the HEALPix~\citep{GOR05} parameter $N_{\rm SIDE}$=64. Each simulation consists of:

\begin{itemize}
\item CMB related maps (convolved with the corresponding pixel window function and a Gaussian beam of 160 arcmin)
\begin{itemize}
\item CMB temperature (T) map, without the ISW contribution
\item ISW map
\item CMB Q and U polarization maps, where Q and U are the Stokes’ parameters
\item From Q and U, the E-mode map is also derived

\end{itemize}
\item LSS tracers
\begin{itemize}
\item lensing potential
\item low redshift survey, generated using a redshift distribution function ($dn/dz$) modelled by a Gamma function, $dn/dz \propto (z/z_0)^\alpha e^{-\alpha z/z_0} $, with $\alpha = 0.4$ and $z_0 = 0.3$
\item intermediate redshift survey, derived using the same $dn/dz$, with $\alpha = 1.0$ and $z_0 = 0.6$
\item high redshift survey, obtained with the same function for $dn/dz$, with $\alpha = 1.8$ and $z_0 = 0.9$
\end{itemize}
\end{itemize}
When appropriate, noise is also added to the different simulated data maps at realistic levels. The considered noise amplitudes as well as other characteristics of the simulations are summarised in Table~\ref{tab:poiss}.

The parameters $\alpha$ and $z_0$ have values such that the three galaxy redshift distributions have the same variance but peak at different redshifts. Moreover, the surveys have been chosen to be representative of different families of sources. Although they do not  exactly correspond to real data, some of their characteristics have been selected to resemble those from current surveys. For instance, the high-z survey can be representative of a radio survey like NVSS \citep{CON98}, the one peaked at intermediate-z has a redshift distribution typical of luminous galaxies from SDSS \citep{AHN12} and, finally, the low-z survey includes near galaxies like the photometrically-selected galaxies from the SDSS \citep{AIH11}. The different levels of noise have also been chosen to be similar to those of the three aforementioned surveys. This has been simulated as Poissonian noise with the corresponding power spectrum amplitude given in Table~\ref{tab:poiss}, that has been obtained as $1/\bar{n}$, being $\bar{n}$ the mean number of galaxies per steradian.

For the CMB polarization, we have simulated Gaussian white noise at a level expected for the Planck data \citep{PLA05}, while we assume that the instrumental noise is negligible for the CMB intensity map, which is a very good approximation for current CMB observations, such as Planck, at the considered scales.  For the lensing potential map, the noise has been generated as a Gaussian field with an amplitude similar to that estimated for the lensing map recovered by Planck \citep[see Fig. 1 of][]{PLA_LENS}. The dispersion of the simulated noise maps for the lensing potential is given in Table~\ref{tab:poiss}.

Finally we have considered the following masks for the different data sets:
\begin{itemize}
\item For the low- and intermediate-z surveys, we adopt a simple galactic cut of $\pm 30^{\circ}$ around the galactic plane.
\item For the high-z survey we use a mask which excludes the areas not observed by NVSS as well as regions with galactic emission, within $14^{\circ}$ from the galactic plane, plus some nearby objects.
\item For the CMB (temperature and polarization) as well as for the lensing map we adopt the WMAP 9-year point source catalogue mask \citep[the one used by the WMAP team to exclude the Galactic plane and the Magellanic cloud regions when building the 9-year point source catalogues,][]{BEN13}.
\end{itemize}
The sky fraction kept for each data set after applying the considered mask is given in Table~\ref{tab:poiss}.

The maps are coherently simulated by assuming that they are Gaussian (which is appropriate even for the Poissonianly distributed galaxy density maps, since the mean number of galaxies per pixel is $\approx 40$ for the worst case) and, therefore, all the required information is given by all the auto- and cross- angular power spectra~\cite[see, for instance,][for details on how to simulate correlated Gaussian fields]{BAR08}. In particular, given two surveys $a$ and $b$, the theoretical cross-angular power spectra between the surveys read as:
\begin{equation}
C^{ab}_\ell = 4 \pi \int \frac{\mathrm{d}k}{k} \ \Delta^2(k) I^a_\ell(k)
I^b_\ell(k) \ ,
\end{equation}
where $\Delta^2(k)$ is the matter power spectrum per logarithmic interval, and $I_{\ell}^a(k)$ is a transfer function represented by the redshift integral:
\begin{equation}
I^a_{\ell}(k) = \int_0^\infty \mathrm{d}z \ W_a(z,k) j_\ell(kr(z)) \ .
\end{equation} 
Here $r(z)$ is the comoving distance as a function of the redshift, and $j_{\ell}$ are the spherical Bessel functions, which project the window function $W_a (z,k)$ into each multipole ${\ell}$ of the power spectrum. In the case of a galaxy survey the window function is independent of k and it is given by
\begin{equation}
W_a(z) = b_a(z) D(z) \frac{\mathrm{d}n_a}{\mathrm{d}z} \ .
\end{equation}
It depends on the galaxy redshift distribution and the bias function $b_a(z)$ of each survey. For simplicity, we have assumed a constant bias equal to 1 for all the surveys\footnote{It is possible to see that choosing a constant bias $b$ different from 1 only affects to the filter in Eq. \ref{eq:rec_n} by rescaling the Poissonian noise by $1/b^2$. In this way, the noise level of each survey in Table~\ref{tab:poiss} can be understood as an effective noise including the bias dependence, given by $1/(\bar{n}b^2)$. In particular note that in a more realistic scenario where $b>1$, for a fixed value of $\bar{n}$, we would have a smaller effective noise than in the case considered here and, therefore, the filter would provide slightly better results.}.The growth factor $D(z)$ in this expression takes into account the linear evolution of the matter perturbations. For the ISW effect the window function involves the evolution of the potential with redshift:
\begin{equation}
W_{ISW}(z,k) = -3 \Omega_m \left( \frac{H_0}{ck} \right)^2
\frac{\mathrm{d}}{\mathrm{d}z} \left[ (1+z) D(z) \right] \ .
\end{equation}
It depends on $k$ due to the Poisson's equation relating the matter and the potential. If the Universe is matter-dominated, then the function $(1 + z)D(z)$ is constant and the ISW vanishes. As mentioned in Section \ref{sec:intro}, the presence of a cosmological constant (or a more general form of dark energy in general) produces a non-zero ISW function, although it is worth recalling that there are other possible sources for generating non-zero ISW effect. Among others, a non-zero spatial curvature~\citep{KAM94}, or models of modified gravity~\citep[e.g.,][]{Munshi2014} will produce a net ISW effect.

The three galaxy redshift distributions used in this work are plotted in Fig. \ref{fig:redshift}. Finally, as an example of the simulated maps we present in Fig. \ref{fig:simu_maps} the simulated ISW (top), CMB temperature and E-mode (upper middle), lensing and the survey peaking at low-z (lower middle), and the surveys peaking at intermediate and high-z (bottom).

\section[]{Results}
\label{sec:rec}

We have assessed the quality of the ISW reconstruction under two different scenarios. First, in Section \ref{sec:allsky} we consider all-sky data. Under these conditions, we study the contribution of each of the LSS tracers considered to the recovery of the ISW fluctuations. We also check how important is to include the CMB information, and, in particular, the polarization anisotropies. In addition, we study how the recovery is affected by the presence of instrumental (for CMB polarization and lensing) and Poissonian (for galaxy-surveys) noises. Second, a more realistic situation is explored in sec. \ref{sec:mask}, where maps with partial sky coverage are analysed. Again, we test the role played by each one of the surveys, as well as the CMB temperature and polarization data, in recovering the ISW map. As before, also the impact of the noise is assessed.

\subsection{Full-sky data}
\label{sec:allsky}
 \begin{figure}
 \centering
   \subfloat{\includegraphics[width=5.0cm, angle=90]{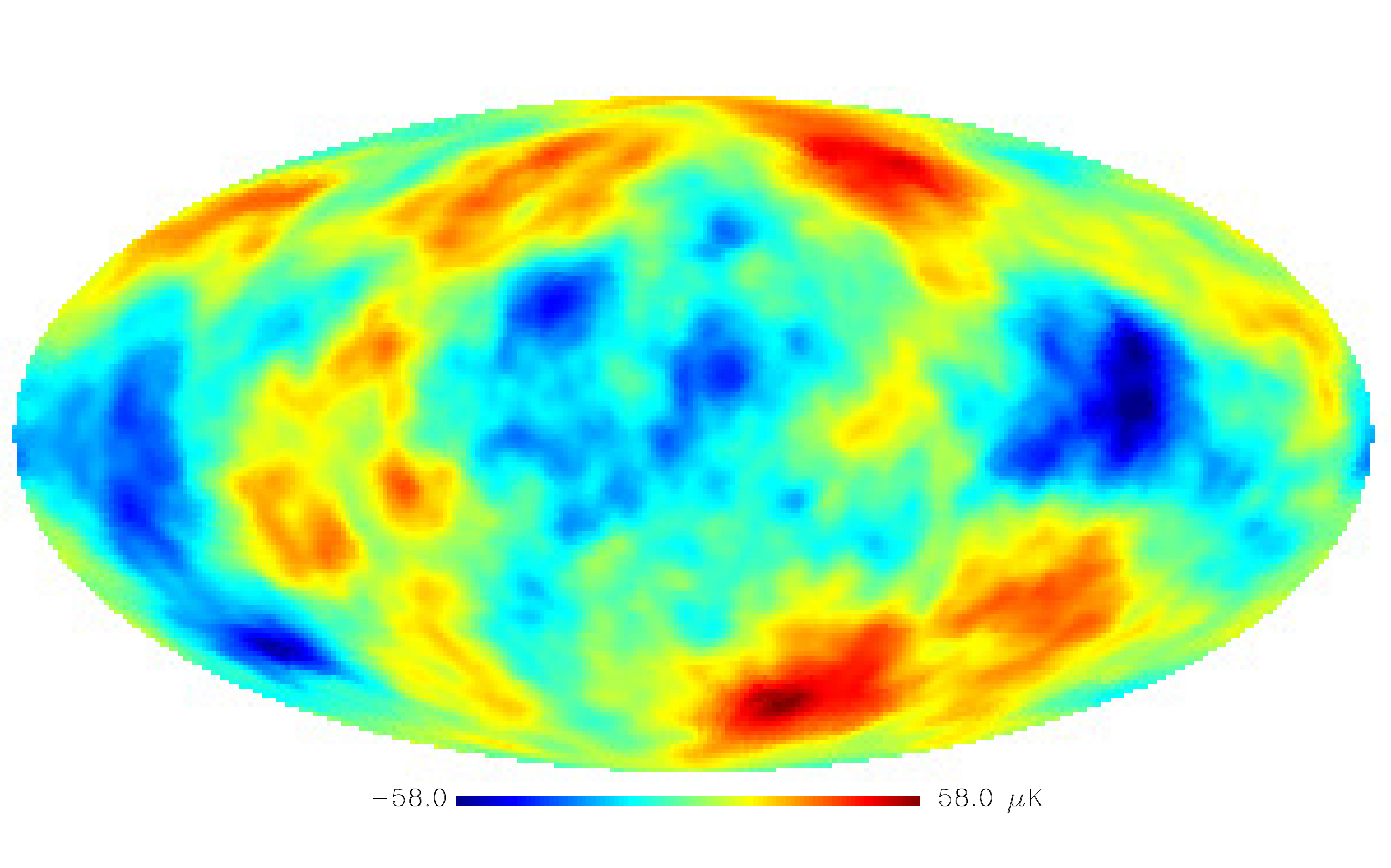}}\\
   \subfloat{\includegraphics[width=5.0cm, angle=90]{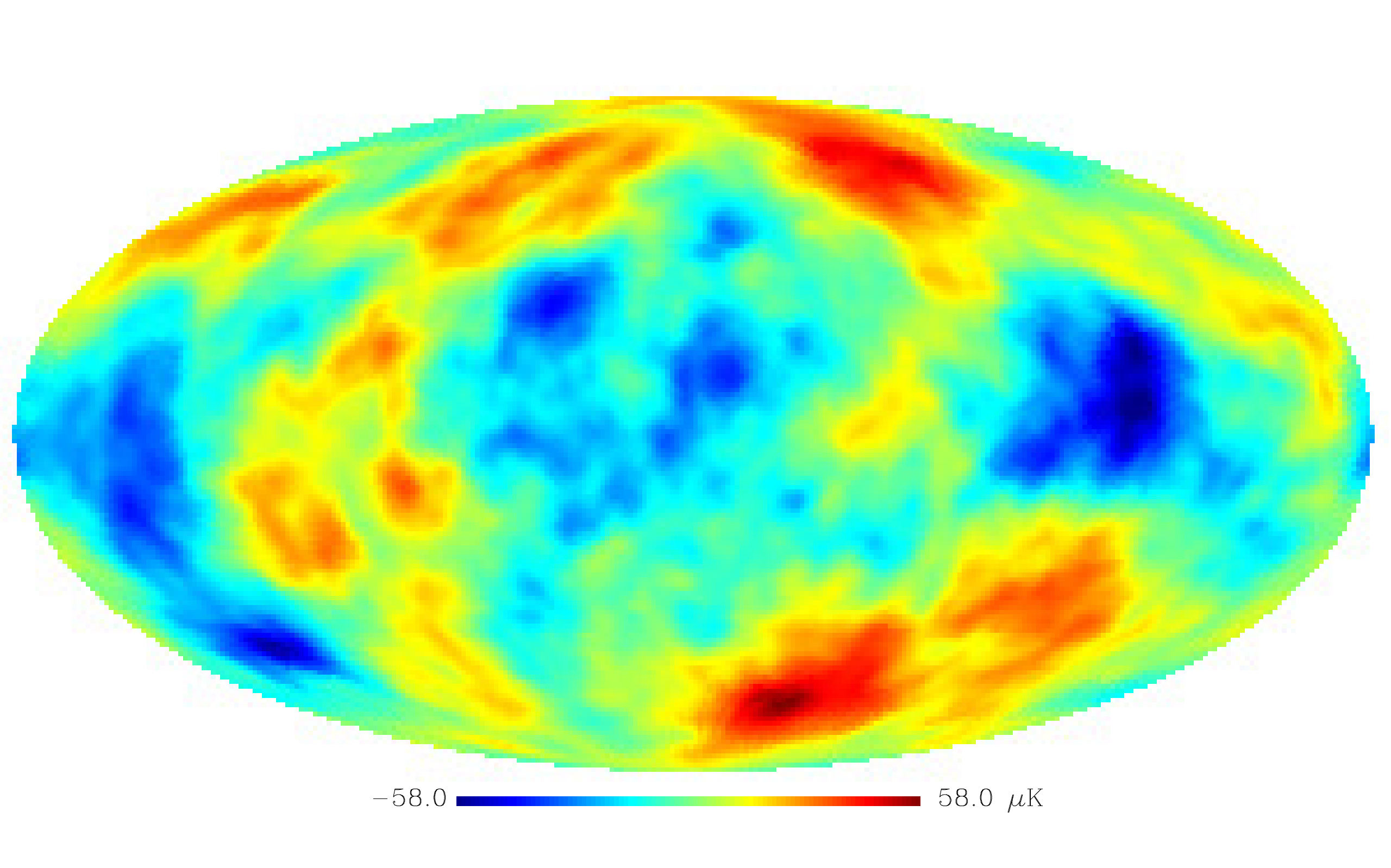}}\\
   \subfloat{\includegraphics[width=5.0cm, angle=90]{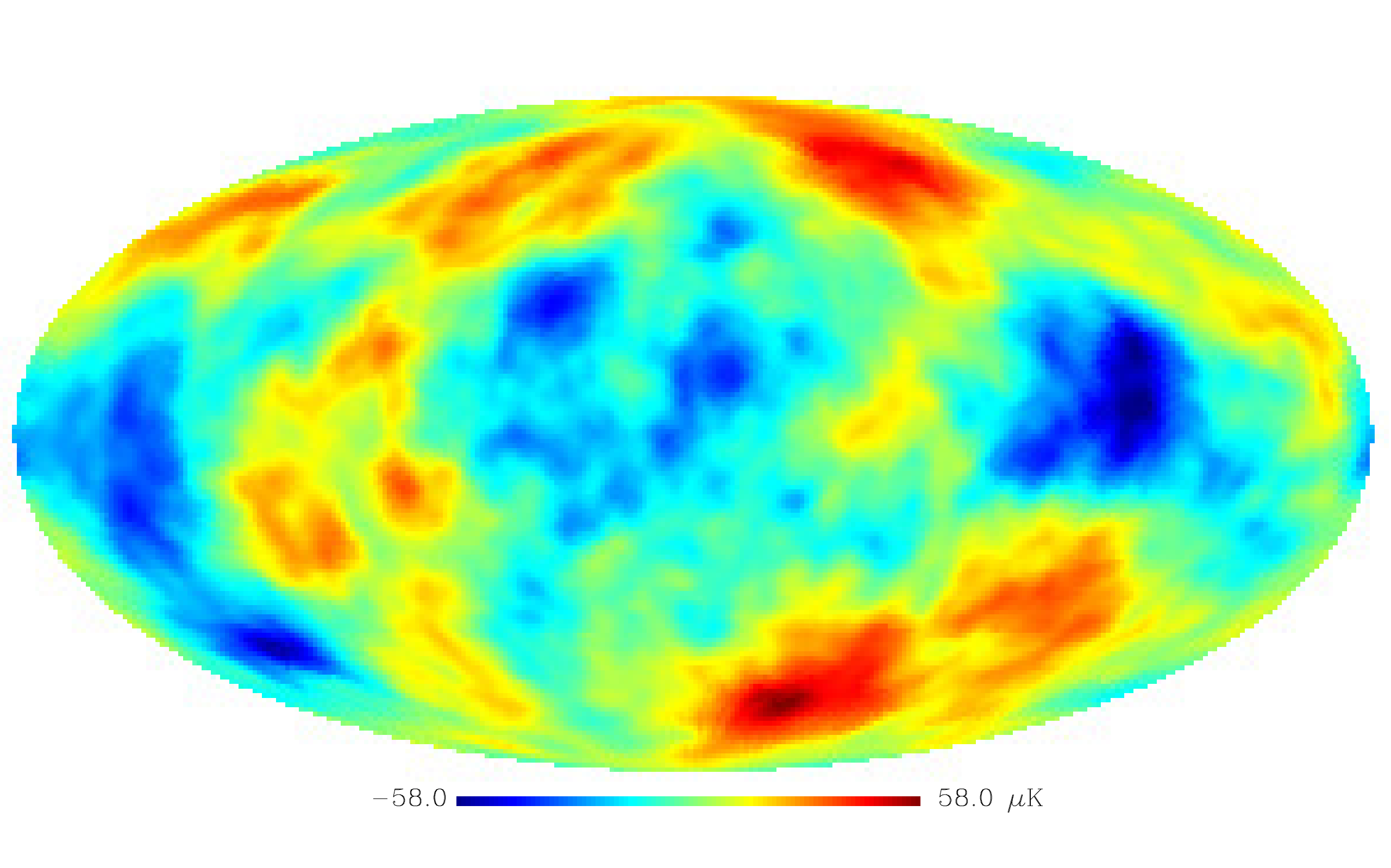}}\\
 \caption{Reconstruction of the ISW signal in the ideal case of full-sky and noiseless for the following cases: with (top) and without (middle) exploitation of polarization information, and without CMB, i.e. only the surveys are used (bottom). The correlation coefficient for this particular simulation is 1.00 for all the three cases.}
 \label{fig:rec_isw_ideal_all}
\end{figure}

\begin{figure*}
 \centering
   \subfloat{\includegraphics[width=5.0cm, angle=90]{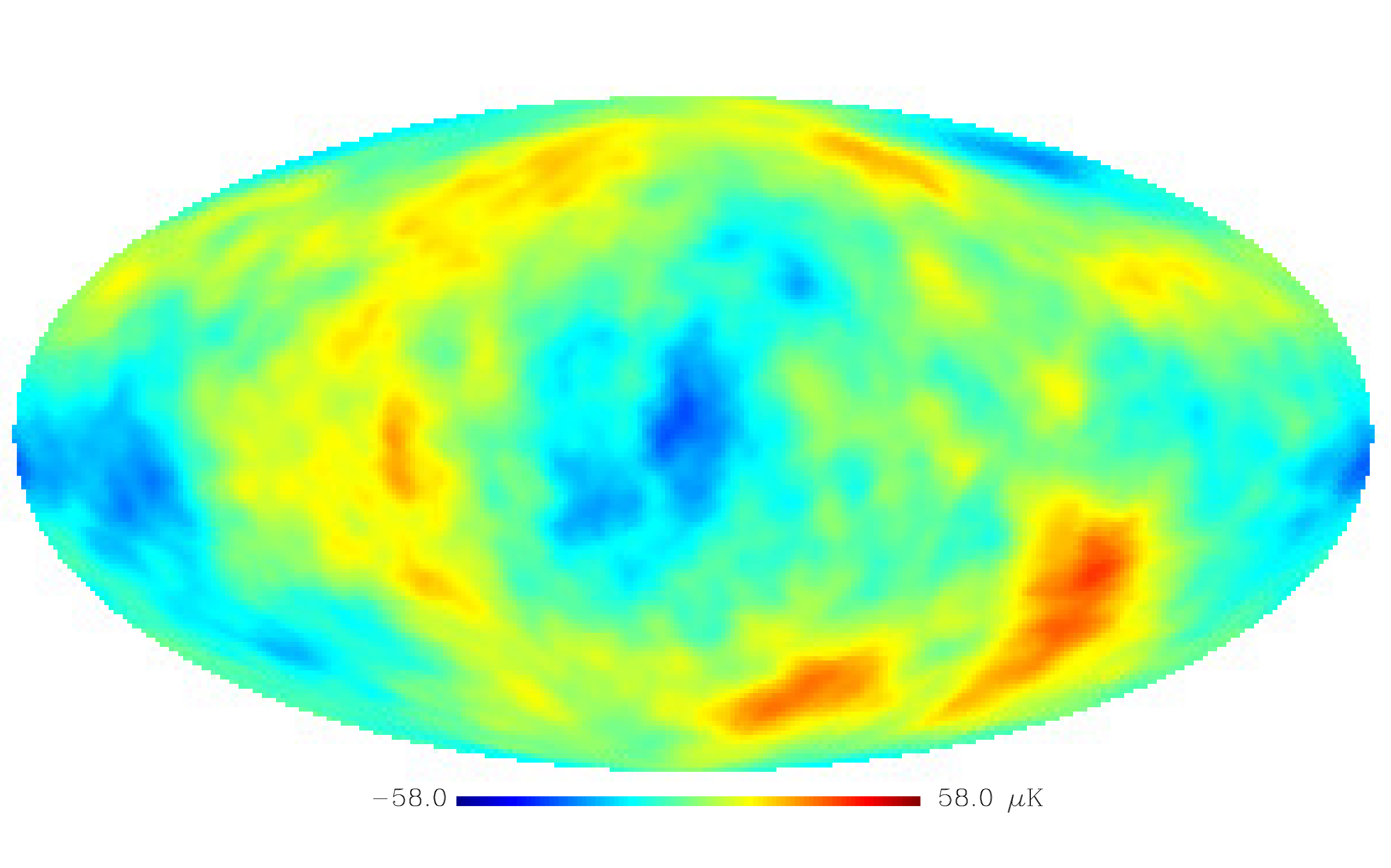}}\quad
   \subfloat{\includegraphics[width=5.0cm, angle=90]{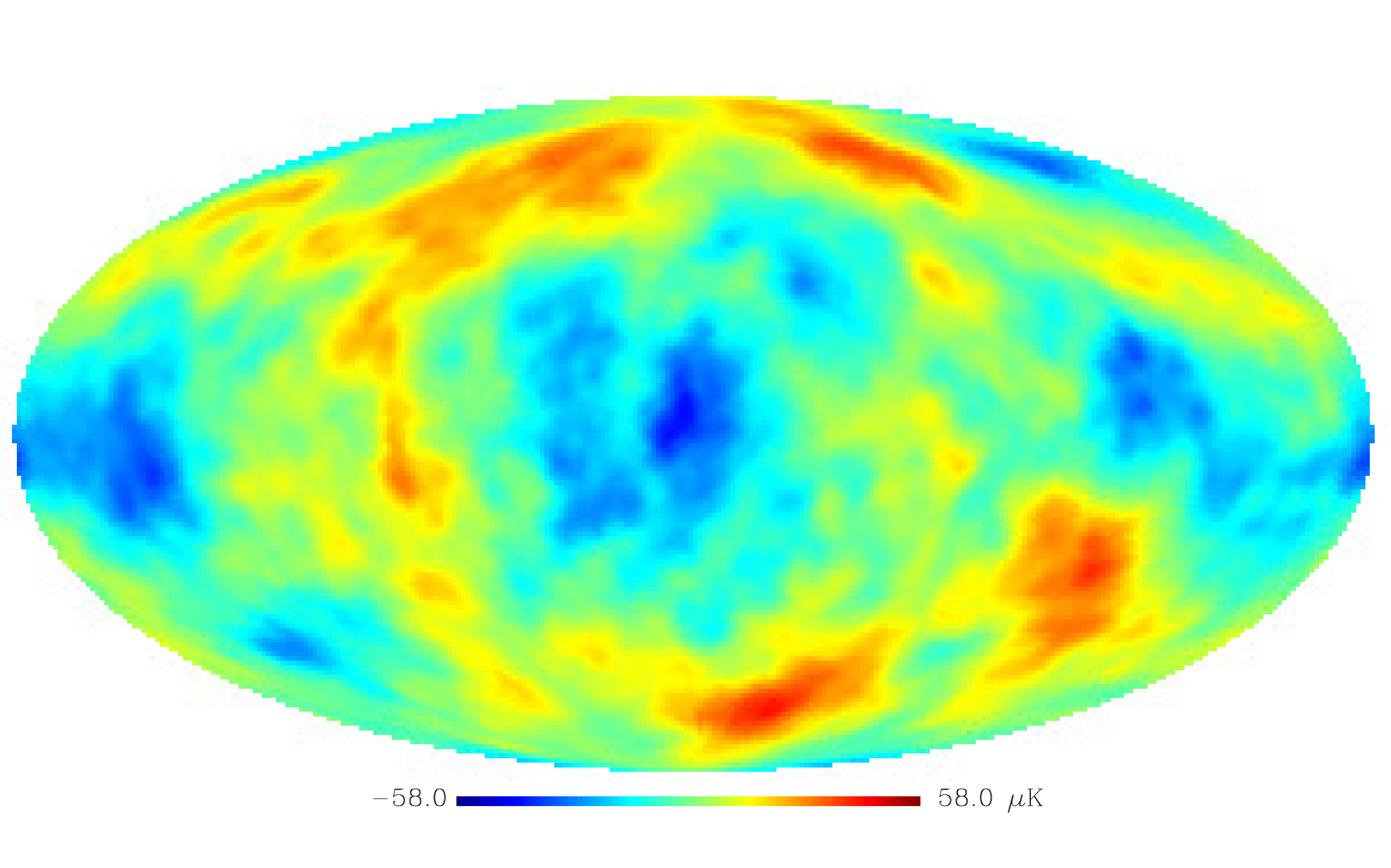}}\\
   \subfloat{\includegraphics[width=5.0cm, angle=90]{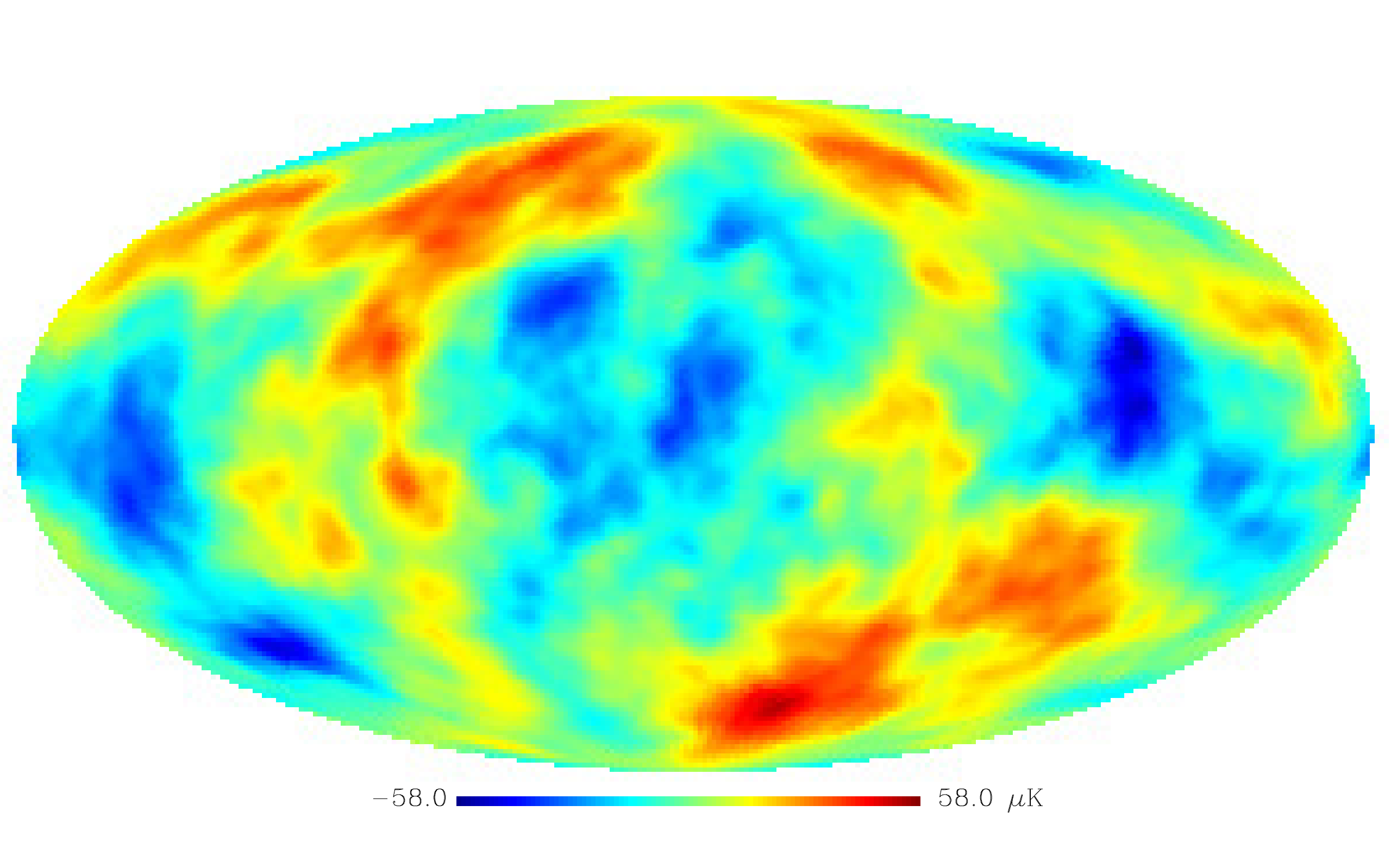}}\quad
   \subfloat{\includegraphics[width=5.0cm, angle=90]{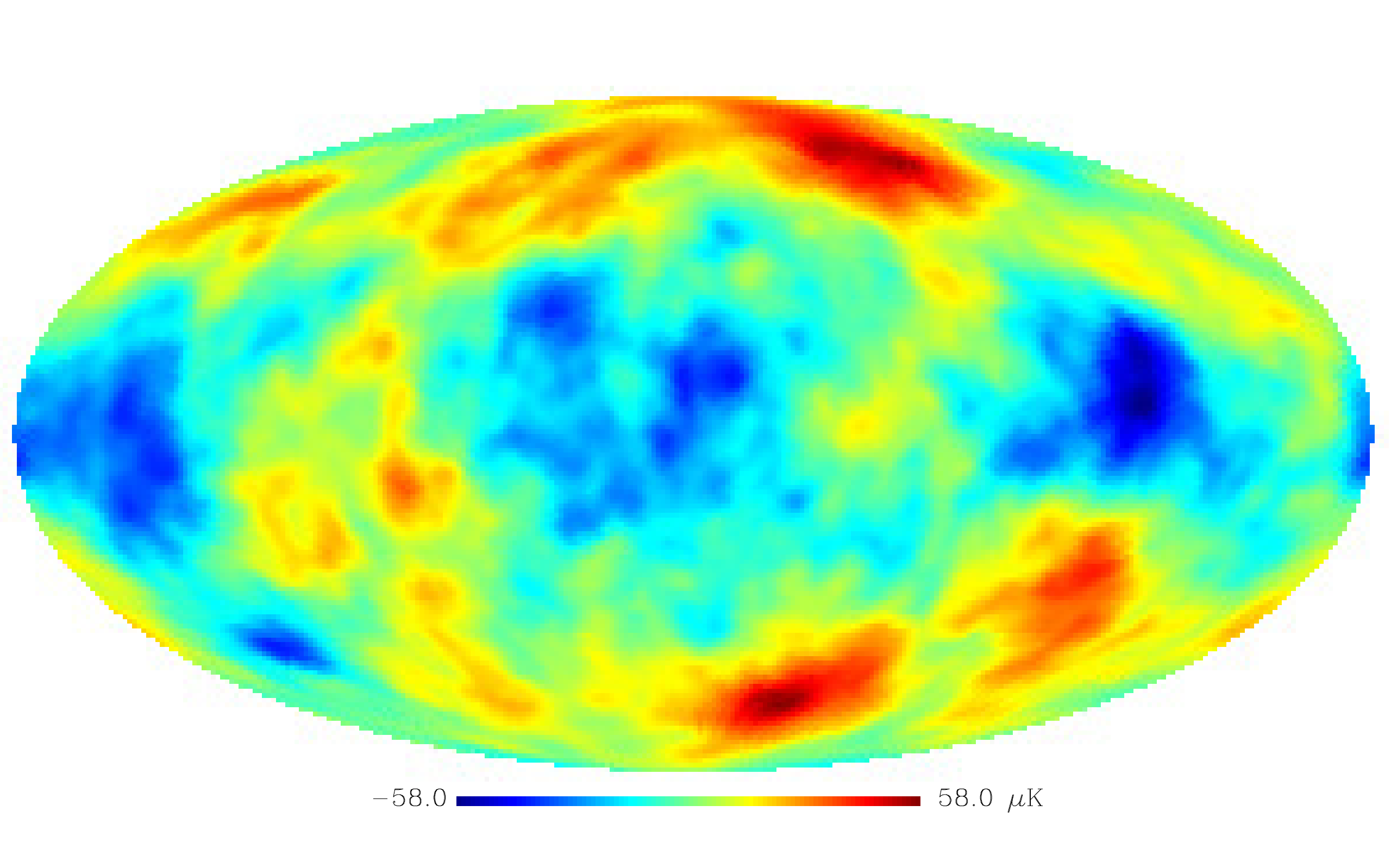}}\\
 \caption{ISW reconstruction for full-sky noiseless data, for different cases: using CMB (temperature and polarization) plus one LSS tracer: low-z (top left), intermediate-z (top right), high-z (bottom left) or lensing (bottom right). The correlation coefficients for this particular realization are 0.71, 0.83, 0.94 and 0.96 respectively.}
 \label{fig:rec_isw_ideal_nopol}
\end{figure*}

\begin{figure*}
 \centering
   \subfloat{\includegraphics[width=5.0cm, angle=90]{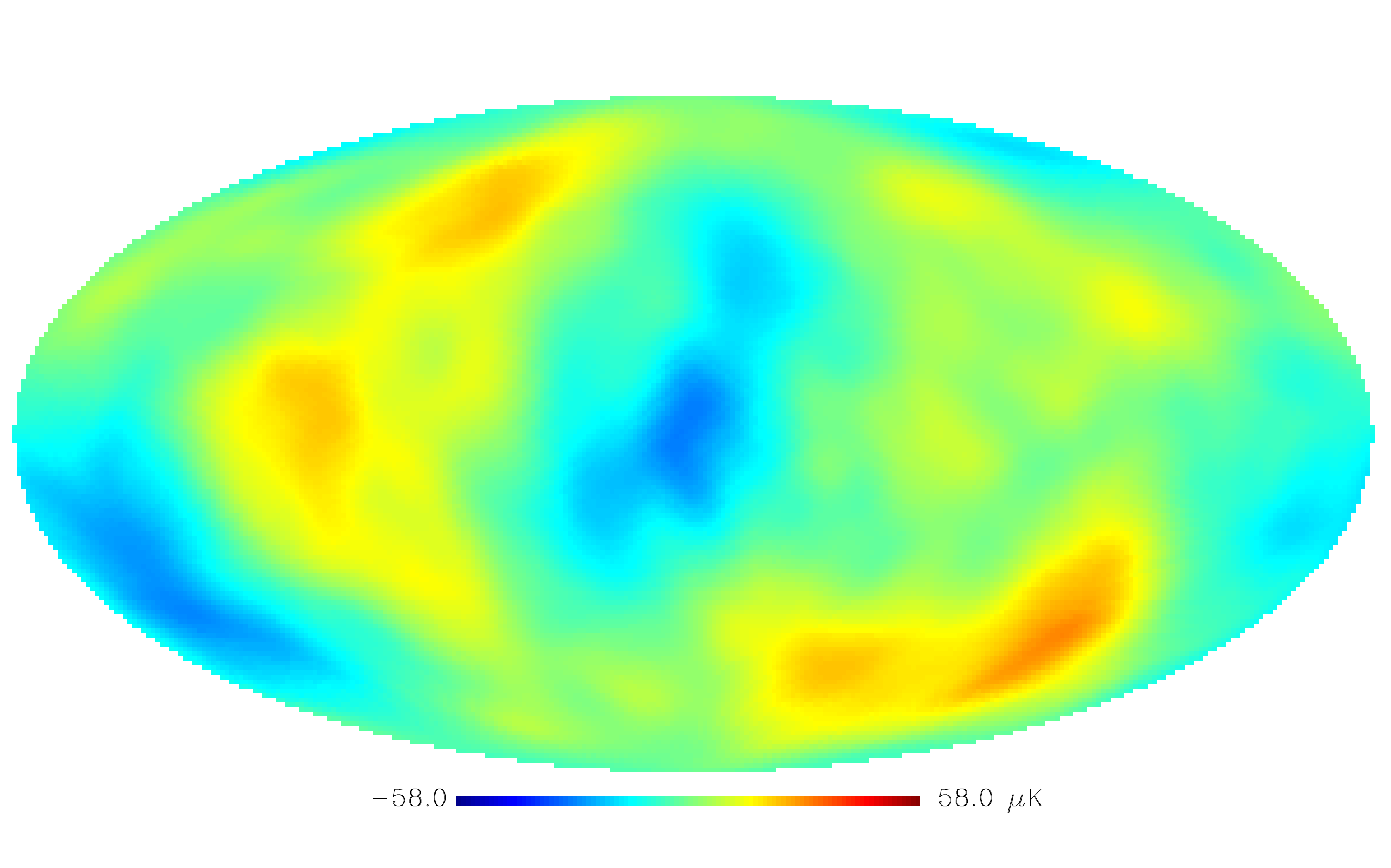}}\quad
   \subfloat{\includegraphics[width=5.0cm, angle=90]{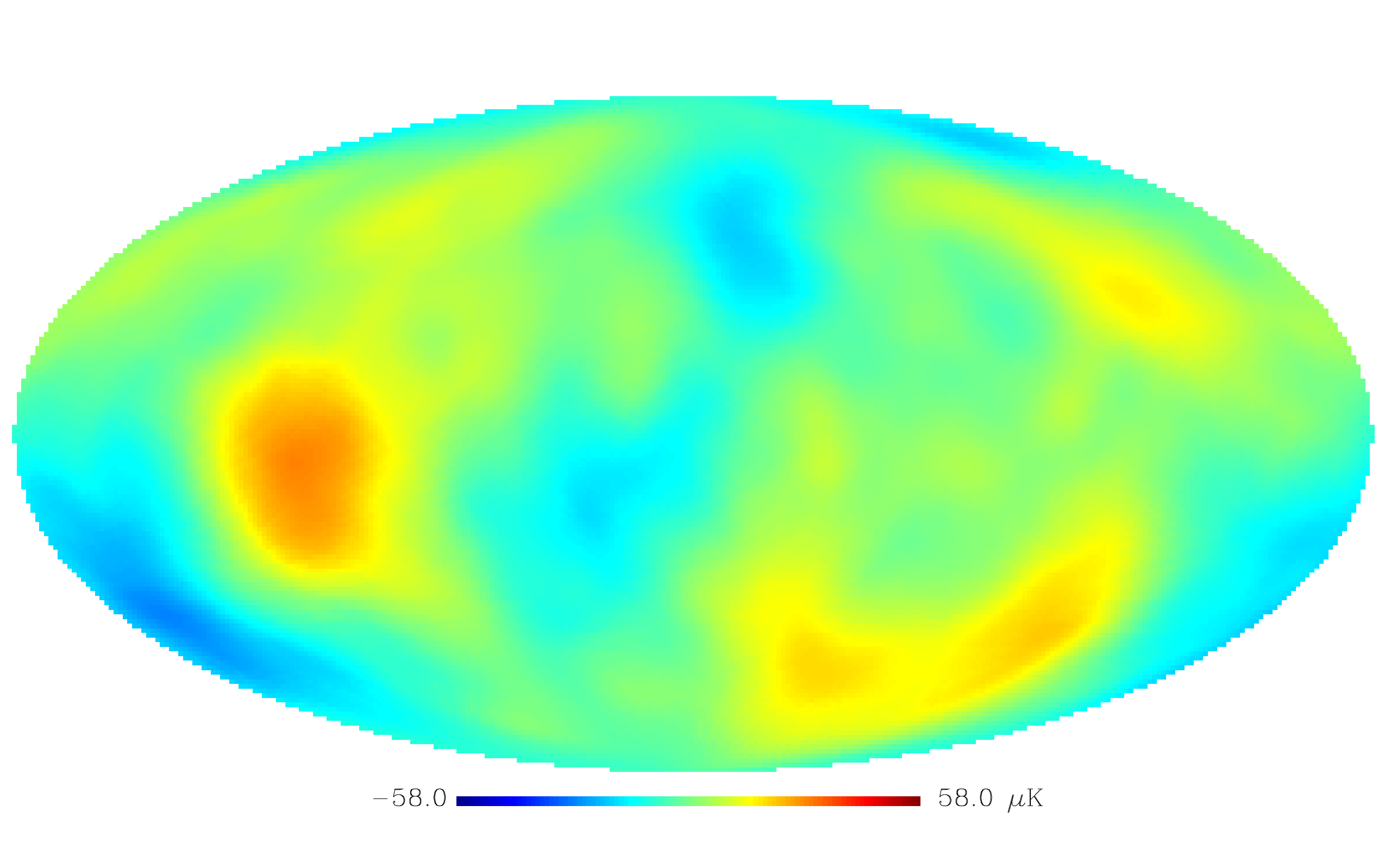}}\\
\caption{ISW reconstruction for full-sky noiseless data, for WF with (left) and without (right) including polarization information. For this particular realization, the correlation coefficients are 0.59 and 0.44 respectively.}
 \label{fig:wf_ideal_nopol}
\end{figure*}

\begin{figure}
 \centering
   \subfloat{\includegraphics[width=8.5cm]{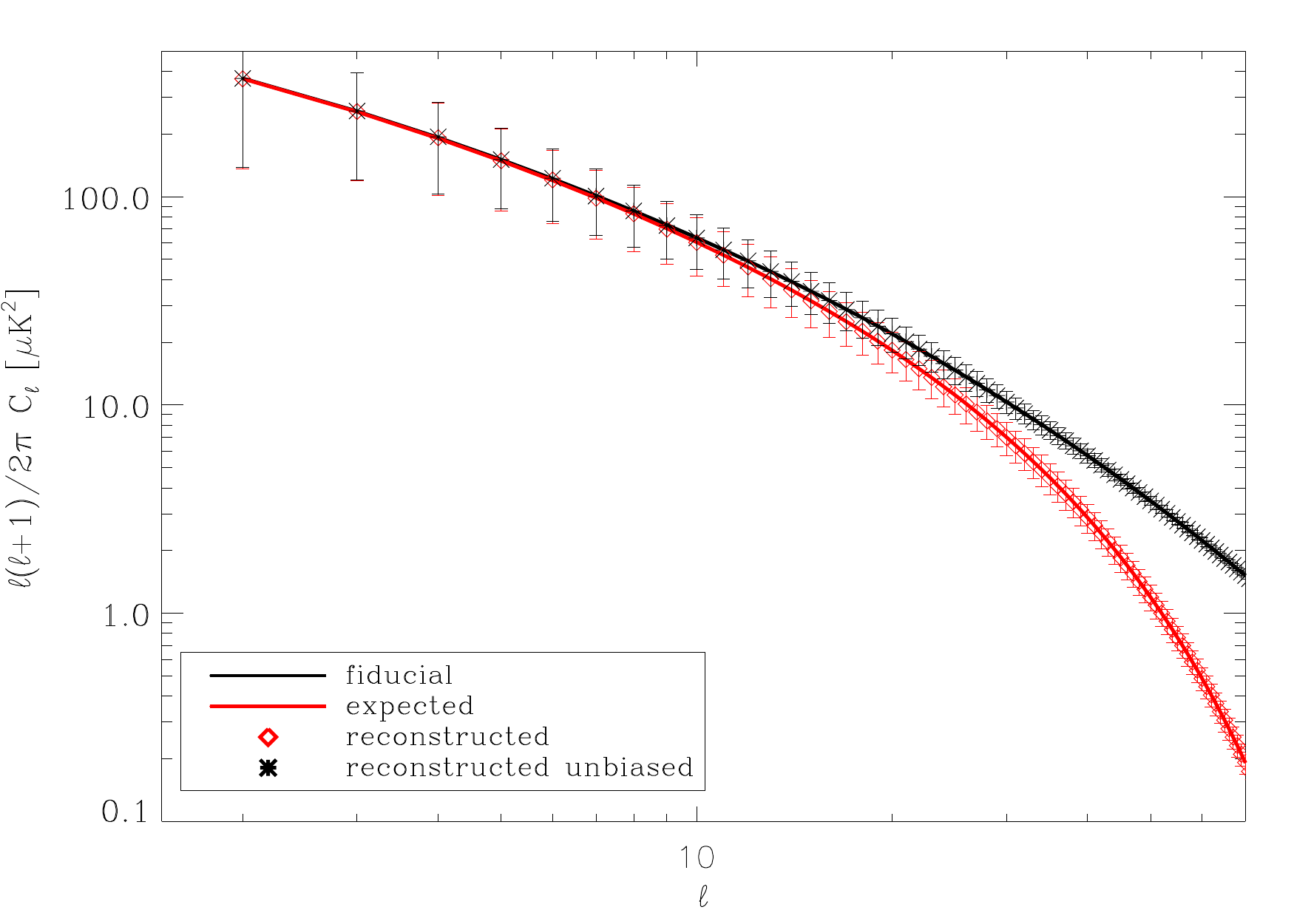}}\\
 \caption{Power spectra of the recovered ISW signal for full-sky and noiseless data including CMB (temperature and polarization) when all three surveys and lensing are used (red diamonds) averaging over 10000 simulations. The error bars correspond to the dispersion of the simulations at each $\ell$. The black line is the fiducial model and the red line corresponds to the expected power spectrum of the reconstruction. The black asterisks are the unbiased reconstructed power spectrum.}
 \label{fig:cls_4comp}
\end{figure}

\begin{figure}
 \centering
   \subfloat{\includegraphics[width=8.5cm]{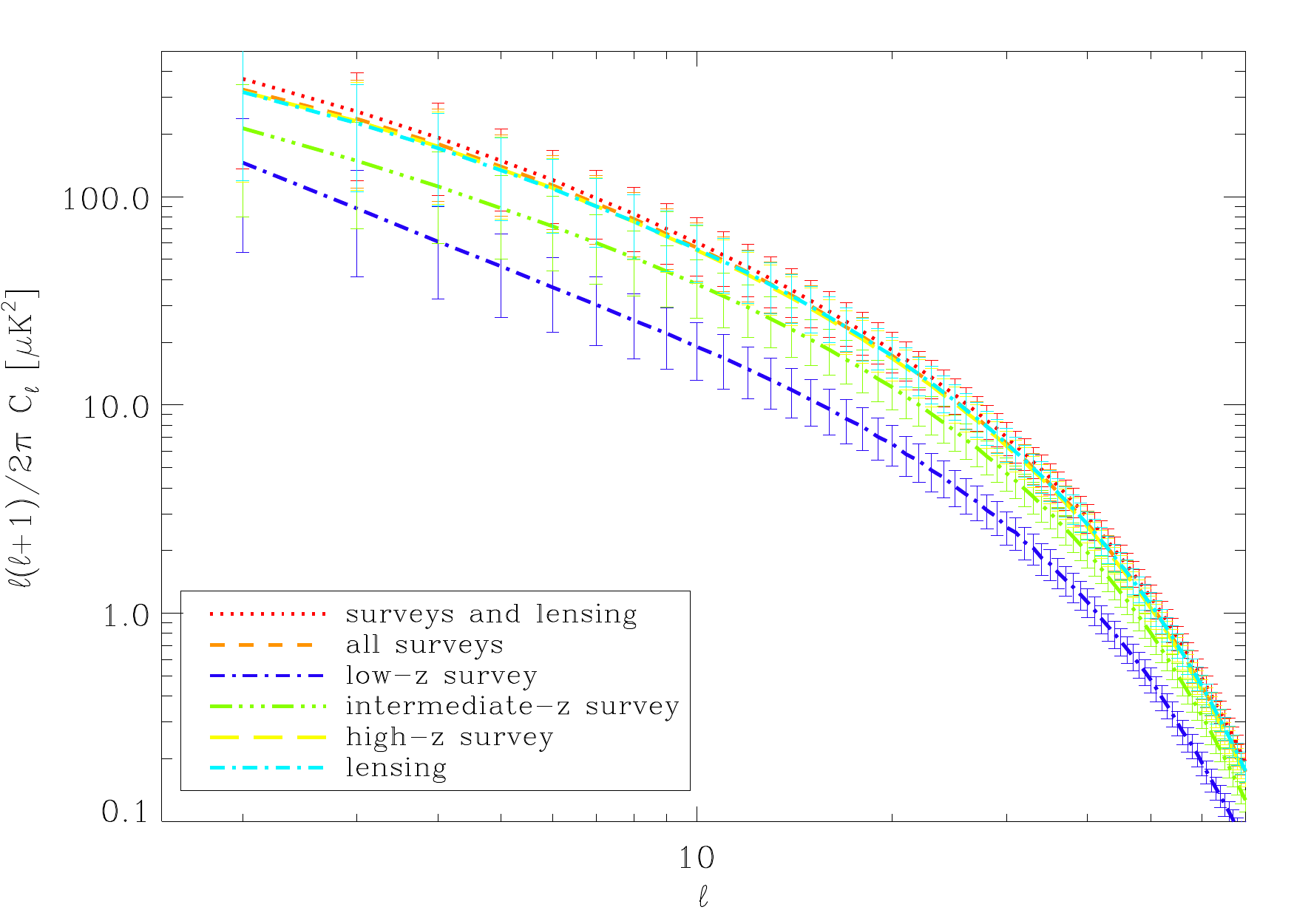}}\\
 \caption{Power spectra of the recovered ISW signal for full-sky and noiseless data including CMB (temperature and polarization) for the following cases: all three surveys and lensing are used (red), all surveys (orange), only low-z (blue), only intermediate-z (green), only high-z (yellow) surveys and only lensing (cyan). The different power spectra have been obtained averaging over 10000 simulations and the error bars correspond to the dispersion of the simulations at each $\ell$.}
 \label{fig:cls_all}
\end{figure}

\begin{figure}
 \centering
 \subfloat{\includegraphics[width=8.5cm]{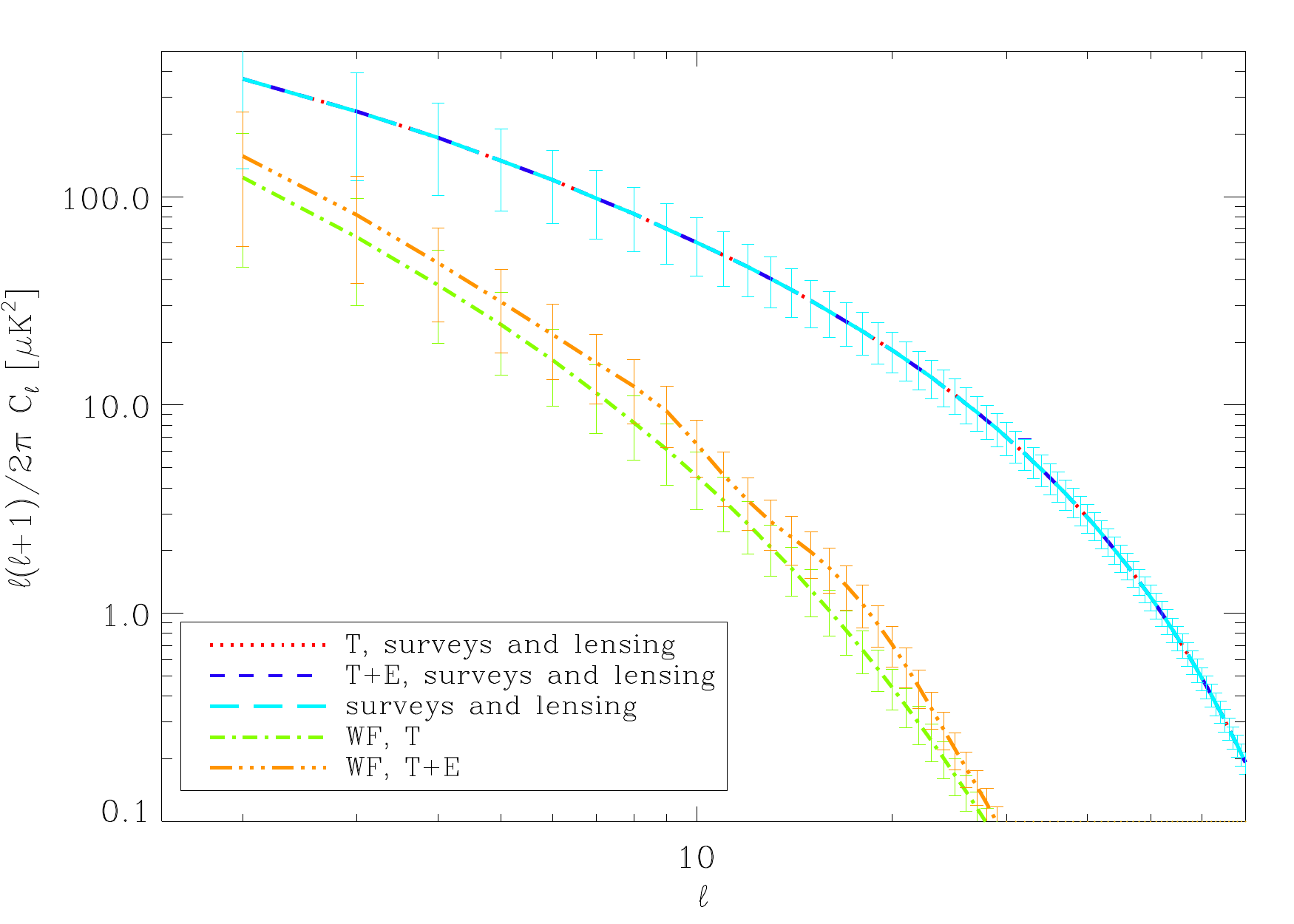}}
 \caption{Power spectra of the recovered ISW signal for full sky and noiseless data for different cases: using CMB temperature, all surveys and lensing with (dashed blue) and without (dotted red) polarization information, using only LSS tracers and lensing (long dashed cyan), WF reconstruction with (orange) and without (green) polarization information. As before, 10000 simulations have been used to obtain the mean power spectrum and the error bars at each case.}
 \label{fig:cls_all_bis}
\end{figure}

\begin{table}
 \centering
 \begin{tabular}{@{}lcccccc@{}}
  \hline
      & \multicolumn{3}{|c|}{$\bar{\rho}$} & \multicolumn{3}{|c|}{$e_r$}\\
\hline  
      &  T+E   & T  &  no CMB & T+E & T & no CMB  \\
\hline
 Noiseless  &    &   &     &  & & \\
\hline
3surveys+lens & 1.00 & 1.00 & 1.00 & 0.04 & 0.04 & 0.04 \\
3surveys      & 0.96 & 0.95 & 0.95 & 0.29 & 0.29 & 0.29 \\
low           & 0.63 & 0.59 & 0.42 & 0.76 & 0.80 & 0.90 \\
int.           & 0.78 & 0.76 & 0.72 & 0.62 & 0.63 & 0.69 \\
high          & 0.94 & 0.94 & 0.94 & 0.32 & 0.33 & 0.33 \\
lensing          & 0.94 & 0.94 & 0.93 & 0.34 & 0.34 & 0.35 \\
Wiener filter & 0.54 & 0.47 & -    & 0.83 & 0.87 & -\\
 \hline
 Noisy  &    &   &     &  & & \\
\hline
3surveys+lens & 0.81 & 0.80 & 0.77 & 0.58 & 0.59 & 0.63\\
3surveys      & 0.79 & 0.78 & 0.74 & 0.60 & 0.61 & 0.66\\
low           & 0.62 & 0.58 & 0.41 & 0.77 & 0.80 & 0.90\\
int.           & 0.76 & 0.74 & 0.69 & 0.64 & 0.66 & 0.71\\
high          & 0.55 & 0.50 & 0.20 & 0.82 & 0.85 & 0.97\\
lensing          & 0.63 & 0.60 & 0.48 & 0.76 & 0.79 & 0.87\\
Wiener filter & 0.53 & 0.47 & -    & 0.83 & 0.87 & -\\
\hline
 \end{tabular}
   \caption{Mean correlation coefficient (columns 2 to 4) and mean relative error of the reconstructed maps (columns 5 to 7) computed over 10000 simulations, for full sky datasets without (upper half) and with noise (bottom half) for different combinations of CMB data and LSS tracers. Results for the cases that include both CMB intensity and polarization (T+E), only intensity (T) or no CMB in combination with all surveys and lensing (3surveys+lens), all surveys (3surveys) or only one tracer at a time (low-z, intermediate-z, high-z or lensing) are considered. The results obtained with the Wiener filter using only CMB, with and without including polarization, are also shown.}
   \label{tab:corr_coeff}
 \end{table}

\begin{figure}
 \centering
 \subfloat{\includegraphics[width=8.5cm]{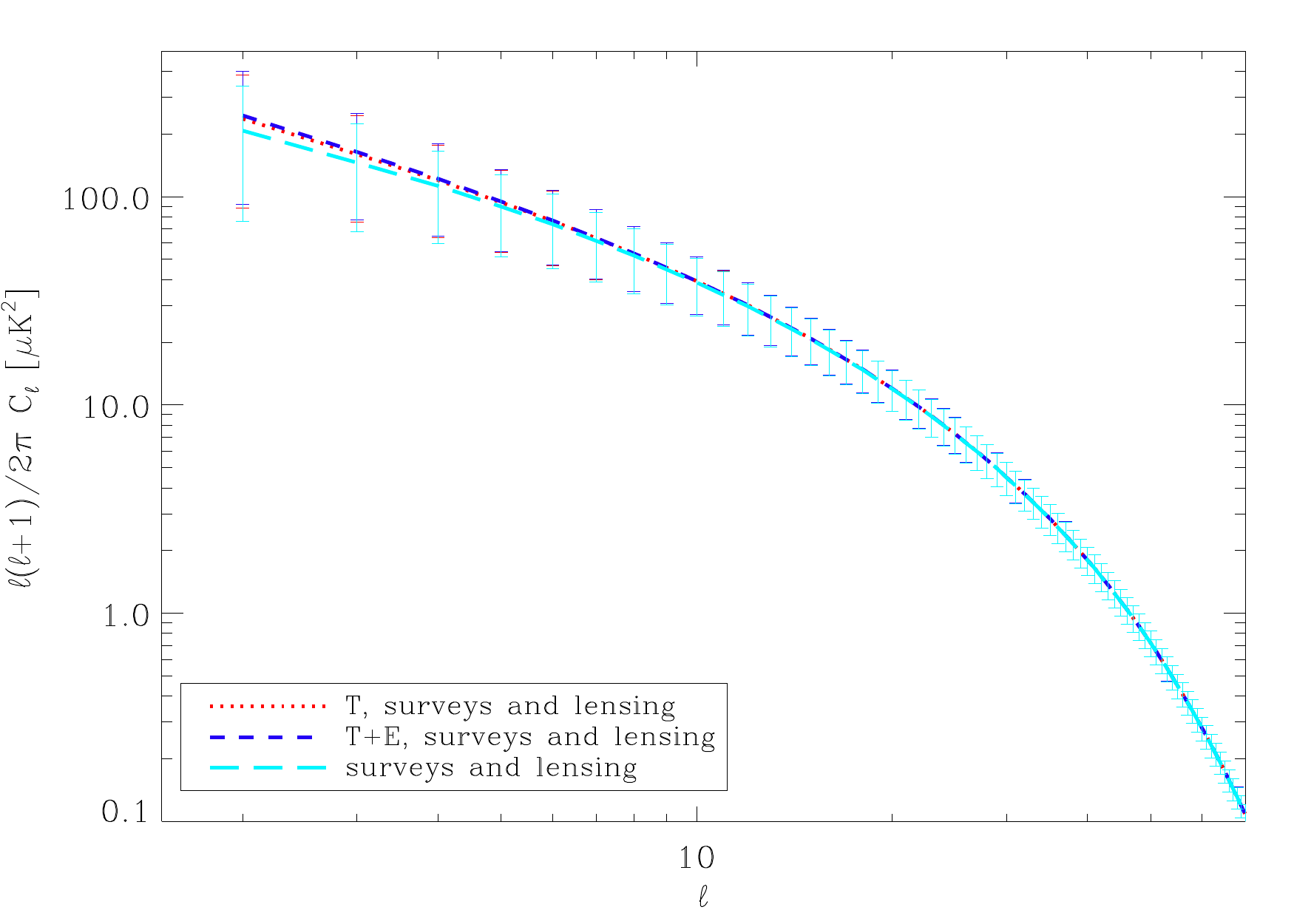}}
 \caption{Power spectra for full-sky case with noise considering CMB temperature and all surveys, without including (dotted red line) and including (short dashed blue) polarization information. For comparison, the case without using CMB data (long dashed cyan) is also shown.}
 \label{fig:cls_all_noise}
\end{figure}

In this section, we study the quality of the recovered ISW using different combinations of all-sky simulated data sets: CMB (intensity and polarization), low-, intermediate- and high-z surveys and lensing. Our aim is to assess the performance of the proposed filter in this ideal case, as well as to show which is the contribution of the different simulated data sets to the final reconstruction. We will also study the effect of instrumental and Poissonian noise in our results.

As an illustration, in Fig.~\ref{fig:rec_isw_ideal_all}, we show the reconstructed ISW obtained for our reference simulations (given in Fig.~\ref{fig:simu_maps}) using all the considered LSS surveys in three different cases: with (top) and without (middle) including polarization information and using only information from the LSS surveys corresponding to the first addend of Eq. \ref{eq:rec_n} (bottom). 
As seen, the three cases are very similar, which indicates that, in this ideal case, the main contribution to the recovered ISW map comes from exploiting the cross-correlation between the ISW signal and the LSS tracers.

We have also tested the contribution of each LSS survey separately. Fig.~\ref{fig:rec_isw_ideal_nopol} presents the results when only one LSS tracer is used in combination with the CMB map (intensity and polarization): low- (top left), intermediate- (top right) and high-z (bottom left) surveys and lensing (bottom right). For comparison, in Fig.\ref{fig:wf_ideal_nopol} it is given the reconstructed ISW obtained by applying a simple Wiener filter to the CMB map (see e.g. \citealt{BAR08} for details) with (left) and without (right) including polarization information.

By comparing the results of Fig.~\ref{fig:rec_isw_ideal_all}, Fig.~\ref{fig:rec_isw_ideal_nopol} and Fig.~\ref{fig:wf_ideal_nopol} with the input ISW (top panel of Fig.~\ref{fig:simu_maps}), it becomes apparent that, as expected, the best reconstruction is obtained when all the surveys are included. In this ideal case, when only the lensing map is used the reconstruction is quite good due to its strong correlation with the ISW at all the redshifts. From the three considered LSS tracers, the high-z survey is the one that contributes most to the ISW reconstruction. This can be understood by looking at the redshift distribution of the galaxies of each survey (Fig.~\ref{fig:redshift}), since the high-z catalogue covers a more suitable redshift range to extract the ISW signal than the other surveys (see e.g. \citealt{AFS04}). Also, as shown in previous works, the LCB filter is clearly superior to a simple WF of the CMB, since the latter does not exploit the information included in the LSS surveys. Regarding polarization, its contribution to the LCB  solution in this ideal case is very modest, while it slightly improves the performance of the WF.

These results can be further quantified by obtaining the average correlation coefficient $\bar{\rho}$ and the average relative error $e_r$ between input and reconstruction using a large number of simulations. To compute these quantities we first calculate the mean residual dispersion map whose pixels values are given by $\bar{\sigma}_i=\frac{1}{N_s}\sum_{j=1}^{N_s}  \sqrt{\left\langle r_{i,j}^{2} \right\rangle - \left\langle r \right\rangle _{i,j}^{2}  }$ where $r_{i,j}$ is the residual map for the $ith$ pixel of the $jth$ simulation obtained by subtracting the reconstructed map $\hat{s}$ from the input ISW simulation $s$, and $N_s$ is the number of simulations. Then we compute the weights map $\omega_i$ and the weighted correlation coefficient as:
\begin{equation}
\label{eq:pesi}
\omega_i=\dfrac{1/\bar{\sigma}_i^{2}}{\sum_{i=1}^{N_{p}} 1/\bar{\sigma}_i^{2}}
\end{equation}
\begin{equation}
\rho=\dfrac{\sum_{i=1}^{N_p}  \omega_i (s_i-\left\langle s\right\rangle )(\hat{s}_i-\left\langle \hat{s}\right\rangle)}{\sigma_s\sigma_{\hat{s}}},
\end{equation}
where $s$ and $\hat{s}$ are the input and reconstructed ISW maps, $\sigma_s$ and $\sigma_{\hat{s}} $ are the weighted dispersions of the input and reconstructed ISW map, $ \left\langle s\right\rangle $ and $\left\langle \hat{s}\right\rangle$ are the weighted mean values for the same maps, and $N_p$ is the number of pixels. Finally we compute the mean correlation coefficient $\bar{\rho}$ as the mean of the weighted correlation coefficient over 10000 simulations. We obtain the average relative error $e_r$ as the ratio between the average over simulations of the weighted dispersion of the residual maps $\sigma_r$ and that of the input ISW maps, i.e.:
\begin{equation}
e_r=\frac{ \frac{1}{N_s}\sum_{j=1}^{N_s} \sigma_r(j) }{ \frac{1}{N_s}\sum_{j=1}^{N_s}  \sigma_s(j)}
\end{equation}
where 
\begin{equation}
\sigma_r(j)= \sqrt{ \sum_{i=1}^{N_p} \omega_i \left(r_{i,j}-\left<r_{i,j}\right>\right)^2  }
\end{equation}
The mean correlation coefficient and average relative error are given in Table~\ref{tab:corr_coeff} for the different considered cases.

When all data are used (CMB intensity and polarization, the three surveys and lensing), the average correlation coefficient is 1 while the relative error is 0.04. This shows that, under ideal conditions, a good reconstruction of the very weak ISW signal can be obtained using these data sets. 
When using all the CMB information with only one survey, the relative error increases to 0.76 (low-z), 0.62 (intermediate-z), 0.32 (high-z) and 0.34 (lensing), confirming that most of the reconstructed signal is achieved by extracting the information from the high-z survey or lensing.
The effect of the CMB information to the final ISW reconstruction is moderate, although its relative importance is larger when less information about the LSS is available. For instance, when combining CMB intensity and polarization with the low-z survey, the relative error of the reconstruction improves by $\approx$ 17 per cent with respect to the case when only the survey information is used.
Adding CMB polarization on top of temperature data is reflected in an improvement of a few per cent in the mean correlation coefficient and relative error in the ISW reconstructed signal, with the major differences found for the WF case, where only CMB data is used (around 15 per cent improvement in the mean correlation coefficient and 5 per cent in the relative error).

As a further test of the quality of the reconstruction, we have also estimated the power spectrum of the recovered ISW averaging over 10000 simulations for each of the considered cases. In Fig.\ref{fig:cls_4comp} we plot the $C_{\ell}$'s of different reconstruction for the case with CMB (intensity and polarization), all surveys and lensing (red diamonds) together with the expected reconstructed power spectrum given by  Eq. (\ref{eq:cl_lbc_n_E})  (red line) and the fiducial model (black line). We also plot the unbiased reconstructed power spectrum (black asterisks): once corrected for the known bias the reconstructed $C_{\ell}$'s agree well with the fiducial model. In Figs.~\ref{fig:cls_all} and \ref{fig:cls_all_bis} we give the uncorrected (i.e. biased) $C_{\ell}$'s of several reconstructions to enhance the differences between the considered cases. In particular, Fig.~\ref{fig:cls_all} shows the results using CMB intensity and polarization in combination with all the LSS tracers or with only one of them while Fig.~\ref{fig:cls_all_bis} shows the comparison between the reconstructed power spectrum when using only LSS information (cyan line), adding CMB intensity (red) and adding CMB polarization (blue). As already shown in the previous results, these three cases are very similar under ideal conditions. The results for the Wiener filter with (orange) and without (green) polarization are also plotted, showing a certain improvement when polarization information is included.
We note that the recovered power spectra of both figures is in agreement with Eqs. \ref{eq:cl_lbc_n} and \ref{eq:cl_lbc_n_E} for the results without and with polarization, respectively. 

We have also studied how the ISW reconstruction is affected by the presence of noise simulated according to the values given in Table~\ref{tab:poiss}.
The results are summarised in the bottom part of Table~\ref{tab:corr_coeff}, while the power spectra of the different reconstructions are given in Fig. \ref{fig:cls_all_noise}. 

The presence of noise degrades significantly the quality of the ISW reconstruction, although not all data sets are equally affected. For the best case, i.e. combining all the CMB, lensing and LSS information, the correlation coefficient is now 0.81 (to be compared with 1 in the ideal case) and the relative error increases to 0.58 (versus 0.04 without considering noise). This is mainly due to the degradation of the information in the high-z survey which is significantly affected by Poissonian noise. 
Note that for the intermediate-z survey the simulated Poisson noise is low (see Table~\ref{tab:poiss}) and, therefore, its effect is almost negligible, as reflected in the values of $\rho$ and $e_r$ in Table \ref{tab:corr_coeff}. Also the lensing seems to be quite affected by noise, the relative error for the lensing-only case goes from 0.35 to 0.87 when adding noise.

Due to the degradation of the LSS information when introducing Poissonian noise, we also find that the relative role of the CMB data in the ISW recovery becomes more significant in this case. Conversely to the noiseless case, where the reconstruction obtained using only LSS data was already close to the one obtained using all the considered data sets, now it becomes apparent that including the CMB improves the quality of the final reconstruction. In particular, in Fig.~\ref{fig:cls_all_noise}, it can be seen that the bias is reduced at large scales (l$< 10$) when adding CMB intensity data to the reconstruction (red line) versus the case when only LSS data are used (cyan line). Moreover, the use of CMB polarization (dark blue line) slightly reduces the bias at the smallest multipoles. Therefore, in a realistic case, the combination of CMB and LSS data to recover the ISW signal becomes especially relevant. 

As a further test, we have also investigated the case in which the reconstruction is obtained using only one survey at a time with the same level of noise added to all the surveys. As Poissonian noise we choose that of the low-z survey, since it is an intermediate value among the ones considered for the three surveys (see Table \ref{tab:poiss}). As expected, we find that in this case the relative performance of the surveys is the same as in the noiseless case, with mean correlation coefficients of 0.78 (high-z survey), 0.75 (intermediate) and 0.62 (low) for the reconstruction obtained using the considered survey plus CMB intensity and polarization.

\subsection{Incomplete-sky case}
\label{sec:mask}

\begin{savenotes}
\begin{table}
 \centering
 \begin{tabular}{@{}lcccccc@{}}
 \hline
      &  \multicolumn{3}{|c|}{$\bar{\rho}$}  &  \multicolumn{3}{|c|}{$e_r$} \\
\hline
      &  T+E   & T & no CMB  &  T+E   & T & no CMB \\
\hline
 Noiseless  &    &   &     &  & & \\
\hline
3surveys+lens. & 0.98 & 0.98 & 0.98 & 0.20 & 0.20 & 0.21 \\
3surveys      & 0.90 & 0.90 & 0.89 & 0.41 & 0.41 & 0.42 \\
low\footnote{\label{nota1} Note that a few cases involving the low-z survey as the only tracer seem to provide slightly better results for $e_r$ when a mask is considered than in the corresponding full-sky case (given in Table~\ref{tab:corr_coeff}), although this does not occur for the correlation coefficient. To solve this apparent inconsistency, one should compare the results from the masked case to those obtained in the same way for the full-sky reconstruction, i.e., considering only those pixels allowed by the union mask and after subtracting the monopole and dipole outside the mask from the input and reconstructed ISW. In this way, we find that the values of $e_r$ are actually the same for both, masked and full-sky cases, up to the considered precision.} & 0.61 & 0.57 & 0.41 & 0.77 & 0.79 & 0.89 \\
int.          & 0.75 & 0.73 & 0.68 & 0.64 & 0.65 & 0.71 \\
high         & 0.89 & 0.89 & 0.89 & 0.44 & 0.44 & 0.46 \\
lensing         & 0.93 & 0.93 & 0.93 & 0.34 & 0.35 & 0.36 \\
Wiener Filter& 0.52 & 0.46 & -    & 0.84 & 0.87 & - \\
 \hline
 Noisy  &    &   &     &  & & \\
\hline
3surveys+lens. & 0.79 & 0.77 & 0.74 & 0.59 & 0.60 & 0.64 \\
3surveys      & 0.76 & 0.74 & 0.70 & 0.62 & 0.64 & 0.68 \\
low$^4$ & 0.60 & 0.57 & 0.41 & 0.77 & 0.80 & 0.89 \\
int.           & 0.73 & 0.71 & 0.65 & 0.65 & 0.67 & 0.73 \\
high          & 0.53 & 0.48 & 0.19 & 0.83 & 0.86 & 0.97 \\
lensing          & 0.62 & 0.59 & 0.48 & 0.77 & 0.79 & 0.87 \\
Wiener Filter & 0.51 & 0.46 & -    & 0.84 & 0.87 & -\\
\hline
 \end{tabular}
     \caption{Mean correlation coefficient (columns 2 to 4) and mean relative error of the reconstructed maps (columns 5 to 7) computed over 10000 simulations, for incomplete sky datasets without (upper half) and with noise (bottom half) for the same combinations of CMB and LSS tracers as those in Table~\ref{tab:corr_coeff}. The values have been computed using only those pixels allowed by the union mask.}
   \label{tab:rms_mask}
 \end{table}
\end{savenotes}

\begin{figure*}
 \centering
 \subfloat{\includegraphics[width=5.0cm, angle=90]{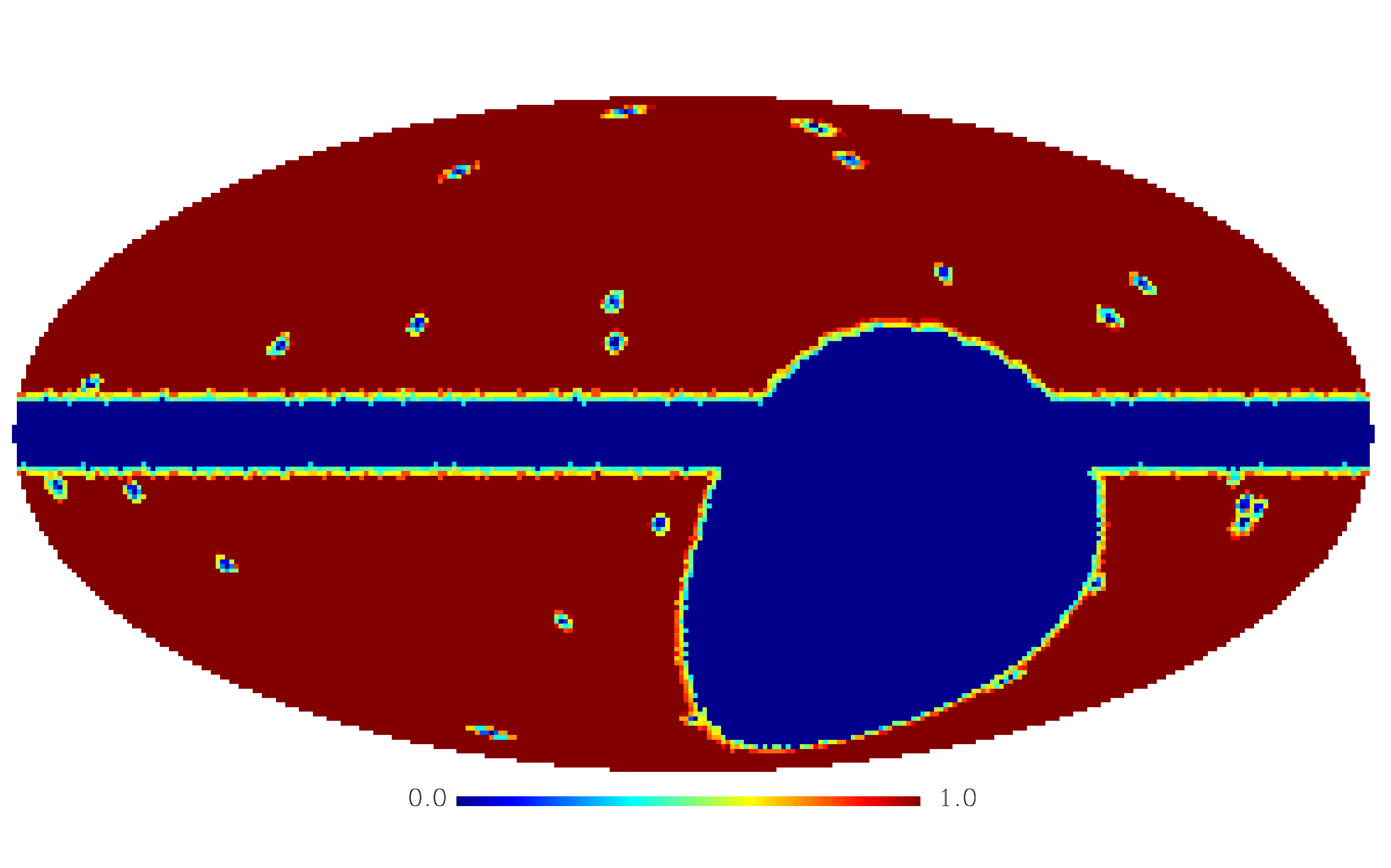}}\quad
 \subfloat{\includegraphics[width=5.0cm, angle=90]{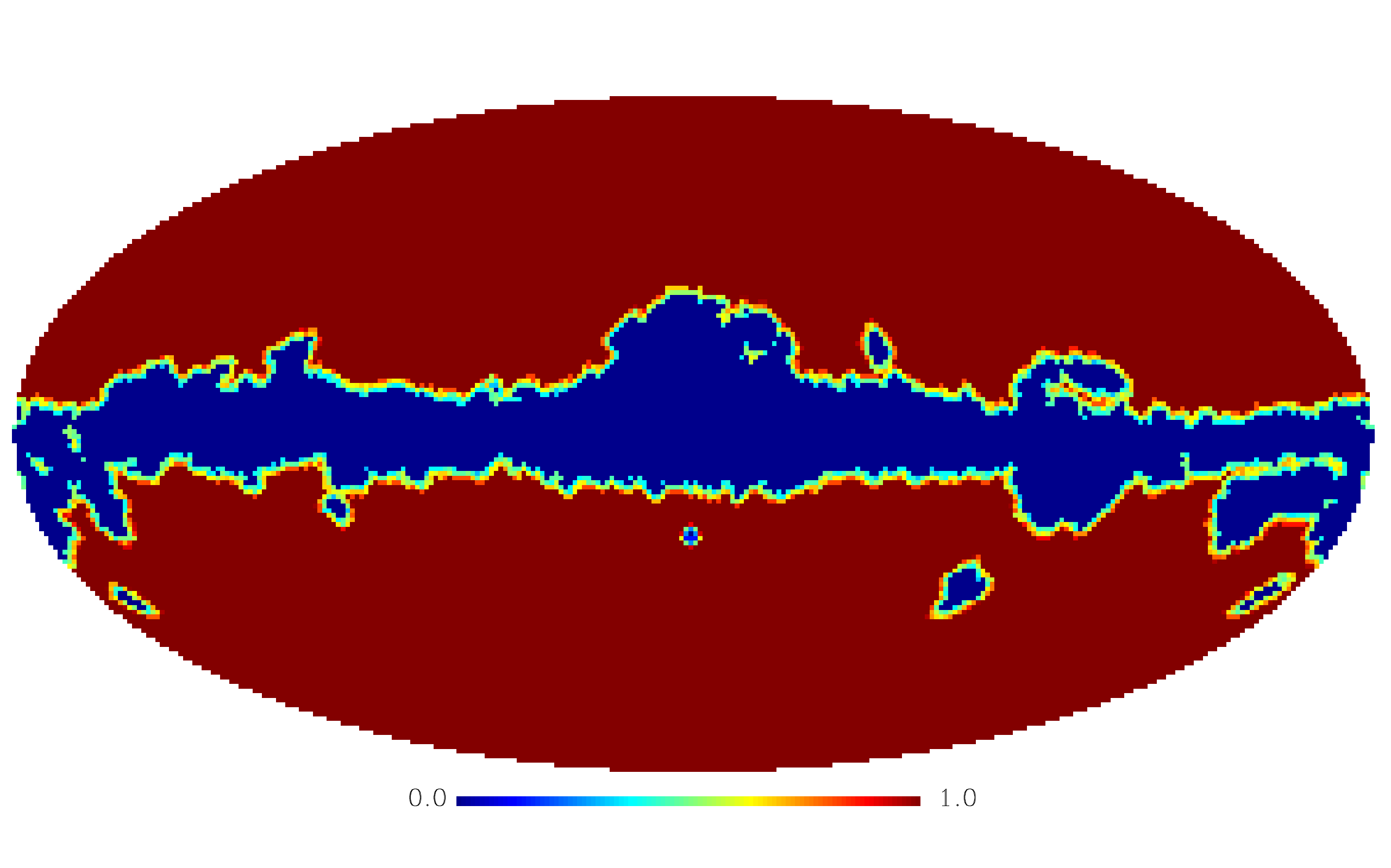}}\\
 \caption{Two of the apodised masks used in this work: high-z mask (left), and WMAP 9-year point source catalogue mask (right).}
 \label{fig:mask}
\end{figure*}

\begin{figure*}
 \centering
   \subfloat{\includegraphics[width=5.0cm, angle=90]{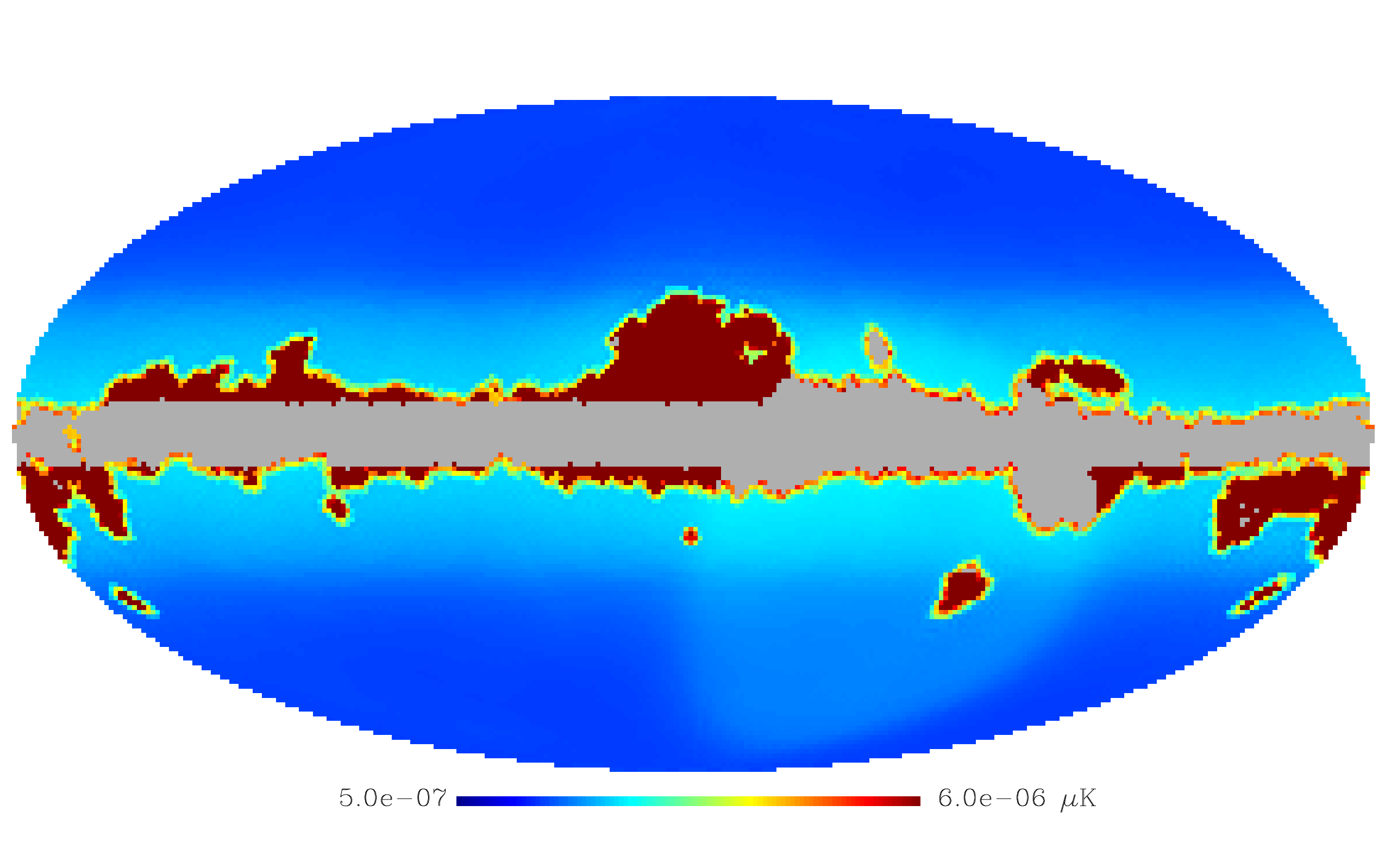}}\quad
   \subfloat{\includegraphics[width=5.0cm, angle=90]{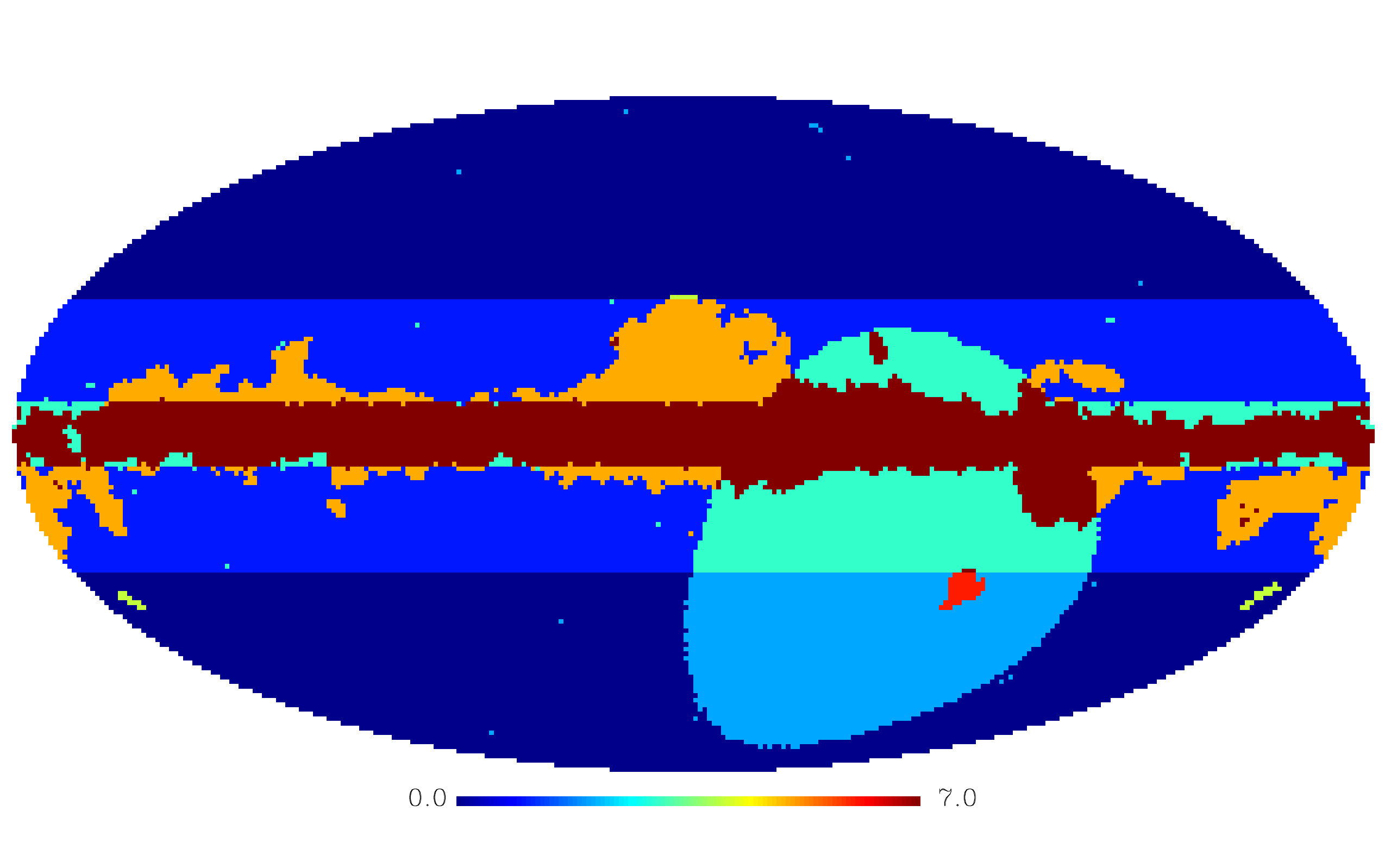}}\\
 \caption{On the left, a map of the dispersion of the residuals for the ISW recovery at each pixel, with the intersection mask applied, is given. It has been obtained from 10000 simulations for the case  with mask and without noise and using all the available information (three surveys, lensing, CMB intensity and polarization). On the right, the intersection of the masks considered for each data set is shown: 7 (dark red) stands for pixels which are not observed by any data set, 6 (red) for the pixels excluded by both high-z and in WMAP masks, 5 (orange) for those both in the WMAP and the galactic cut masks, 4 (yellow) for the pixels that are only in the WMAP mask, 3 (turquoise) for those both in high-z and the galactic cut masks, 2 (azure) for those covered only by the high-z mask, 1 (blue) for the pixels masked only by the galactic cut and 0 (dark blue) for pixels present in all data sets.}
 \label{fig:rms_isw_intermask_nonoise}
\end{figure*}

\begin{figure*}
 \centering
   \subfloat{\includegraphics[width=5.0cm, angle=90]{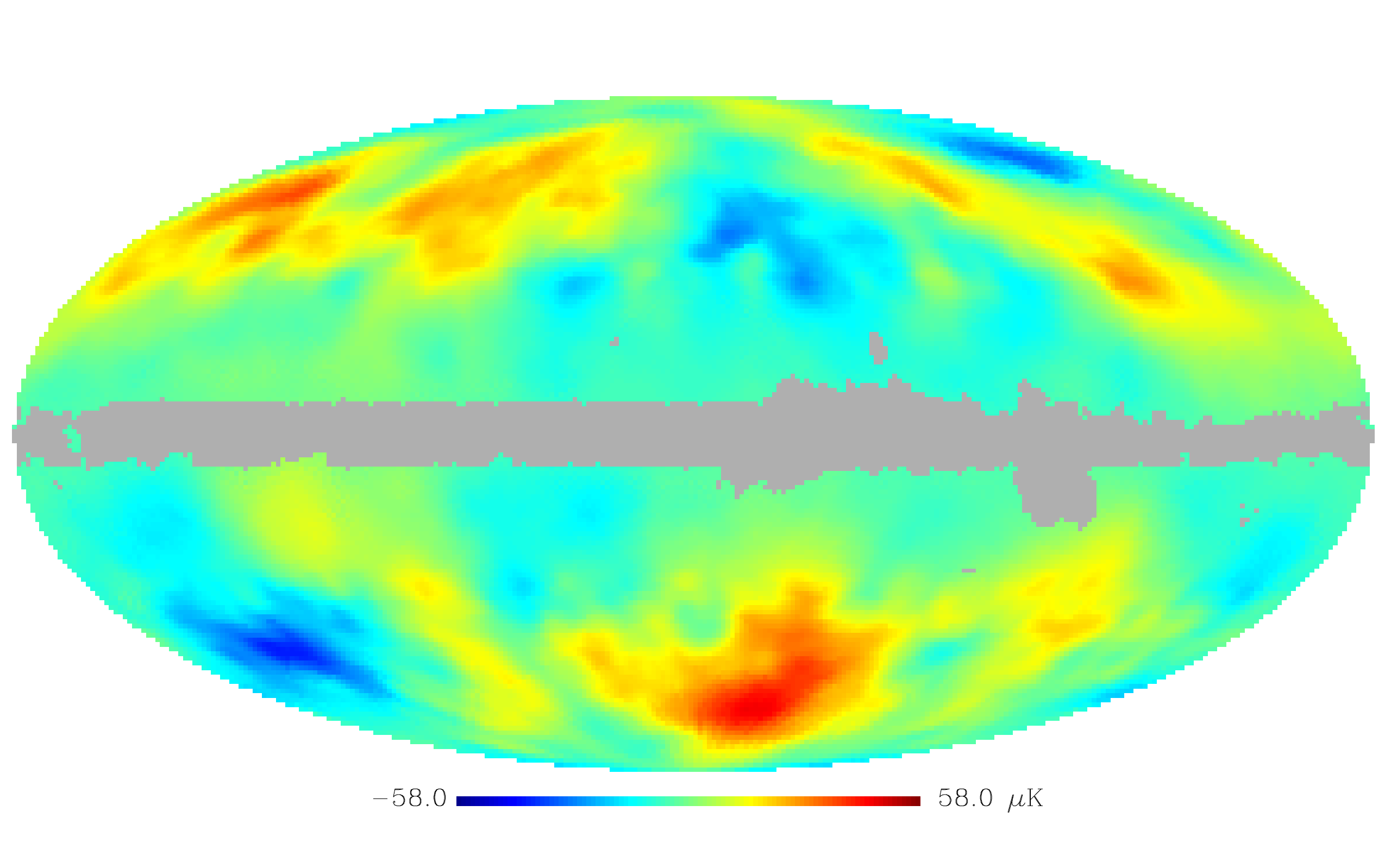}}\quad
     \subfloat{\includegraphics[width=5.0cm, angle=90]{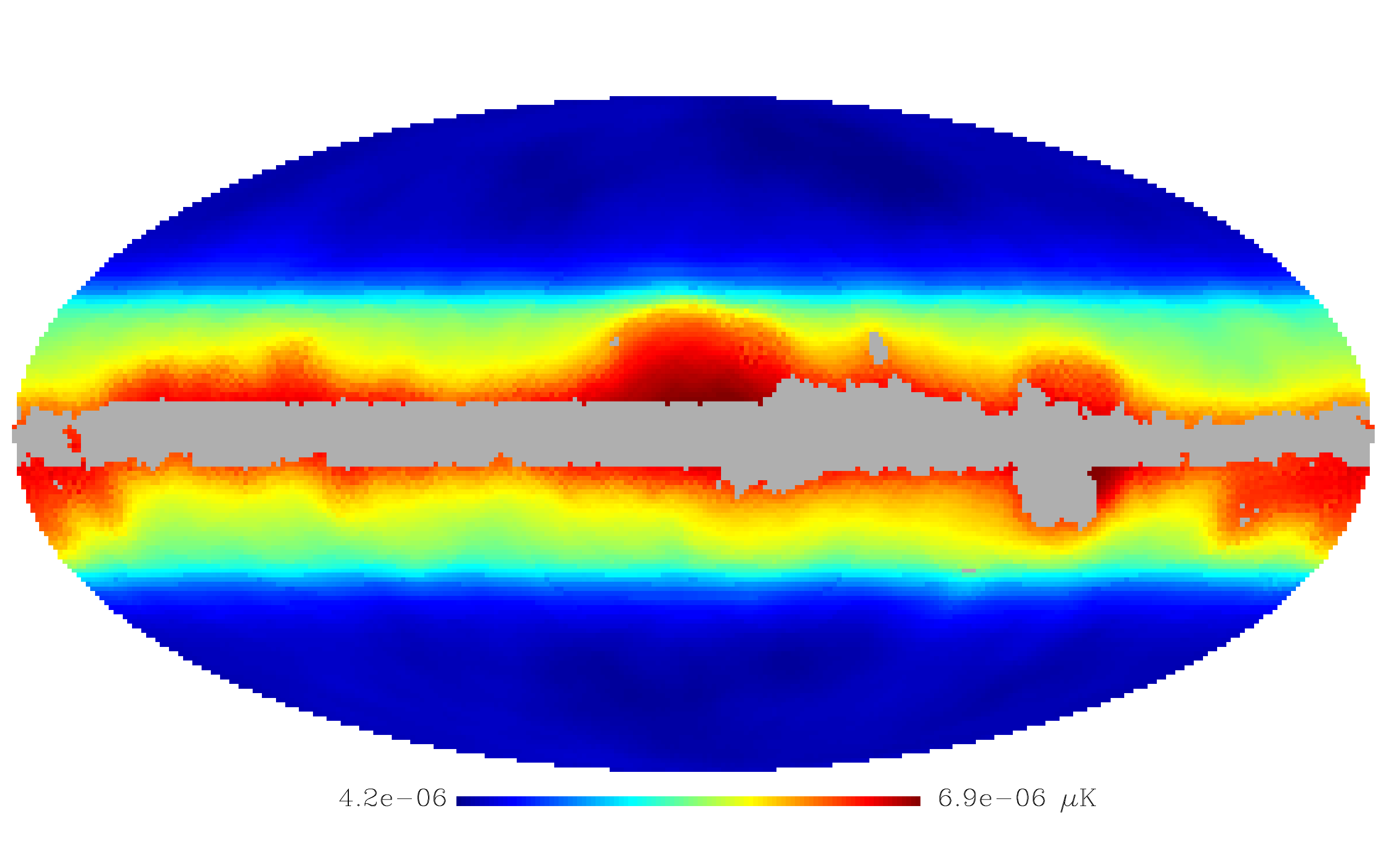}}\\
   \subfloat{\includegraphics[width=5.0cm, angle=90]{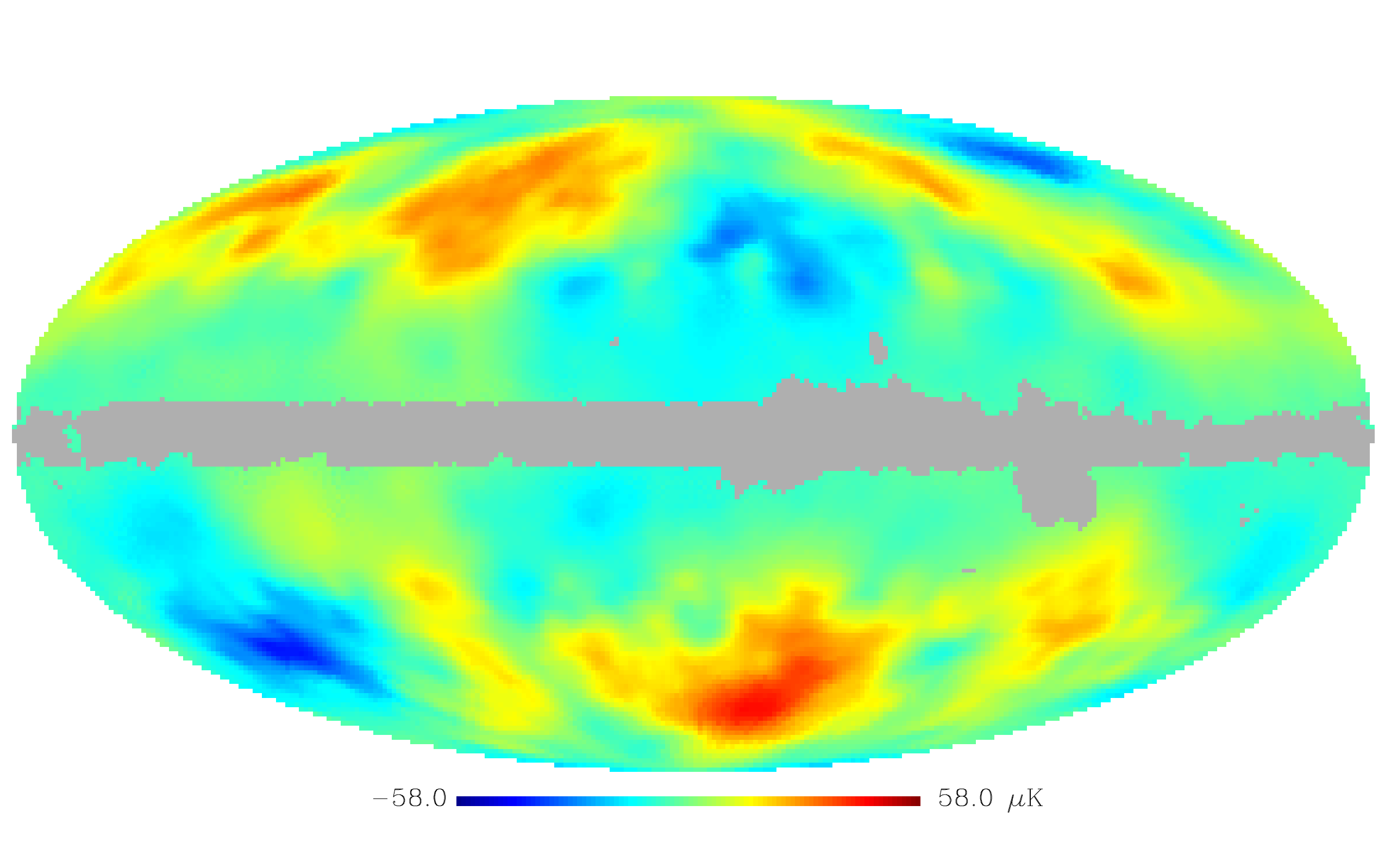}}\quad
   \subfloat{\includegraphics[width=5.0cm, angle=90]{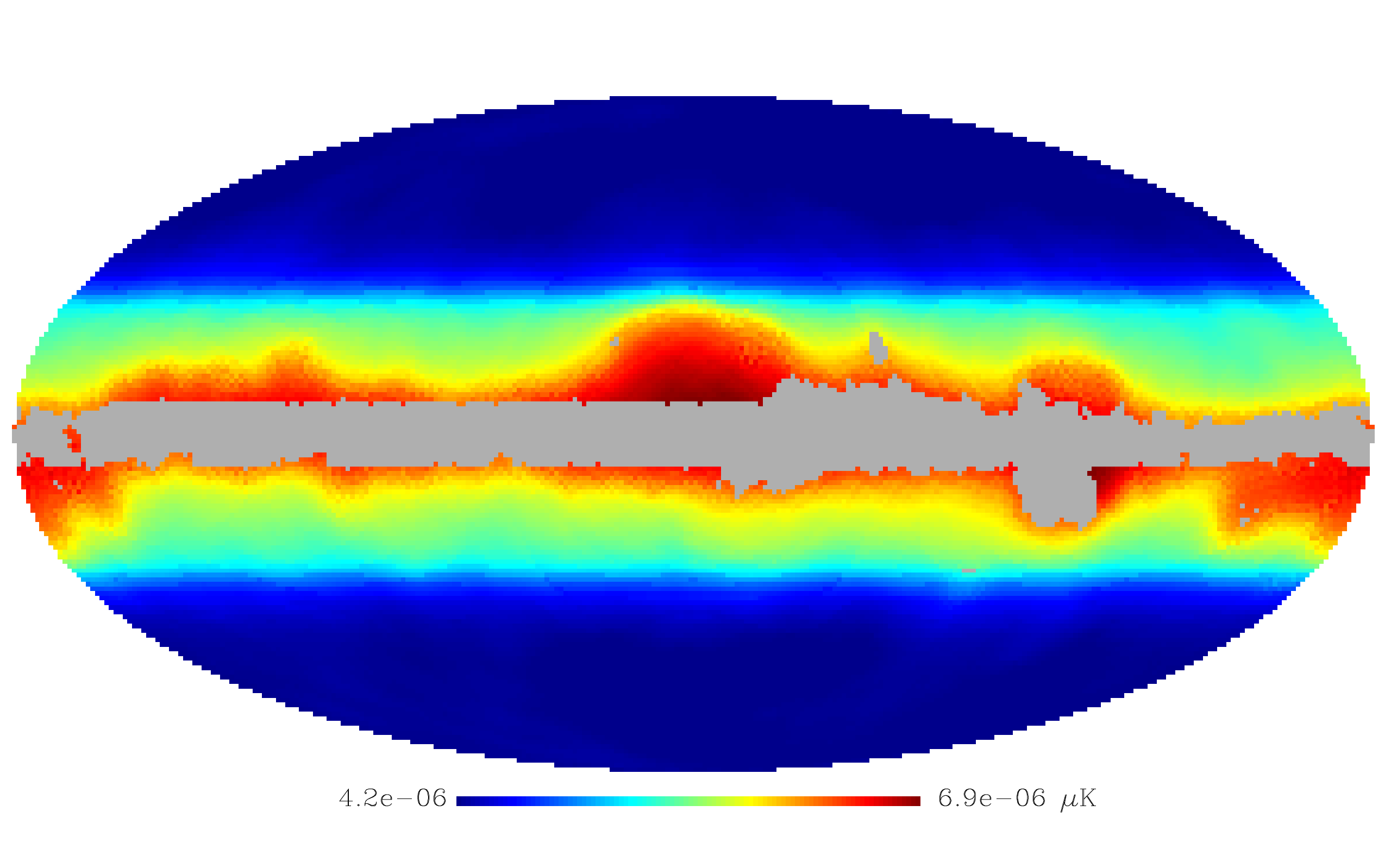}}\\
   \subfloat{\includegraphics[width=5.0cm, angle=90]{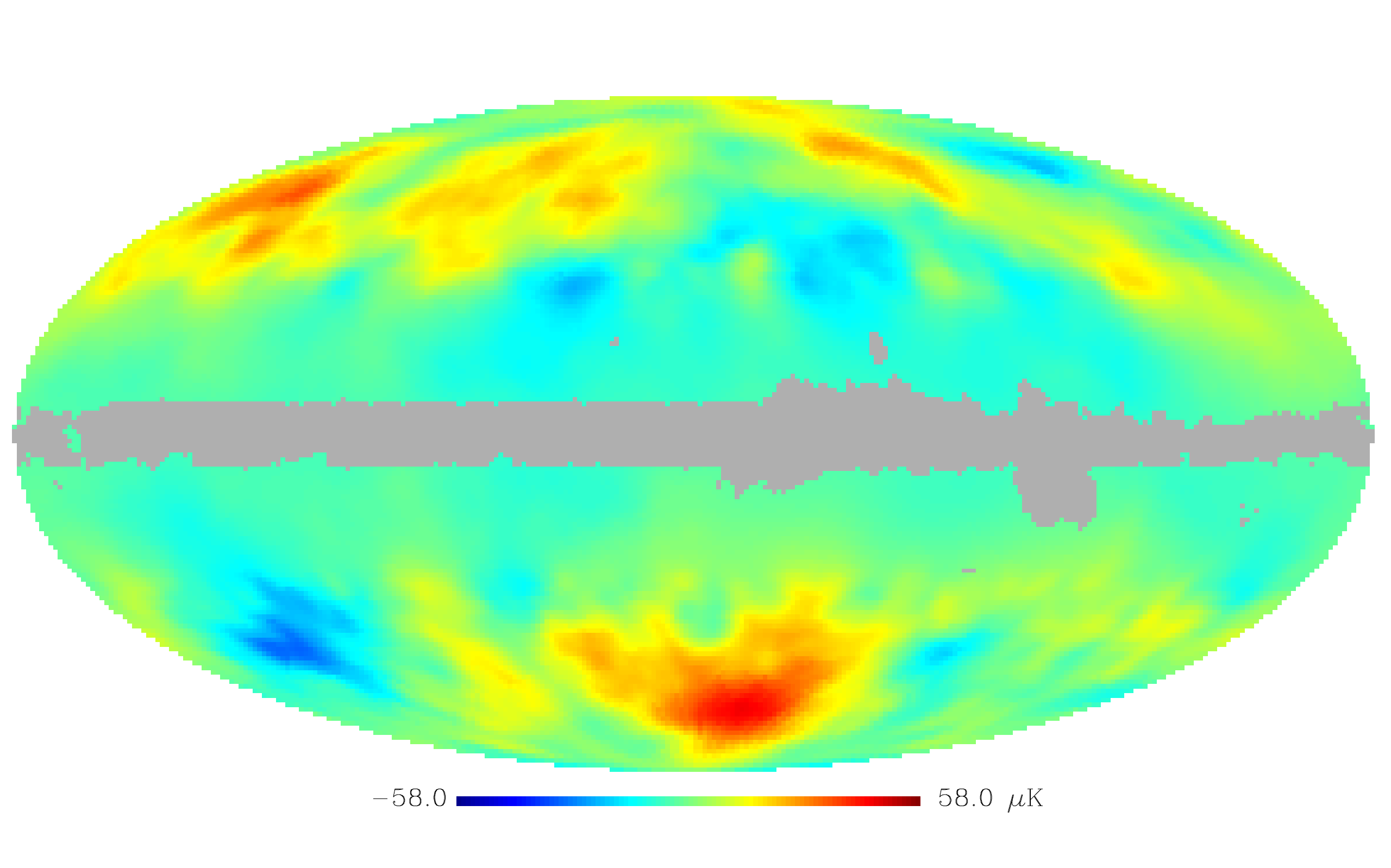}}\quad
   \subfloat{\includegraphics[width=5.0cm, angle=90]{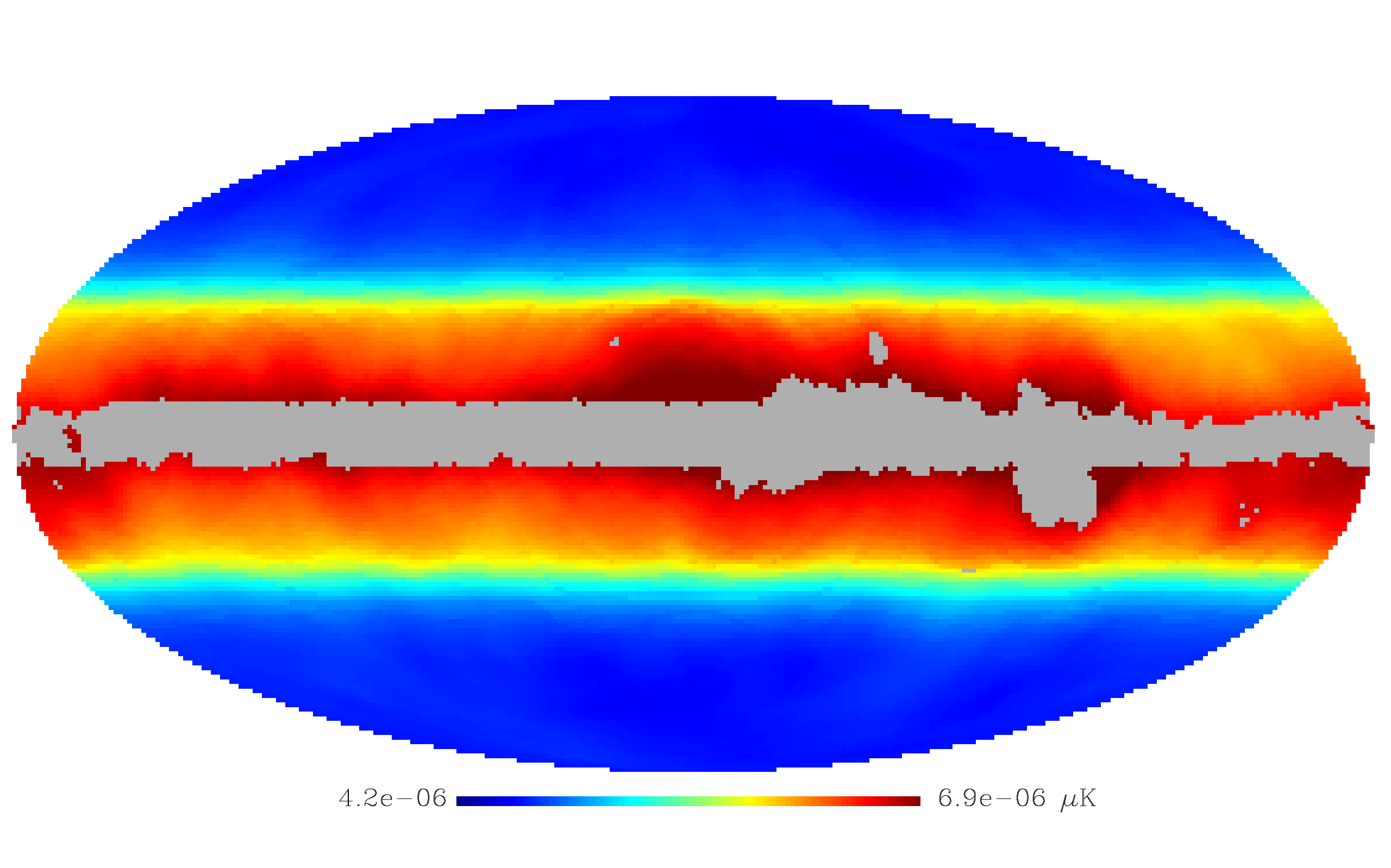}}\\
 \caption{Reconstruction of the ISW (left column) and its dispersion map (right column) in the realistic case of incomplete-sky and noise with (top) and without (middle) polarization information. The bottom panels show the plots for the case without CMB. For this particular simulation the correlation coefficients are 0.80, 0.76 and 0.74 respectively.}
 \label{fig:rec_isw_mask}
\end{figure*}

\begin{figure}
 \centering
   \subfloat{\includegraphics[width=8.5cm]{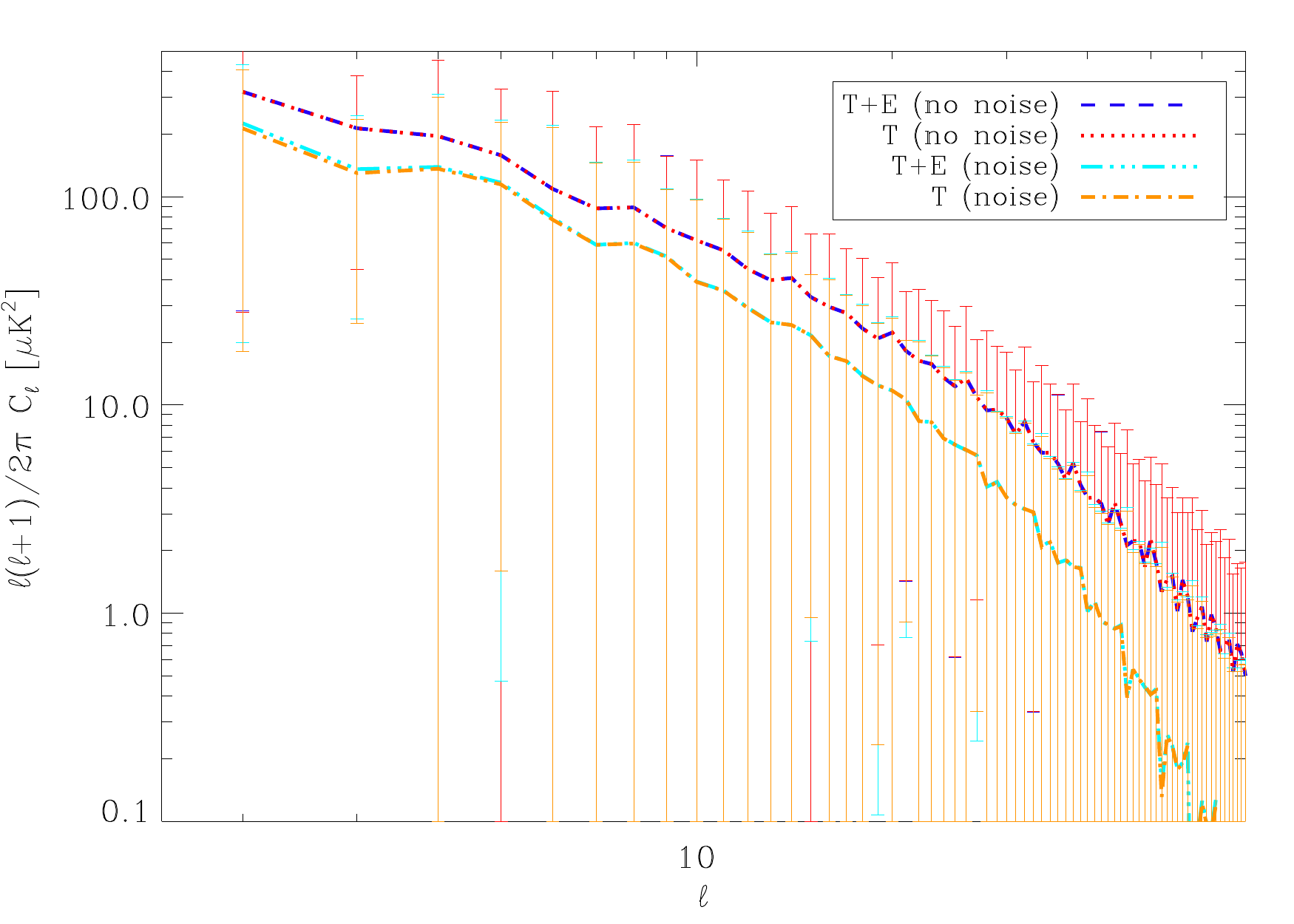}}
\caption{Power spectra computed with the union mask and corrected {\it a la} MASTER for the incomplete sky case when CMB temperature, all the surveys and lensing are used for cases with and without noise and with and without polarization. As before, it has been obtained averaging over 10000 simulations.}
 \label{fig:cls_mask}
\end{figure}

In this section, we extend the previous study to the more realistic situation in which the considered data sets are not full-sky. 
In order to separate the effect due to the presence of a mask from that of the noise, first we will present the results for noiseless incomplete sky data and then we examine the effect that introducing noise has in the recovered ISW signal.

Incomplete or contaminated data implies the presence of a mask, which degrades the estimation of the spherical harmonics coefficients of the map. The masks used are described in section \ref{sec:simu}. To mitigate the spurious correlations introduced in the harmonic coefficients due to incomplete sky data, we use an apodised version (with a cosine function) of the masks, which are obtained by extending 3 pixels the boundary of each mask, where values between 0 and 1 are used (see Fig.~\ref{fig:mask}). To compute the correlation coefficient and the mean dispersion of the residual map we apply for each case the mask obtained as the union of the masks used for that reconstruction (i.e., a pixel is considered only it it has been observed by all the data sets used for that particular reconstruction). Prior the calculation of these statistics, the monopole and dipole is removed outside the union mask for both the input and recovered ISW map.

Table~\ref{tab:rms_mask} gives the mean correlation coefficient between input and reconstructed ISW and the mean dispersion of the residual map for the noiseless (top) and noisy (bottom) incomplete sky case. These values are obtained averaging over 10000 simulations. To allow for a straightforward comparison, the same data combinations as in Table~\ref{tab:corr_coeff} are given. 

As seen, the presence of a mask degrades, but only slightly, the quality of the reconstruction outside the union mask. In particular, the correlation coefficient between input and reconstruction is 0.98 (versus 1 for all-sky) when all the data are included and no noise is considered, with the lensing survey giving the maximum contribution. It is interesting to point out that, conversely to the full-sky case, using only lensing is even better than combining the three surveys: the correlation coefficient is 0.93 against 0.90 and the relative error is 0.34 against 0.41 for the case when CMB (both intensity and polarization) is used. This can be understood taking into account that only one mask is needed for the former case, while three different masks are considered for the latter. Therefore, the reconstruction using the three surveys is more affected by having incomplete sky data sets than the one obtained from lensing. We find similar results for the case without CMB. 
Regarding the contribution of the CMB polarization, the same conclusions applied as for the full-sky case, i.e., its addition improves, but only slightly, the quality of the reconstruction. Again, the contribution of the polarization to the final reconstruction is more important in those situations where the recovery of the ISW is more difficult, such as in the presence of noise or when less information from LSS tracers is available.

We have also studied in more detail the performance of the method in those regions which are allowed by the intersection mask (mask which excludes only those pixels which have not been observed by any data set) but not by the union mask, i.e., on pixels that have been observed only by some of the considered data sets. Fig. \ref{fig:rms_isw_intermask_nonoise} shows on the left, the map of the dispersion of the residuals for the ISW recovery at each pixel, with the intersection mask applied. It has been obtained from 10000 simulations for the case with mask and no-noise, including CMB temperature and polarization, all the surveys and lensing. This can be compared to the map in the right side of the figure, which gives the intersection of the different considered masks, showing the areas that are reconstructed using only part of the data, going from 0 (pixels present in all the data, i.e. allowed by the union mask) to 7 (pixels that are not observed by any data set, corresponding to the intersection mask). As one would expect, the structure of the map of residuals resembles that given by the intersections of the masks, obtaining lower residuals in those areas where more data are available.

When the effect of noise is taken into account, the quality of the reconstruction is degraded in a similar way as that of the full-sky (see bottom part of Table~\ref{tab:rms_mask}). For the reconstruction obtained using all the information, the correlation coefficient decreases to 0.79 (to be compared with 0.81 when all-sky noisy data are considered) while the relative error slightly increases. The relative contribution of each survey to the final reconstruction is again modified with respect to the noiseless case.

In Fig. \ref{fig:rec_isw_mask} we plot the reconstructed ISW (left), for our reference simulation (given in Fig.~\ref{fig:simu_maps}), and its corresponding mean dispersion of the residuals (right) for incomplete sky coverage and with the presence of noise using all surveys and lensing for three different cases: including both CMB intensity and polarization (top), including CMB intensity only (middle)  and without CMB information (bottom). The intersection mask is applied to the maps. It can be seen that leaving out the CMB increases the error in the reconstruction, confirming again the fact that the contribution of the CMB (both temperature and polarization data) is more important when non-idealities are taking into account. In particular, for this cases, the relative errors are 0.59 (top panel), 0.60 (middle), 0.64 (bottom).
Comparing the right column with Fig. \ref{fig:rms_isw_intermask_nonoise} the different contribution of each component is less evident due to the presence of noise. It is also noticeable that the quality of the reconstruction in the region where only the high-z survey is masked (corresponding to the round masked region in the left panel of Fig.~\ref{fig:mask}, located mainly in the southern Galactic hemisphere) is very similar to that of the region where all data are available, reflecting that the inclusion of the high-z survey is not having a significant effect in the reconstruction in this case, due to its highest noise with respect to the other surveys.
We have also computed the mean angular power spectrum for the estimated ISW signal, after applying the apodised version of the union mask, using 10000 simulations for different cases (see Fig.~\ref{fig:cls_mask}). The effect of the mask on the power spectrum has been corrected {\it a la} MASTER~\citep{HIV02}.
Conclusions are very similar to those found when using all-sky data. 

\begin{figure*}
 \centering
 \subfloat{\includegraphics[width=5.7cm]{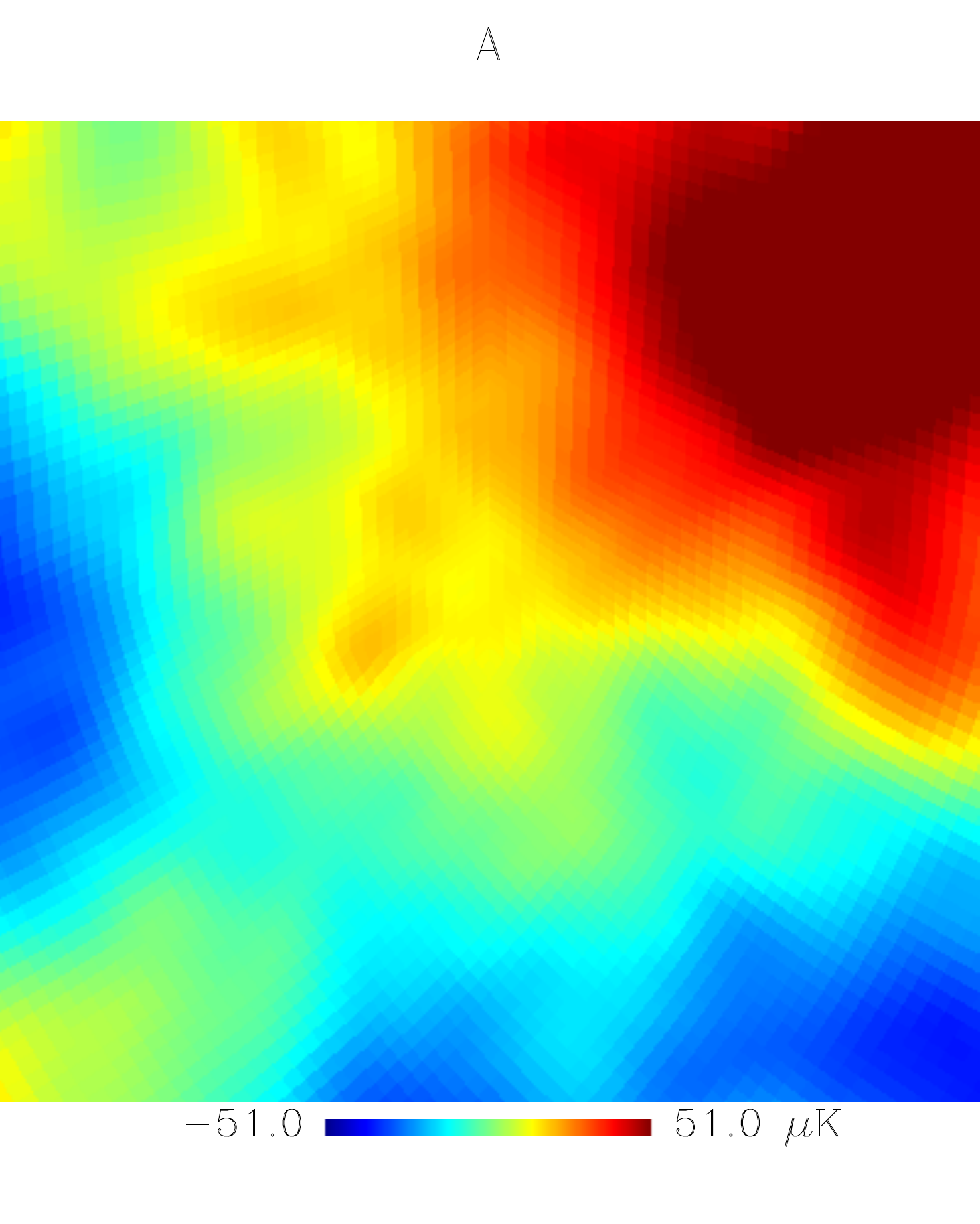}}\quad
 \subfloat{\includegraphics[width=5.7cm]{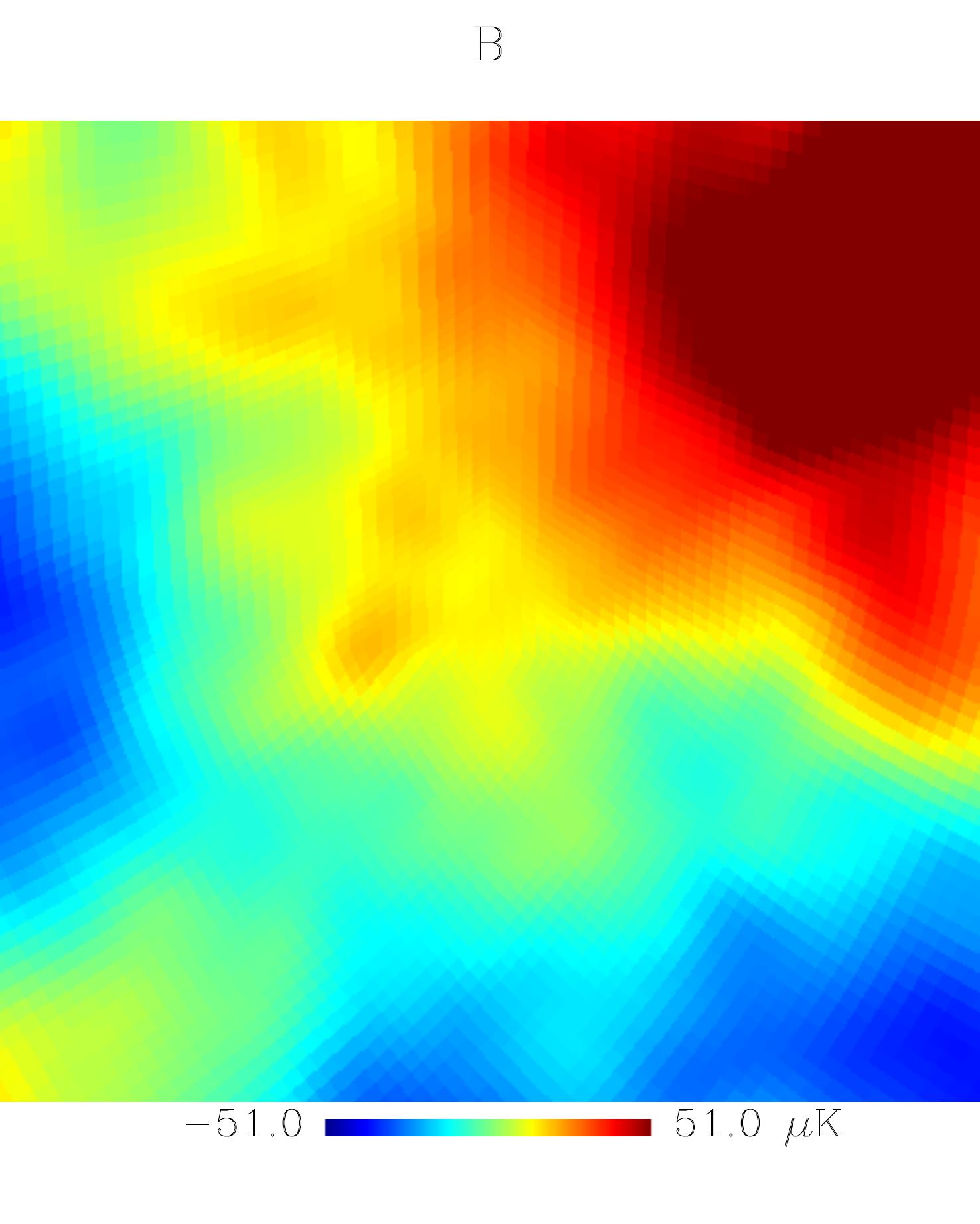}}\quad
 \subfloat{\includegraphics[width=5.7cm]{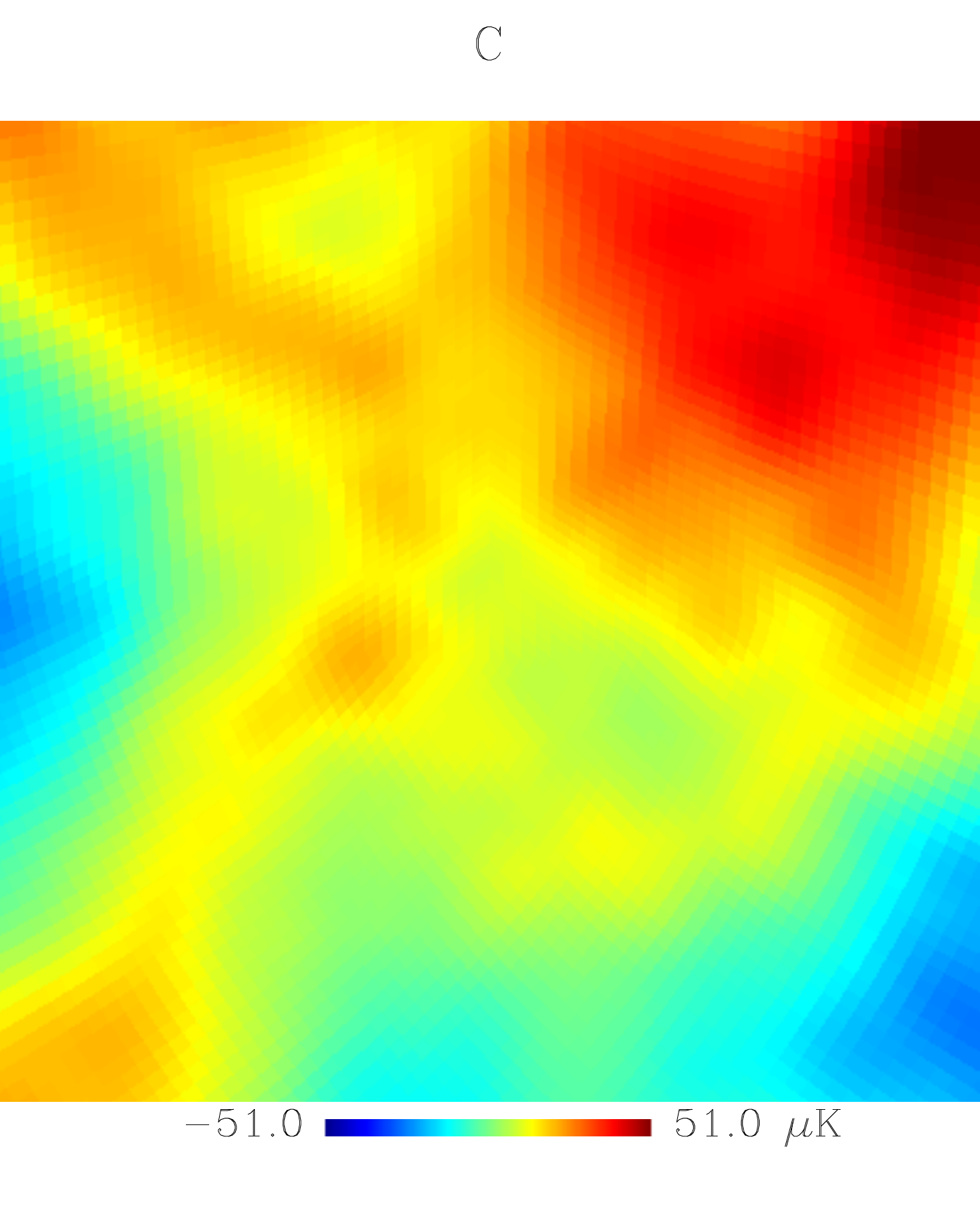}}\\
 \subfloat{\includegraphics[width=5.7cm]{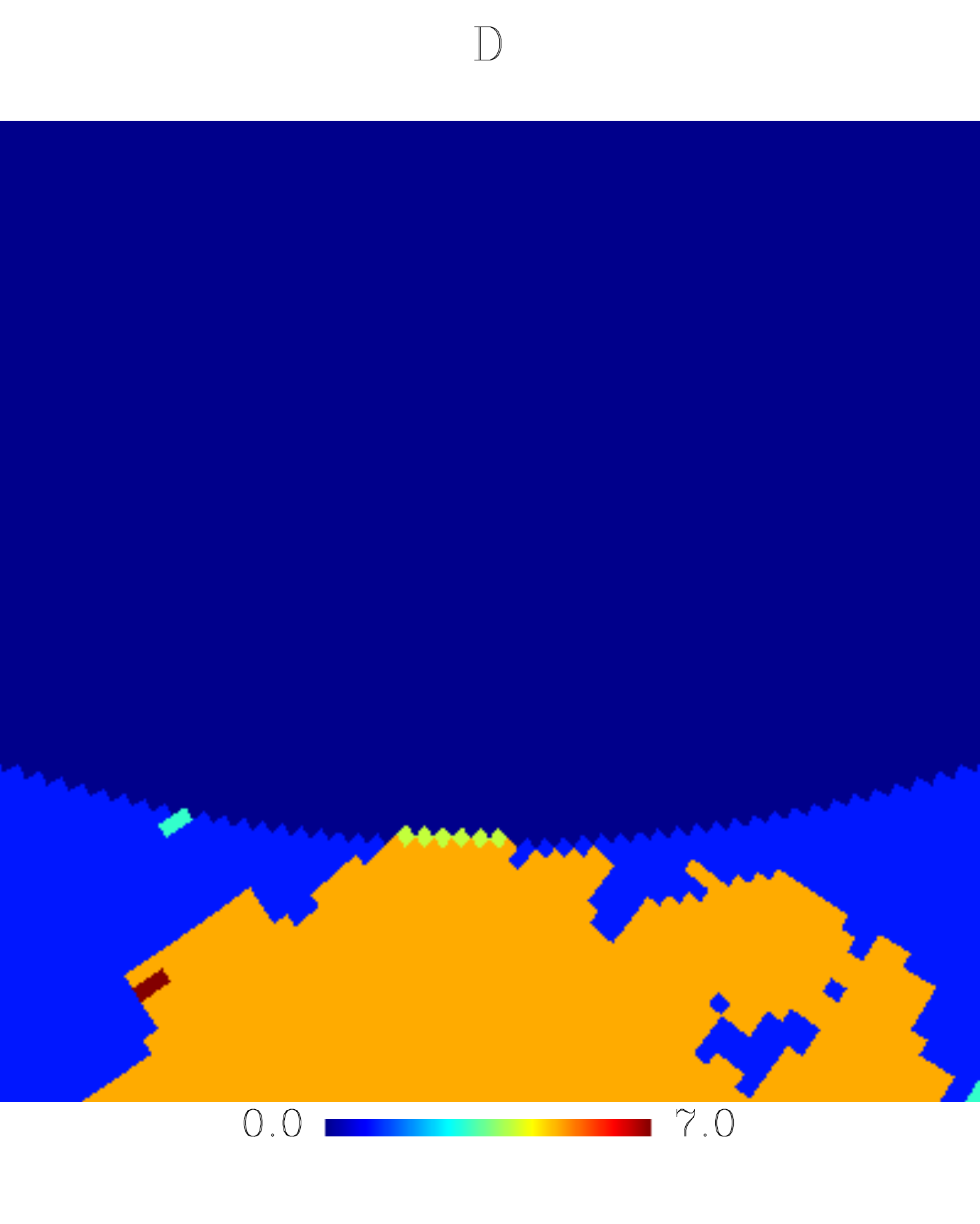}}\quad
 \subfloat{\includegraphics[width=5.7cm]{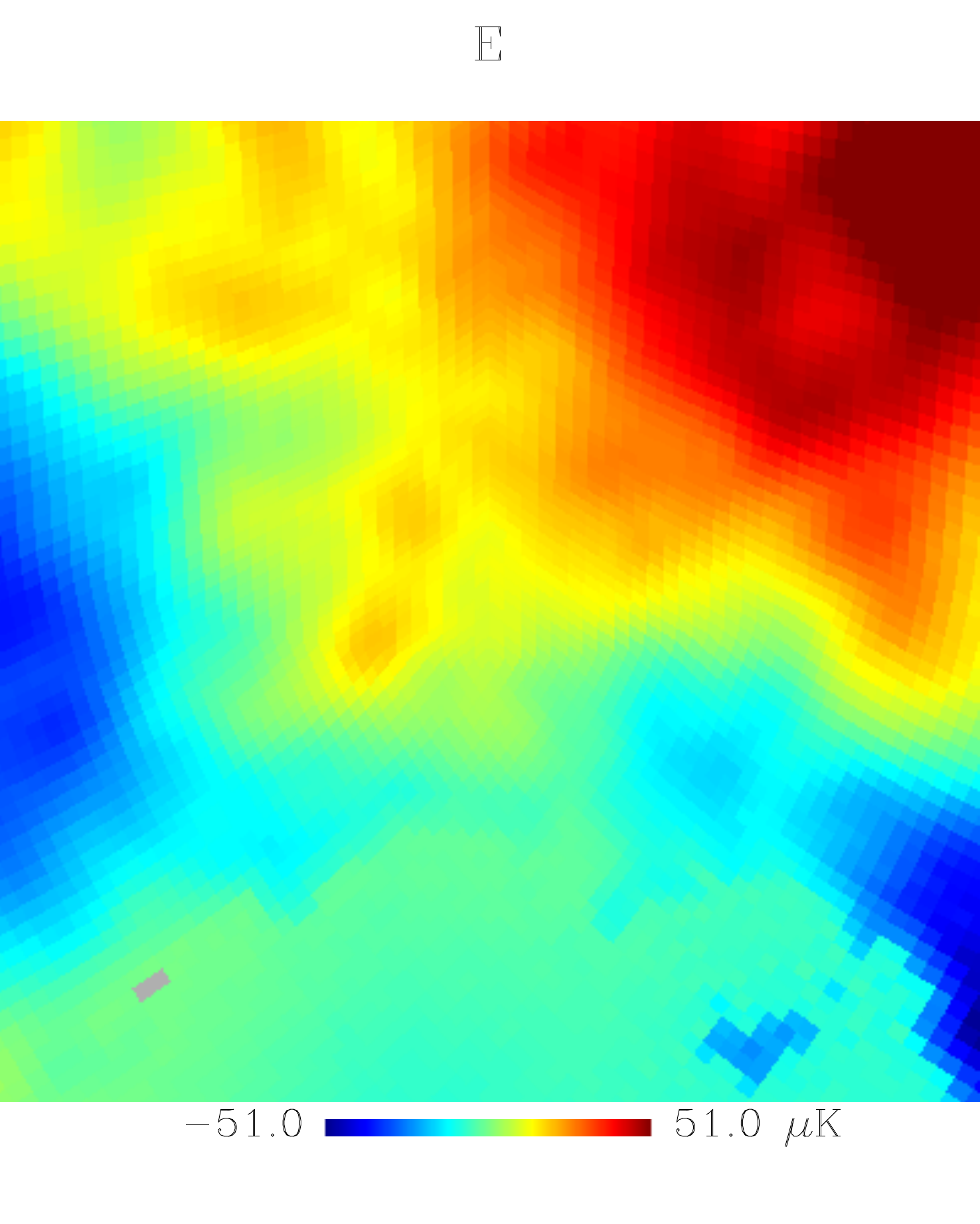}}\quad
 \subfloat{\includegraphics[width=5.7cm]{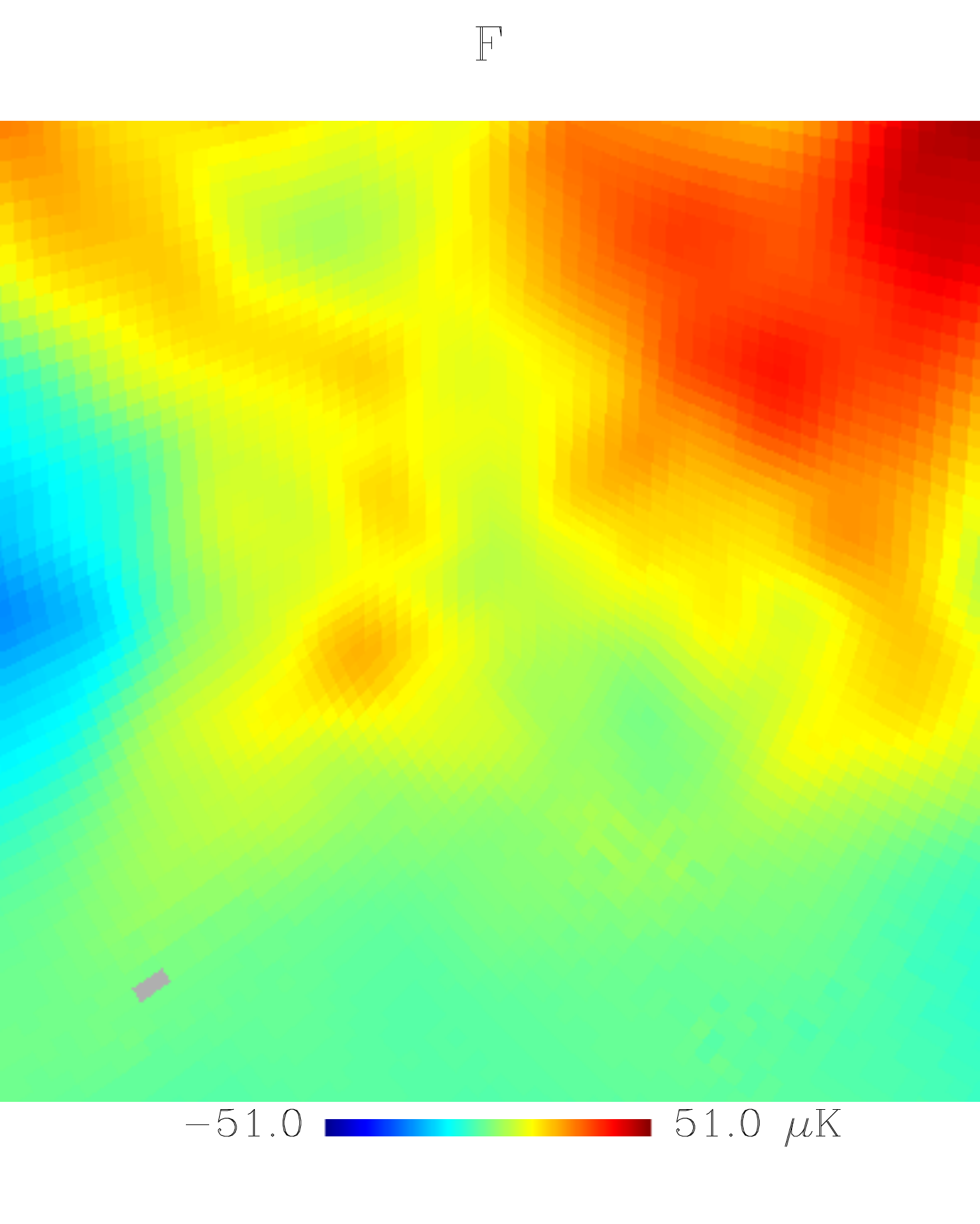}}\\
 \subfloat{\includegraphics[width=5.7cm]{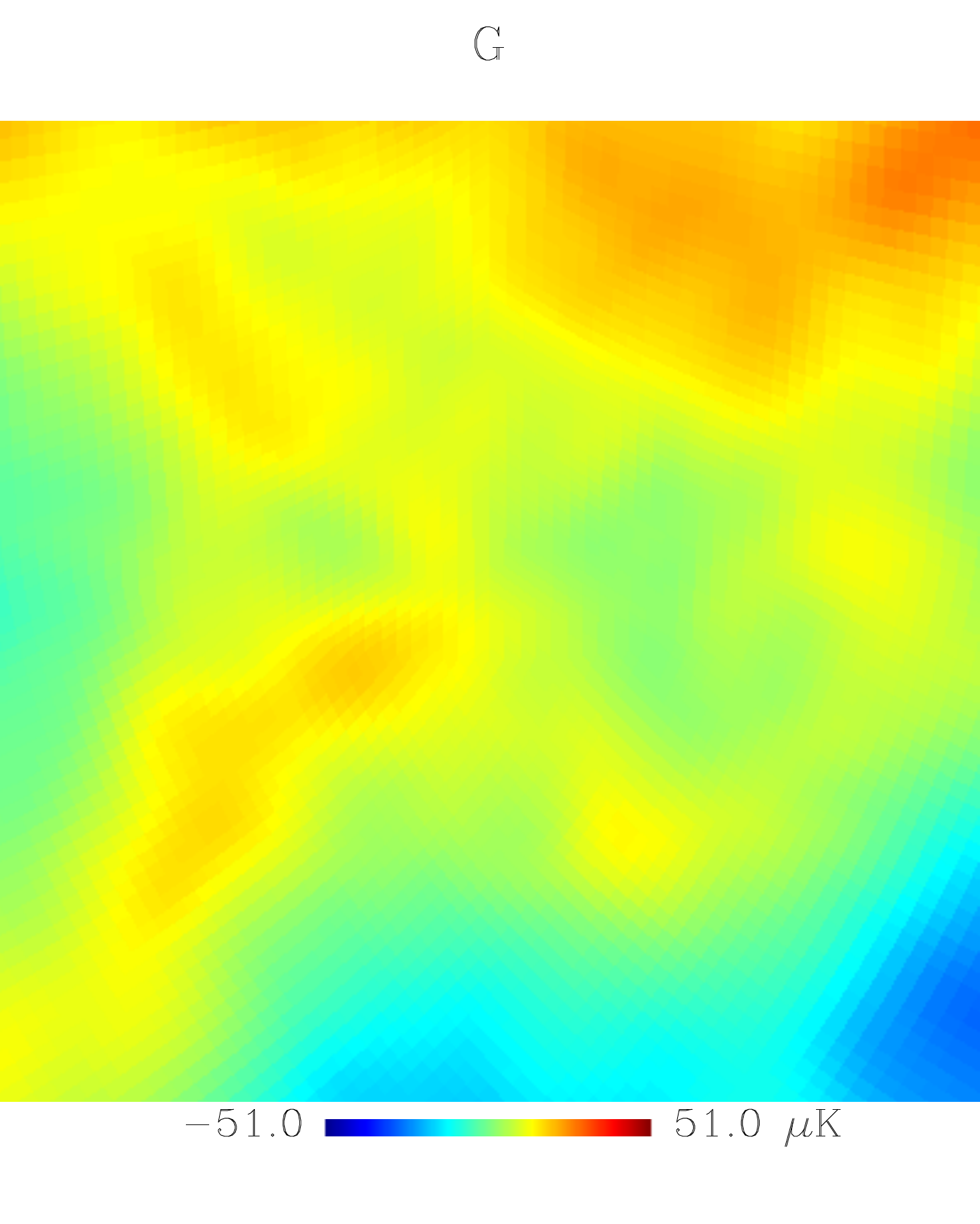}}\quad
 \subfloat{\includegraphics[width=5.7cm]{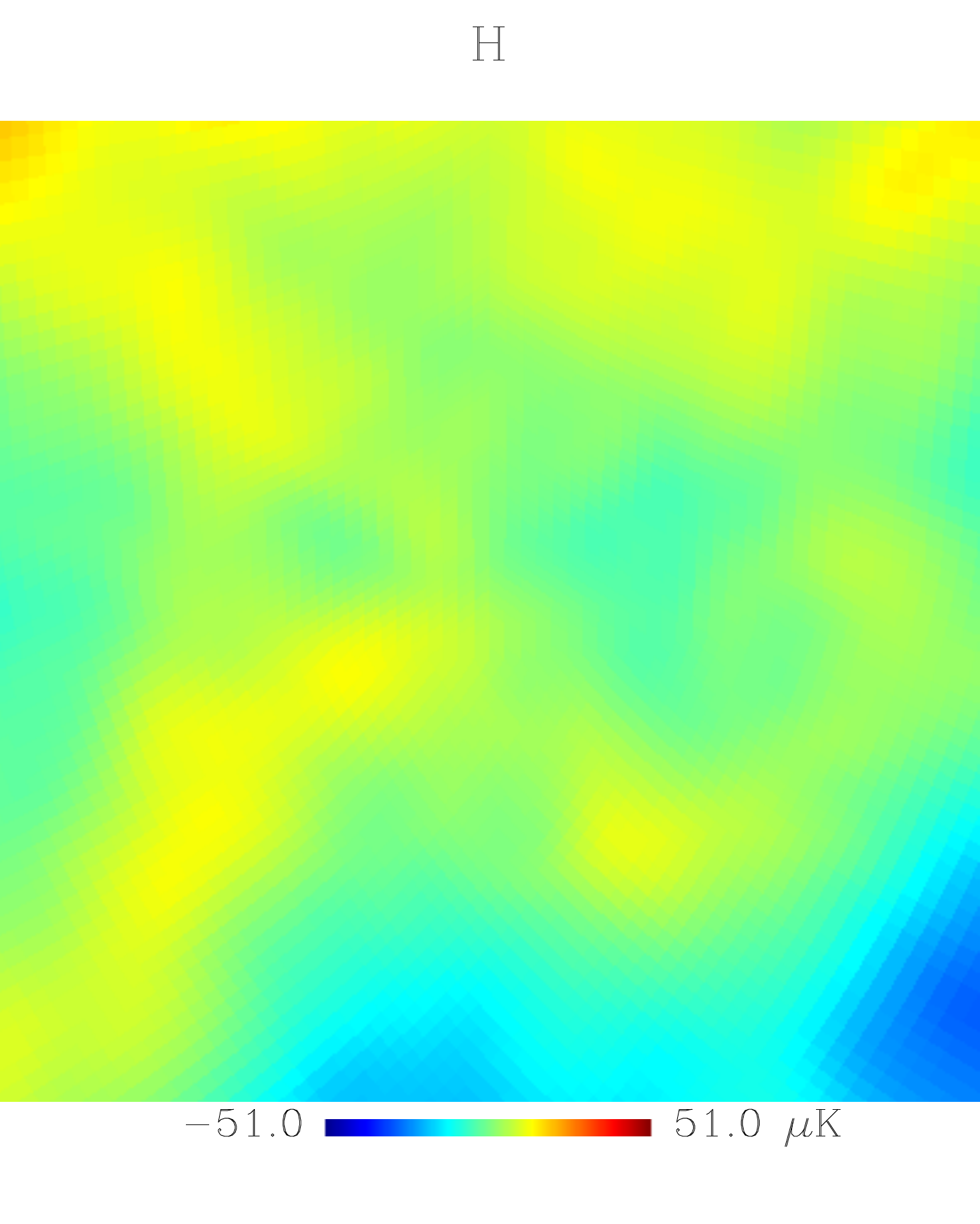}}\quad
 \subfloat{\includegraphics[width=5.7cm]{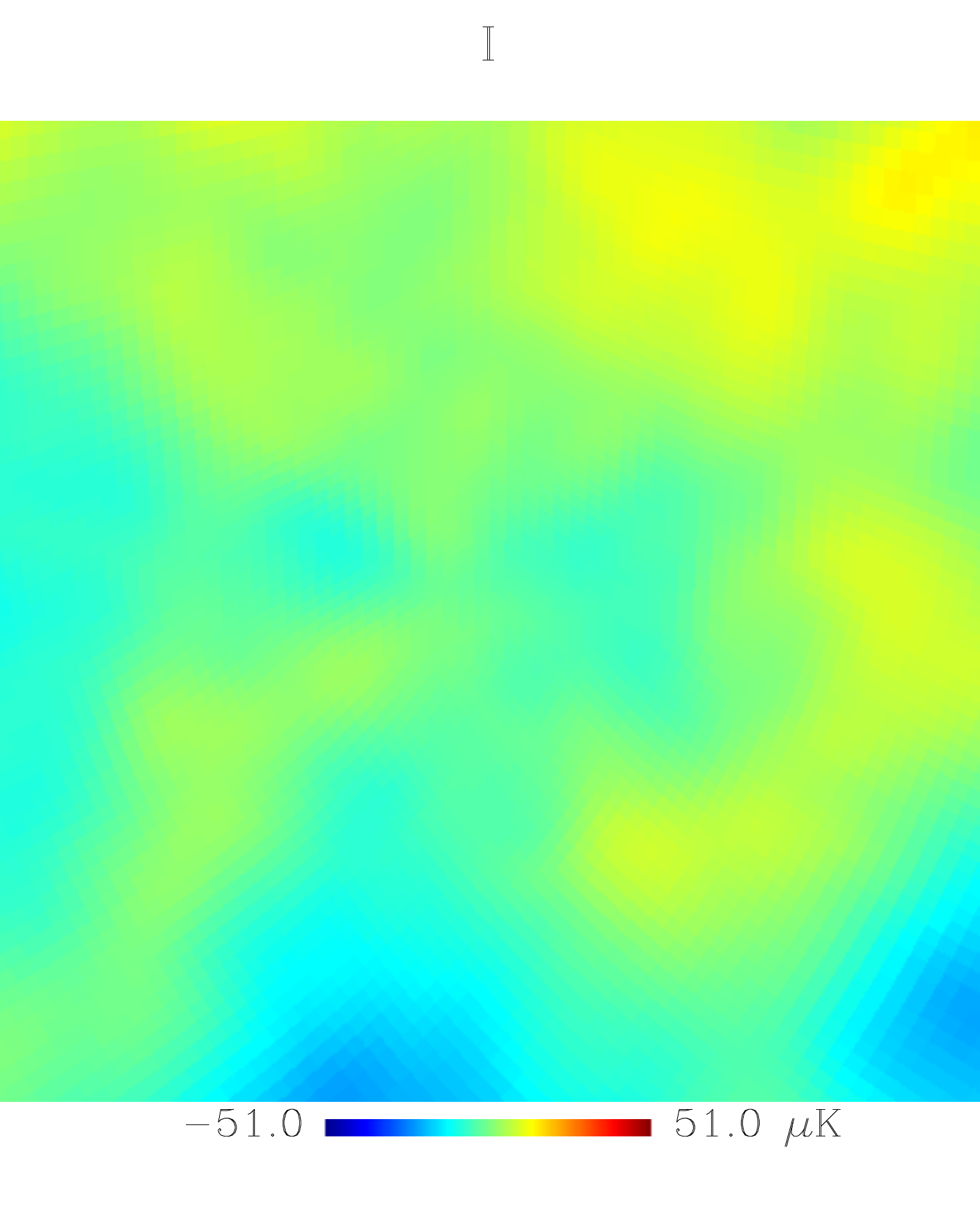}}\\
  \caption{Detail of the reconstructed ISW for one simulation for different cases in a patch of the sky centred at $l=0.0, b=45.0$: simulated ISW (A), reconstructed full-sky ISW using all available information without noise (B) and with noise (C), intersection of the masks in the considered region (D), reconstruction with all the available information in the masked case without noise (E) and with noise (F), full-sky reconstruction in presence of noise with the low-z survey and CMB both intensity and polarization (G), without polarization (H) and without CMB (I).}
 \label{fig:patch_fullsky}
\end{figure*}

In order to visualize in more detail the differences between the attained reconstructions, which may not be evident in the full-sky plots, we also show, for a given simulation, a comparison between the input and reconstructed ISW for several cases in a patch of the sky. In particular, Fig. \ref{fig:patch_fullsky} gives the reconstructed ISW for a given simulation for different cases in a patch centred at $l=0.0, b=45.0$. The patch covers an area of approximately 67 square degrees and has been chosen as representative of a region which is in the boundary of the union mask. This allows one to see better the effect of working with incomplete sky. In particular panel A is the simulated ISW signal and panel B is the reconstructed ISW in the most optimistic case when all the data are used (all the three surveys, lensing and the CMB both intensity and polarization) under ideal conditions: as stated in Table~\ref{tab:corr_coeff} the reconstruction is very good. Comparing with panel C (same as B with the addition of noise) we can see that the presence of noise reduces the quality of the reconstruction, although the main features of the ISW signal are still present in the reconstruction. 

Panel D shows the intersection of the masks in the selected region, corresponding to a detail of the right panel of Fig. \ref{fig:rms_isw_intermask_nonoise}. Note that pixels in dark blue are observed by all data sets, while those in orange are observed only by the high-z survey.
Panels E and F show the same cases as B and C but considering incomplete sky coverage. In the masked case, the ISW reconstruction in the pixels belonging to the orange region is significantly affected, due to the fact that the information in this region is more limited than in the rest of the patch but also to the effect of the mask. Conversely, the reconstruction in the region observed by all data sets is only slightly affected by the presence of masked areas, and the result is close to the one obtained in the same case when considering the full-sky. The reconstruction given in panel F is representative of which is the expected reconstruction for a realistic case when noise and incomplete sky coverage are taking into account. The comparison between these different cases confirms visually what is also inferred from the values of the correlation coefficient given in Tables~\ref{tab:corr_coeff} and \ref{tab:rms_mask}, that, when all the data sets are used, the main problem for the ISW reconstruction in a non-ideal case is the presence of noise rather than the mask, provided we restrict ourselves to pixels allowed by the union mask.

The last row of Fig. \ref{fig:patch_fullsky} illustrates the effect of including CMB intensity and polarization in the more difficult situation in which only one LSS survey is available. In particular, we show the reconstruction obtained for the full-sky case in the presence of noise using CMB intensity and polarization and the low-z survey (panel G), using CMB intensity and the survey (H) and keeping only the survey information (I). As seen, it becomes apparent that the inclusion of CMB intensity and polarization is important in order to improve the quality of the reconstruction although, even in this situation, the ISW signal is only very roughly recovered. The inclusion of CMB information is especially relevant for this case, since we have used the survey that peaks at low-z, whose information is less suited to recover the ISW effect. 

\subsection{Discussion on other non-idealities}

In addition to the non-idealities considered previously (sky masks, instrumental and Poissonian noise), other factors can affect the quality of the reconstruction, such as the presence of systematics in both LSS and CMB data or uncertainties in the knowledge of the cosmological model. In particular, the previous results assume a perfect knowledge of the correlation matrix for all the considered data sets. Given the many ways in which the assumed spectra can deviate from the underlying one in a general case as the one considered here, it is complicated to carry out an exhaustive study on how these uncertainties can propagate into the recovered signal. However, some tests can be performed to answer, at least partially, this question. In particular, this point was already addressed in \cite{BAR08}, where the quality of the ISW reconstruction was studied assuming a cosmological model different from the one used in the simulations. It was found that the correlation coefficient was only slightly affected by these uncertainties, although some differences were found in the amplitude of the reconstructed ISW,  since by using a wrong covariance matrix, the recovered map is not properly scaled by the filter.

In the current work, we have performed an additional test to study the effect of the uncertainties in the bias model. In particular, we have performed an additional set of simulations, where all parameters are left unchanged except for the high-z survey, which has been simulated assuming a redshift-dependent bias (based on the NVSS model of \citealt{MAR13}) and normalised to have a mean amplitude of the auto-power spectrum lower (around a 30 per cent) than our reference case. We have studied the quality of the reconstruction for two different cases: one assuming the correct underlying model and a second one, which uses the model of our reference data set (meaning that all elements in the covariance matrix related to the high-z survey differ from the ones used in the alternative simulations). When noise is included in the data, the difference between both cases is very small for both the correlation coefficient and the relative error, since the error introduced by the noise clearly dominates over the uncertainties in the model. In the ideal case (full-sky, no noise), we find that the correlation coefficient is only slightly affected by the assumption of a wrong model, whereas this is not the case for the relative error. For instance, when considering the reconstruction obtained by combining the CMB and the high-z survey, the relative error increases around a factor of two, when comparing the case reconstructed with the correct model versus that where a different model is considered. As in \cite{BAR08}, this indicates that the overall structure of the ISW signal is reasonably well reconstructed but its amplitude is more affected by possible errors in the model. 

It is also worth remarking that the LSS simulations used in this paper do not account for some of the systematics that are typically observed in galaxy catalogues, such as inhomogeneous sensitivity (which can be due to several effects), stars contamination, or galactic emissions, among others. As described in several works~\cite[e.g.,][]{Hernandez2010,Hernandez2014}, to deal with these systematics is not a trivial task, and it requires a deep knowledge of the instrumental characteristics of the experiment producing the data. In the best scenario, the proper treatment of these effects would be translated into a degradation on the ISW detection level and, in the worst case, into a bias on its determination. An exhaustive analysis of the influence of these systematics on the recovery of the ISW signal goes beyond the scope of this paper. In particular, because, as mentioned above, such systematics depend very much on the particular properties of the LSS tracer under study and, as we already discussed, the surveys simulated on this work do not pretend to reflect the particular properties of a given data set, but rather the global characteristics associated with the redshift distribution of the galaxy populations. Nevertheless, since the method relies on simulations to characterise the uncertainties of the recovered ISW map, the impact of the systematics of a given data set on the ISW detection, could be addressed as far as they could be simulated.

In addition, we have only considered the effect of instrumental noise when using CMB polarization information. However, the current release of Planck polarization data is significantly affected by systematics at large scales. Given that these are the scales relevant for the reconstruction of the ISW, the current Planck polarization maps are not suitable to perform this task. Fortunately, a new release of Planck data is planned for 2016, where this problem is expected to be largely palliated. Even in this case, some small level of residual systematics may be present in the final maps, and, if this were the case, the effect of these residuals in the reconstructed ISW should be quantified. This is important to ensure that the (moderate) improvement obtained by using polarization data is not hampered by the presence of systematic residuals.

Finally, we would like to point out that for a particular application, some consistency tests could be performed in order to test the validity of the considered model and the presence of systematics, such as comparing the recovered ISW power spectrum with the one expected for the considered cosmological models (similarly to Fig.~\ref{fig:cls_4comp}) or calculating the cross-correlation between the recovered ISW and the different galaxy surveys. Although small departures would be difficult to detect given the weakness of the signal and the presence of cosmic variance, a systematic departure with respect to the expected values would indicate the need to reconsider the model assumptions and/or to check for the presence of systematics in the data set.

\section{Conclusions}
\label{sec:con}

In this work we present an extension of the LCB filter \citep{BAR08} which was originally designed to reconstruct the ISW signal by combining CMB intensity data and one LSS tracer. On the one hand, the method is now able to deal with several large-scale structure tracers at the same time; on the other hand it makes use of CMB polarization data to reduce the cosmic variance.

The performance of the method is tested on coherent simulations whose fiducial angular power spectra are obtained with a modified version of CAMB. In addition to CMB intensity and polarization, we simulate three different surveys, whose galaxy redshift distribution peaks at low, intermediate and high-z and a lensing map. In particular, we obtain our results averaging over 10000 simulations.

Firstly we perform the reconstruction under the ideal all-sky and noiseless scenario, finding that the mean cross-correlation between the ISW signal and reconstruction is as large as 1.00 when all the information is included. The mean relative error is 0.04. For this ideal case, the main contribution to the reconstruction comes from exploiting the information of the LSS tracers, and including the CMB produces only a moderate improvement of the recovered ISW signal. The relative importance of the CMB increases when less information about the LSS is available. Regarding the role of each of the considered surveys, we find that the one peaked at high-z gives the major contribution to the reconstruction, since this catalogue covers a redshift range that is more suitable to extract ISW information than the other surveys. A similar level of performance to that of the high-z survey is also obtained when using the lensing map.

We have also studied the effect in the results of including instrumental Gaussian (for CMB polarization and lensing) and Poissonian (for LSS surveys) noise, at the level expected in current data sets. In this case, the quality of the ISW reconstruction is degraded, although not all data sets are equally affected. In particular, when all information is included,  we find an average correlation coefficient of 0.81 and a relative error of 0.58. Another interesting point is that the contribution of the CMB, both temperature and polarization, becomes more important when noise is included, mainly improving the recovery of the largest angular scales. The relative contribution of each of the LSS tracers in the ISW reconstruction changes with respect to the ideal case, due to the fact that we are considering different levels of noise for each survey. The level of noise has been chosen in order to cover a range of realistic noise amplitudes as well as to test the performance of the method. In this case the intermediate-z catalogue, which has been simulated with a lower level of Poissonian noise, becomes the one giving the major contribution to the ISW reconstruction.

Finally, we have considered the more realistic case of incomplete sky coverage. We find that introducing a mask degrades, but only slightly, the performance of the filter in those regions observed by all data sets, finding in the best case, a mean correlation coefficient between input and reconstruction of 0.98 for the noiseless case and of 0.79 when noise is present (versus 1.00 and 0.81 for the full-sky reconstruction). As one would expect, the quality of the reconstruction is degraded, by different amounts, on those areas where observations are not available for some of the data sets. 

We would also like to point out that this paper focuses in the performance of the proposed filter in different situations and how this is affected by noise or incomplete sky. The discussion regarding the contribution of each of the considered LSS tracers should be taken as an illustration of the method rather than as a conclusion on the optimal choice of catalogues for the ISW recovery. When working with real data, other non-idealities, such as the presence of systematics, can play an important role and affect the decision on which LSS tracers are best suited for that particular case.

Different extensions of the work presented here can be considered. In particular,
a further generalisation of the method would be its implementation in the real space, which would allow one to deal with all type of masks in a straightforward way, enabling the use of surveys with a relatively small sky coverage (e.g. the SDSS, \citealt{AHN12}). It is also interesting to point out that the ISW signal reconstructed with this method could be useful to study the large scale anomalies found in the CMB and its possible relationship with this secondary anisotropy. Finally, the redshift information that will be provided by future surveys, such as e.g. Euclid \citep{LAU11}, LSST \citep{LSS09} or J-PAS \citep{BEN14}, could be used to reconstruct the ISW signal in different redshift slices. These topics will be the subject of future works. 

\section*{Acknowledgements}
We acknowledge partial financial support from the Spanish \textit{Ministerio de Econom\'ia y Competitividad} Project AYA2012-39475-C02-01 and Consolider-Ingenio 2010 CSD2010-00064. The authors thank Enrique Mart{\'\i}nez-Gonz\'alez for useful discussions and Francesco Paci for providing an apodisation code. The HEALPix package~\citep{GOR05} was used throughout the data analysis.


\bsp

\label{lastpage}

\end{document}